\documentclass[a4paper,12pt]{article}

\usepackage[utf8]{inputenc}
\usepackage{fullpage}
\usepackage{graphicx}
\usepackage[usenames,dvipsnames]{xcolor}
\usepackage{amsmath}
\usepackage{amsthm}
\usepackage{amssymb}
\usepackage{breakcites}
\usepackage[left]{lineno}
\usepackage{blindtext}
\usepackage{subfig}
\usepackage{hyperref}
\usepackage{lineno}
\usepackage{url}
\usepackage{multicol}
\usepackage[affil-it, auth-lg]{authblk}

\theoremstyle{definition}
\newtheorem{definition}{Definition}

\theoremstyle{remark}

\newtheorem{exam}{Example}

\newcommand{\D}{\mathcal{D}}
\newcommand{\mG}{\mathcal{G}}
\newcommand{\mH}{\mathcal{H}}
\newcommand{\mP}{\mathcal{P}}
\newcommand{\mL}{\mathcal{L}}

\newcommand{\EG}{E_{\mathcal{G}}}
\newcommand{\EH}{E_{\mathcal{H}}}
\newcommand{\EP}{E_{\mathcal{P}}}

\newcommand{\tin}{\mathrm{in}} 
\newcommand{\out}{\mathrm{out}}

\newcommand{\inci}{\mathcal{I}^{\tin}}
\newcommand{\incip}{\mathcal{I}^{\tin}_{\mP}}
\newcommand{\inco}{\mathcal{I}^{\out}}
\newcommand{\incop}{\mathcal{I}^{\out}_{\mP}}

\newcommand{\RR}{\mathbb{R}}
\newcommand{\NN}{\mathbb{N}}

\newtheorem{prop}{\small\bf Proposition}
\newtheorem{thm}{\small\bf Theorem}
\newtheorem{rem}{\small\bf Remark}

\title{Pangraphs as models of higher-order \\
interactions}

\author[1]{Mateusz Iskrzyński\thanks{Correspondence: mateusz.iskrzynski@ibspan.waw.pl}}
\author[2]{Aleksandra Puchalska}
\author[1]{Aleksandra Grzelik}
\author[3]{G\"okhan Mutlu} 
\affil[1]{\normalsize Systems Research Institute, Polish Academy of Sciences, Warsaw, Poland}
\affil[2]{\normalsize Institute of Applied Mathematics and Mechanics, University of Warsaw, Warsaw, Poland}
\affil[3]{\normalsize Department of Mathematics, Gazi University, Ankara, Turkey}

\begin{document}

\maketitle

\begin{abstract}
Graphs depict pairwise relationships between objects within a system. Higher-order interactions (HOIs), which involve more than two objects simultaneously, are common in nature~\cite{GOLUBSKI2016344, Moleon_scavenging, Levine_2017, Mayfield2017}. Such interactions can change the stability of a complex system~\cite{Grilli2017}. Hypergraphs can represent an HOI as an arbitrary subset of vertices. However, they fail to capture the specific roles of the vertices involved, which can be highly asymmetric, particularly in the case of interaction modifications.

We introduce pangraphs, a robust and quantitative generalisation of graphs that accurately captures arbitrarily complex higher-order interactions. We demonstrate that several higher-order representations proposed in the literature are specific instances of pangraphs. Additionally, we introduce an incidence multilayer digraph representation of a pangraph, referred to as Levi digraph. We adapt degree and Katz centrality measures to the pangraph framework and show that a consistent generalisation of recursive graph measures cannot be simplified to a Levi digraph of a pangraph.

We construct a pangraph for a real-world coffee agroecosystem~\cite{GOLUBSKI2016344} and compare Katz centrality between its dihypergraph and pangraph representations, both analytically and numerically. The choice of representation significantly affects centrality values and alters vertex ranks. Additionally, we emphasise the use of real-valued incidence matrices to quantify interaction strengths and the roles of vertices within the system.

\end{abstract}

\tableofcontents

\section{Introduction and empirical motivation}

    A proper representation of the interaction structure within a complex system is crucial for its understanding. Graphs (networks), which represent the pairwise relationships between system components, have been instrumental in studying system stability, comparing the importance of various components, and identifying potential substructures (communities). They also enable simulations of disturbances that affect complex systems, as well as the study of their emergence and evolution.

     Higher-order interactions involving more than two components simultaneously are ubiquitous in nature~\cite{BATTISTON20201, hoi_book} and can significantly alter conclusions regarding how the entire ecosystem operates. Many of these interactions can modify other existing interactions. To address this,  we propose a mathematically consistent graph generalisation - called a pangraph - that faithfully captures the roles of all vertices and allows for arbitrarily nested interactions. We highlight the importance of real-valued incidence matrices and Levi graphs, which facilitate the easy and broad application of pangraphs to the problems discussed in the scientific literature. 

     A great variety of empirically observed higher-order interactions have been described in biology and chemistry, and as a result, we facilitate our examples from these fields. The simultaneous interactions of more than two species in the ecological competition were studied in~\cite{levine2017, letten_stouffer_non_additivity}. One of the most striking examples of non-pairwise interactions is lichens, which are symbiotic organisms such that pairings of individual species are non-exclusive~\cite {lichen_duran_nebreda}. The interactions of lichens require all species that make up the lichen organism. Similarly, many chemical reactions involve multiple substrates and  products~\cite{39_chemistry_Jost_2019}.

    Simultaneous relationships involving multiple vertices are typical in brain activity~\cite{37_brain_Giusti_2016, 38_brain_Faskowitz_2021} or climate observations~\cite{40_climate_Boers_2019, 41_climate_Su_2021}. The group character of the dynamics of real-world social interactions has a significant role. The possible impact of HOIs in game theory was studied in~\cite{53_Alvarez-Rodriguez_2020}.      

    Dynamical models offer the most literal examples of higher-order interactions. As bilinear terms in equations correspond to pairwise interactions, higher-degree polynomials correspond to higher-order interactions. These were included, for example, in the context of plant communities~\cite{Mayfield2017}. However, higher-order interactions can arise through more complex terms. In a system with more than one consumer of the same prey, their abundances collectively reduce the success rates of each consumer's hunts. These terms are known as Holling type II functional responses~\cite{Holling_1959} and were explicitly described as HOIs in predator-scavenger interactions~\cite{Getz_2011, Moleon_scavenging}.

    Network science discussed impacts of HOIs hypergraph representations on synchronization~\cite{42_Millán_2019, 43_Skardal_2019, 44_Zhang_2021, 45_Mulas_2020} and contagion dynamics~\cite{47_St-Onge_2021, 48_Arruda_2020, 49_Iacopini_2018, 50_Arruda_2020, 51_Sun_2021, 52_Taylor_2015}. The authors of~\cite{46_random_walk_Carletti_2020} studied random walk dynamics. In the random walk context, ~\cite{Rnd_walk_edge_vertex_weights} employed a concept of edge-dependent vertex weights, which can easily be realised by real-valued incidence matrices as proposed in this article.

    Including higher-order interactions in a model can change the core conclusions of a study, as illustrated in an analysis of system stability~\cite{Grilli2017}. Understanding the stability of complex systems is necessary to draw insights into their evolution and resilience. The complexity-stability debate~\cite{Landi2018} in theoretical ecology attempts to explain the apparent discrepancy between their mathematical models' often chaotic properties and the observed systems' slow changes~\cite{May1972}. This proves the high stakes of a proper representation of higher-order interactions. 

    Most of the existing studies~\cite{GOLUBSKI2016344, BATTISTON20201, hoi_book} represented higher-order interactions through hypergraphs and their particular cases, such as simplicial complexes. Hypergraphs allow an edge to be an arbitrary subset of the set of vertices. This representation underlines the necessity of the simultaneous presence of all vertices involved in an interaction for it to exist. A directed hypergraph (dihypergraph) enables us to distinguish the roles of influencers/sources and influenced/targets in an HOI.

    Many empirical higher-order interactions can be regarded as modifications of other interactions. These were studied in mutualistic networks~\cite{Vandermeer_2010, Perfecto_coffe_agroeco, GOLUBSKI2016344}, behavioural interactions like intimidation by possible predators~\cite{preisser_intimidation}, or inhibition of defensive strategies in a phototroph-predator microbial community~\cite{Mickalide_microbial}. Interactions between microbes may shape their vulnerability to antibiotics~\cite{bac_31_kelsic2015counteraction, bac_32_perlin2009protection, bac_33_abrudan2015socially}, leading even to four-way interactions~\cite{bac_34_reading1977clavulanic}. Chemical catalysts and inhibitors can influence reactions that do not change their amount. Thus, some studies have explicitly added HOIs as additional digraph vertices acting upon edges. They have observed the relevance of the HOIs on such network properties as percolation~\cite{Sun_2023_triadic_Bianconi}. 
    
    If a vertex modifies an interaction between two other vertices, then the three of them may have very asymmetric roles. It is indeed the case in a network model of a coffee agroecosystem~\cite{Vandermeer_2010, Perfecto_coffe_agroeco, GOLUBSKI2016344}. It represents the feeding and mutualistic interactions of organisms studied over 15 years in a 300-hectare organic coffee plantation in southern Mexico. These interdependencies were shown to be crucial in agricultural pest control, an economically important ecosystem service. Unlike the other two vertices, a vertex modifying a feeding relationship through a
behavioural interaction does not gain or lose matter. This
asymmetry is also more substantial than in an analogous case of chemical catalysis.

    We propose a graph generalisation that can represent causal and quantitative roles and relations in systems with arbitrarily nested higher-order interactions and call it a pangraph. Its basic premise is to extend the notion of an ordered set representing a directed edge using further nested ordered sets. It is based on the notion of an (unweighted) ubergraph proposed in a purely mathematical study~\cite{Joslyn2017_ubergraph}. We propose the name pangraph underlying its holistic character and to avoid unnecessary and tragic historical and political contexts~\footnote{Genocides relied on dividing humans into superior and inferior, historically abusing the word \emph{\"Ubermensch}.}.

    Pangraphs offer a consistent graph generalisation of directed weighted hypergraphs and can represent arbitrarily nested higher-order interactions. They enable a holistic description of ecosystems by tying together food webs~\cite{food_webs_pimm}, mutualistic networks~\cite{Bascompte_Mutualistic_networks}, and multilayer ecological networks~\cite{Pilosof_2015, Lurgi_2020, Hutchinson2019}. They contain edge-edge interactions, enabling a faithful representation of dynamics in ecological or transport networks. Pangraphs generalise Petri nets~\cite{Petri_thesis, Petri_Peterson_book, Baez_open_petri_2017} and metabolic graphs, both applied in chemical reaction network analysis.
    
The article is organised as follows. Section \ref{sec:graph_theo} briefly summarises notions of weighted, multilayer digraphs and weighted dihypergraphs (resp. Subsections \ref{sec:m_graph} and \ref{sec:H_graph}). The theoretical introduction to weighted pangraphs, including basic notions, is covered in Section~\ref{sec:pangraph}, and the definition of its Levi representation in Section~\ref{sec:Levi_pangraph}. We show how pangraphs arise in different kinds of dynamical models in Section ~\ref{sec:pan_dynamics}.
Next, in Section~\ref{sec:other_structures}, we discuss the relation of food webs, mutualistic networks, multilayer ecological networks, Petri nets, and metabolic graphs with pangraphs. In Section~\ref{sec:measures}, we generalise popular graph measures to the case of a pangraph. In Section \ref{sec:real_exam}, we compare the hypergraph and pangraph representations of a coffee agroecosystem. Final remarks can be found in Section \ref{sec:discussion}.

\section{Graph theory toolbox}\label{sec:graph_theo}

In this section, we introduce the notation and recall the definitions of  multilayer weighted digraphs and weighted dihypergraphs. We do not classically introduce them, but
in a form that allows their most natural generalisation.

\subsection{Weighted multilayer digraph}\label{sec:m_graph}

We start with a weighted digraph, see \protect\cite[Sec.~2.1]{KivArena2014} for the classical definition. Let us denote by $\mathbb{P}^*(\cdot)$ the power set (the family of all nonempty subsets of a given set)  and by $\mathbb{M}_{m\times n}(A)$ the space of $m\times n$ matrices having entries in the set $A$.

\begin{definition}
    \label{def:WeightedDigraph}
     \textbf{A weighted digraph} is a 4-tuple $\mG= (V,\EG, \inci_\mG,\inco_\mG)$ where
    \begin{enumerate}
        \item $V = \{v_i\; | \; i \in I\}$, $|I|=n$, is a set of vertices;
        \item $\EG = \{(e_k^{\tin},e_k^{\out})\; | \; k \in K\}\subseteq \mathbb{P}^*(V) \times \mathbb{P}^*(V)$, $|K|=m$, is a set of directed edges such that each $e=(e^{\tin},e^{\out})\in \EG$ satisfies
            \begin{equation}\label{eq:edge}
                |e^{\tin}|=|e^{\out}|=1;
            \end{equation}
        \item $\inci_{\mG}=((\inci_{\mG})_{ij})_{i\in K,j\in I}, \hspace{0.3cm} \inco_{\mG}=((\inco_{\mG})_{ij})_{i\in K,j\in I}\in \mathbb{M}_{m\times n}([0,\infty))$ are respectively incoming and outgoing incidence matrices, which encode the weights of edges in the following way:
        \begin{itemize}
            \item $(\mathcal{I}^{\mathrm{x}}_{\mG})_{ij}>0$ is the weight of a directed edge satisfying the condition $v_{j} \in e_{i}^{\mathrm{x}}$, for $\mathrm{x}=\mathrm{in},\mathrm{out}$, 
            \item $(\mathcal{I}^{\mathrm{x}}_{\mG})_{ij}=0$ informs that $v_{j} \notin e_{i}^{\mathrm{x}}$, for $\mathrm{x}=\mathrm{in},\mathrm{out}$.
        \end{itemize}
    \end{enumerate} 
\end{definition}

We note that compared to the classical definition of the weighted digraph, in this approach, an edge can have two weights, each associated with one end. In such a situation the following condition does not have to hold
\begin{equation}\label{eq:multi-weights}
(\inci_{\mG})_{ki}= (\inco_{\mG})_{kj},\qquad \text{for } e_k=(\{v_i\},\{v_j\})\in \EG.
\end{equation}
Additional weights will appear naturally when considering non-trophic interactions in ecosystems, see Section \ref{subsec:mutualistic_nets}, or chemical reactions, see Section \ref{subsec:petri_nets}. In order to define the relation between  vertices in a network, we need also  unweighted incidence matrices $\overline{\mathcal{I}}_{\mG}^{\mathrm{x}}=((\overline{\mathcal{I}}_{\mG}^{\mathrm{x}})_{ij})_{i\in K,j\in I},\, \in \mathbb{M}_{m\times n}(\{0,1\})$, for $\mathrm{x}=\mathrm{in},\mathrm{out}$, which translate the non-zero entries of incidence matrices to ones, namely
\begin{equation*}
(\overline{\mathcal{I}}_{\mG}^{\mathrm{x}})_{ij}=1 \qquad \text{iff }\quad (\mathcal{I}_{\mG}^{\mathrm{x}})_{ij}\neq 0;\qquad \text{for }\mathrm{x}=\mathrm{in},\mathrm{out}.
\end{equation*}
We say that a digraph is \textbf{simple} if it has no multiple edges, namely there are no edges $e_j,e_k\in E_{\mG}$ such that $e_j^{\textrm{x}}=e_k^{\textrm{x}}$, for $x=\tin,\out$. In this article, we consider only simple networks, but for the sake of simplicity, we call them digraphs.  Following \protect\cite[Eq.~2.7]{Mugnolo2013}, we define \textbf{in} and \textbf{out weighted adjacency matrices} $\mathcal{A}_{\mG}^{\tin},\mathcal{A}_{\mG}^{\textrm{out}} \in \mathbb{M}_{n\times n}([0,\infty)) $  by
\begin{equation}\label{eq:adj_and_inc}
\mathcal{A}_{\mG}^{\tin}=(\overline{\mathcal{I}}^{\textrm{out}}_{\mG})^T \inci_{\mG}\qquad \qquad \mathcal{A}_{\mG}^{\out}=(\overline{\mathcal{I}}^{\textrm{in}}_{\mG})^T \inco_{\mG},
\end{equation}
respectively. We note that $(\mathcal{A}_{\mG}^{\tin})_{ij}$ describes the weight of an edge $e_k=(\{v_j\},\{v_i\})$ associated with vertex $v_j$, 
whereas $(\mathcal{A}_{\mG}^{\out})_{ij}$ describes the weight associated with vertex $v_i$. 

More precisely, as digraph is simple, for $\mathcal{A}_{\mG}^{\textrm{x}}=((\mathcal{A}_{\mG}^{\textrm{x}})_{ij})_{i,j\in I}$, $\textrm{x}=\textrm{in},\textrm{out}$ we have
 \begin{eqnarray}\label{eq:adjacency-in}
(\mathcal{A}^{\tin}_{\mG})_{ij}&=&\left\{\begin{array}{cc}
(\inci_{\mG})_{kj}, &\text{if there exists } e_k=(\{v_j\}, \{v_i\})\in \EG\\
0,&\text{otherwise}\end{array}\right.\\ \label{eq:adjacency-out}
(\mathcal{A}^{\out}_{\mG})_{ij}&=&\left\{\begin{array}{cc}
(\inco_{\mG})_{ki}, &\text{if there exists }e_k=(\{v_{j}\}, \{v_{i}\})\in \EG\\
0, &\text{otherwise}\end{array}\right. 
\end{eqnarray}
Furthermore, if \eqref{eq:multi-weights} holds, then $\mathcal{A}_{\mG}^{\textrm{out}}=\left(\mathcal{A}_{\mG}^{\textrm{in}}\right)^T$. In this article, we sometimes use the notation $\mathcal{A}^{\textrm{x}}_{\mG}$, $\textrm{x}=\textrm{in},\textrm{out}$, to underline that adjacency matrix comes from a digraph $\mG$; but we omit the subscript when it does not bring ambiguity.

A \textbf{walk} $W$ of length $l_{W}$ in an unweighted digraph $\mG=(V,\EG)$ is a sequence $W=(w_0,e_1,w_1,\ldots e_{l_W-1},w_{l_{W}})$ such that $w_i\in V$, for $i=0,\ldots,l_{W}$; $e_k\in E_{\mG}$ for $k=1,\ldots,l_W-1$ and 
    \begin{equation*}
e_{s}^{\tin}=\{w_{s-1}\},\quad e_{s}^{\out}=\{w_{s+1}\},\quad \text{for }s=1,\ldots,l_W-1.
\end{equation*}

We say that a digraph is \textbf{d-partite} if there exists a partition of the set $V$ into $d$ subsets $V_1, V_2, \ldots, V_d$ such that for any $e=(\{v_i\},\{v_j\})$, $d'\in \{1,\ldots,d\}$ and $p\in \{i,j\}$, $q\in \{i,j\}\setminus \{p\}$, we have
\begin{equation}\label{partition}
v_p\in V_{d'}\quad \textrm{and} \quad v_q\notin V_{d'}.
\end{equation}

 Let us define the vertex \textbf{in-- and out-- degrees} $\kappa^{\textrm{x}}:V\to \mathbb{R}$, $\mathrm{x}=\mathrm{in},\mathrm{out}$, using both incidence and adjacency matrices. In order to underline its connection to a digraph, we sometimes denote it by $\kappa_{\mG}^{\textrm{x}}$. For any $v_i\in V$ we define
\begin{equation}\label{eq:degree}
\kappa^{\textrm{x}}_{\mG}(v_i)=\sum_{e_j\in E_{\mG}}(\mathcal{I}_{\mG}^{\textrm{x}})_{ji}=\sum_{v_j\in V}(\mathcal{A}^{\textrm{x}}_{\mG})_{ji},\qquad \mathrm{x}=\mathrm{in},\mathrm{out}.
\end{equation}
Finally, let us transform a digraph into a multilayer network by adding layers.

\begin{definition}\label{def:multi_layer_dirgaph}
  Let $\mathcal{G}=(V, E_{\mG},\inci_{\mG},\inco_{\mG})$ be a weighted digraph and $L=\{L_j\subset V\; | \; j \in J\}$ be a family of disjoint subsets of $V$. We say that the graph is \textbf{multilayer} and we denote it by $\mG_L= (V,L, E_{\mG_L},\inci_{\mG_L},\inco_{\mG_L})$. Moreover, we call $L$ a \textbf{set of layers} of cardinality $d=|J|$ and $l_j=|L_j|$ the number of vertices in layer $L_j$, $j\in J$.
 
\end{definition}

Note that no vertex can belong to more than one layer in this approach. According to the standard nomenclature, it is a special type of multilayer digraph called \textbf{layer-disjoint}. It is also equivalent to explicitly treating the instances of a vertex in different layers as different vertices in any multilayer digraph. If the graph in question consists of just one layer, we call it a weighted digraph.
Finally, if the condition \eqref{partition} is satisfied for each layer in a multilayer digraph, then we say that the multilayer digraph is \textbf{$d$-partite with respect to the layers}.

\subsection{Weighted dihypergraph}\label{sec:H_graph}
One can also allow each edge to have more than one head and/or more than one tail. 
\begin{definition}\label{def:dihypergraph}
We say that  $\mH = (V,\EH,\inci_\mH,\inco_\mH)$ is a \textbf{weighted dihypergraph} if it satisfies all conditions of Definition \ref{def:WeightedDigraph} modifying \eqref{eq:edge} with 
\begin{equation}\label{eq:edge2}
        |e^{\tin}|,|e^{\out}|\geq 1.
        \end{equation}
\end{definition}
We note that the incidence matrices in Def.~\ref{def:WeightedDigraph} generalise naturally to dihypergraphs. In order to distinguish elements from a set $\EG$ from those in $\EH$, we call the latter \textbf{hyperedges}. This notion has already been used in the ecological context in the undirected version, see \protect\cite{Bretto2013}. Similarly to the digraph case, a weight is associated with each vertex in a hyperedge. So, the weight of an edge is a vector, but this time, their dimension is not fixed. Namely, for $e_k=(e_k^{\mathrm{in}},e_k^{\mathrm{out}}) \in \EH$ there are $|e_k^{\mathrm{in}}|+|e_k^{\mathrm{out}}|$ weights.

 We say that a hyperedge $e=(e^{\tin}, e^{\out})$ is a hyperloop if $e^{\tin}\cap e^{\out}\neq \emptyset$. In particular, a hyperedge is called a loop if $e^{\tin}=e^{\out}$. 
 
 Unlike in a digraph, there may be several different hyperedges all containing the same vertex as a head and/or a tail. Consequently, defining the adjacency matrices in analogy to equation \eqref{eq:adj_and_inc} by
 \begin{equation}\label{eq:adj_and_inc2}
\mathcal{A}_{\mH}^{\tin}=(\overline{\mathcal{I}}^{\textrm{out}}_{\mH})^T \inci_{\mH} \qquad \textrm{and}\qquad \mathcal{A}_{\mH}^{\out}=(\overline{\mathcal{I}}^{\textrm{in}}_{\mH})^T \inco_{\mH},
\end{equation}
the actual formulas differ compared to \eqref{eq:adjacency-in} -- \eqref{eq:adjacency-out}. Namely, we have
\begin{eqnarray}\label{eq:adjacency-in-out2}
(\mathcal{A}_{\mH}^{\tin})_{ij}=\sum_{\{e_k \in E_{\mH}:\,\, v_j\in e_k^{\out}\}}
(\inci_{\mH})_{ki},\qquad (\mathcal{A}_{\mH}^{\out})_{ij}=\sum_{\{e_k \in E_{\mH}:\,\, v_j\in e_k^{\tin}\}}(\inco_{\mH})_{kj}. 
\end{eqnarray}

This time there is no one-to-one correspondence between a dihypergraph and its adjacency matrix.
The hypergraph literature contains proposals for other generalisations of adjacency matrix, e.g., an adjacency tensor \protect\cite{Michoel2012}, a degree-normalized k-adjacency tensor \protect\cite{COOPER2012}, an eigenvalues normalized k-adjacency tensor \protect\cite{Hu2013}, etc.

Unlike equation \eqref{eq:degree}, when defining vertex degree for a dihypergraph it makes a difference whether we calculate the weights of all hyperedges incident to a fixed vertex $v$, or we calculate the weights of all edges such that fixed vertex $v$ is one of its beginnings. In the second case, one calculates several times the weights of these edges that start at $v$ and terminate in more than one vertex. Consequently, we obtain \textbf{in--} and \textbf{out-- incidence and adjacency vertex degrees}, defined by the the formulas for any $v_i\in V$

\begin{equation} 
    \kappa_{\mH, I}^{\mathrm{x}}(v_i):=\sum_{e_j\in E_{\mH}}(\mathcal{I}^{\textrm{x}}_{\mH})_{ji},\qquad \kappa_{\mH,A}^{\textrm{x}}(v_i):=\sum_{v_k\in V}(\mathcal{A}^{\textrm{x}}_{\mH})_{ki}\qquad x=\textrm{in},\textrm{out},
\end{equation}
respecively.

\section{Introduction to weighted pangraphs}\label{sec:pangraph}

 In this section, we present a generalisation of a dihypergraph, called a pangraph, and characterise its basic properties. Unlike results from the literature, see \protect\cite{Joslyn2017_ubergraph}, we propose a new network structure to map directed and weighted relations between objects.
 
\subsection{Main definitions} 

The major novelty in the new approach that we propose, compared to the standard dihypergraph definition, is to allow an edge to connect any number of objects that can be vertices as well as other edges. The new recursive definition of sets $e^{\tin}$, $e^{\out}$ in Def. \ref{def:WeightedDigraph} creates a pangraph. The object of this kind first appeared under the name ubergraph in \protect\cite{Joslyn2017_ubergraph}. We generalise it by adding directions and weights and making the definition more coherent. 

Given a set of vertices $V$, we introduce a sequence $(P_k(V))_{k\in \NN}$\footnote{We correct the definition of \protect\cite{Joslyn2017_ubergraph} where $P_0$ was inconsistent with the recursive formula and add directions and weights of panedges.} recursively 
\begin{equation}
       P_0(V) = V, \quad P_{k}(V) =\mathbb{P}^* \left( \bigcup_{i=0}^{k-1} P_i(V)\right) \times \mathbb{P}^* \left(\bigcup_{i=0}^{k-1} P_i(V)\right) , \quad k\geq 1.
\end{equation}
Hence, $P_k(V)$ is a set of all ordered pairs such that each element of a pair is a subset of $ \bigcup_{i=0}^{k-1} P_i(V)$. The power set of any set is contained in a power set of its superset, consequently, for $k\geq 1$, it follows 
\begin{align*}
    P_k(V)&=\mathbb{P}^* \left( \bigcup_{i=0}^{k-1} P_i(V)\right) \times \mathbb{P}^* \left(\bigcup_{i=0}^{k-1} P_i(V)\right) \\
    &\subset \mathbb{P}^* \left( \bigcup_{i=0}^{k-1} P_i(V) \cup P_k(V) \right) \times \mathbb{P}^* \left(\bigcup_{i=0}^{k-1} P_i(V) \cup P_k(V) \right)=P_{k+1}(V).
\end{align*}

We start by defining an unweighted pangraph. 

\begin{definition} 
\label{def:weighted_pangraph}
\textbf{An unweighted $k_0$-depth pangraph}, for $k_0\geq 1$, is a pair $\mP= (V,\EP)$, where
    \begin{enumerate}
        \item $V = \{v_i\; | \; i \in I\}$, $|I|=n$, is a set of \textbf{fundamental vertices};
        \item $E_{\mP} = \{e_j =(e_{j}^{\tin},e_{j}^{\out}) \; | \; j \in K\}\subseteq P_{k_0}(V)$, is a set of \textbf{panedges} such that heads and tails of all panedges consist of fundamental vertices or other panedges, namely
        \begin{equation}\label{eq:uEdge}
          (e^{\tin},\, e^{\out}) \in \EP \quad \Rightarrow \quad e^{\tin},\, e^{\out}  \subset V \cup \EP.
        \end{equation} 
    \end{enumerate} 
\end{definition} 

Condition \eqref{eq:uEdge} ensures that heads and tails of any element from $E_{\mP}$ are included either in the set of panedges $E_{\mP}$ or in the set of fundamental vertices $V$. Let us consider an example which visualises the construction of panedges.

\begin{figure}
    \centering
    \includegraphics[width=0.7\linewidth]{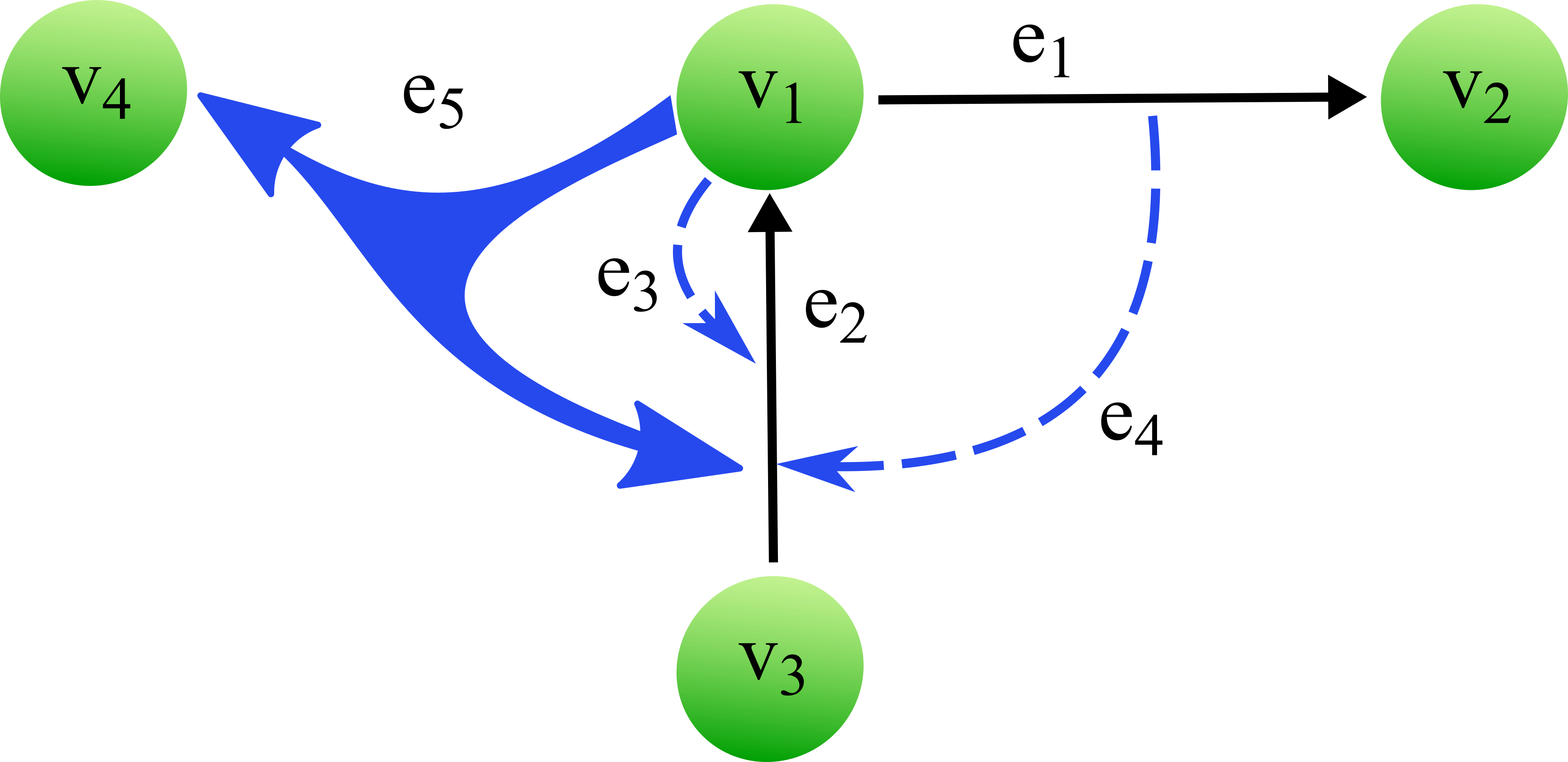}
    \caption{Pangraph in Example 1.}\label{fig:example_1}
\end{figure}

\begin{exam}\label{exam:pangr}
Consider $\mP=(V,\EP)$ such that $V=\{v_1,v_2,v_3,v_4\}$ is a set of fundamental vertices, 
\begin{eqnarray*}
E_{\mP}=\left\{ e_1 = (\{v_1\},\{v_2\}), e_2 = (\{v_3\},\{v_1\} ), e_3 =(\{v_1\}, \{e_2\} ), e_4 =( \{e_1\} , \{e_2\} ),\right. \\
\left.e_5 = (\{v_1\},\{v_4,e_2\}),e_6=(\{v_1\}, \{(\{v_2\},\{v_3\})\}\right\}
\end{eqnarray*} 
is a set of panedges.
Notice that in this case, we have
$$
e_1, e_2 \in P_1(V),\qquad e_3, e_4, e_5, e_6\in P_2(V). 
$$
Furthermore $\{v_1\},\{v_2\},\{v_3\},\{v_4\}, e_1 , e_2, \{v_4,e_2\} \in V \cup \EP $, but $ (\{v_2\},\{v_3\})\notin V \cup \EP $ hence $\mP$ is not a pangraph. Considering $\mP'=(V,\EP \setminus \{e_6\})$, $\mathcal{P}'$ is a $2$-depth pangraph. Pangraph $\mathcal{P}'$ is  presented in Fig.~\ref{fig:example_1}. 
\end{exam}

Based on Definition \ref{def:weighted_pangraph}, we easily note that a $1$-depth pangraph is a dihypergraph. All hyperedges can only contain vertices; hence, condition \eqref{eq:uEdge} is always satisfied. A digraph is obviously also a $1$-depth pangraph.\newline

Since the family $(P_k(V))_{k\geq 1}$ is ascending, for any $e_j\in \EP$ in a $k_0-$depth pangraph $\mP$, there exists a parameter $\D(e_j)$ such that
\begin{equation}
        \mathcal{D}(e_j)=\text{min}\,\{k \in \mathbb{N}\;|\; e_j \in P_{k}(V)\}\leq k_0.
    \end{equation}
    
We call $\D(e_j)$ the \textbf{depth of the panedge} $e_j$ and interpret this parameter as the nestedness of an edge in a pangraph. The depth of a panedge depends only on the largest depth among its constituents, hence can be defined recursively for any $(e^{\tin},\, e^{\out})\in E_{\mP}$:
    \begin{equation}\label{eq:recursive_depth}
        D((e^{\tin},\, e^{\out}))=\text{max}(\{D(w)\;|\; w \in e^{\tin}\cup e^{\out} \} )+1, \qquad \text{assuming}\,\, D(w)=0,\,\, w\in V. 
    \end{equation}
To characterise the order of interaction, one needs to count the number of all fundamental vertices involved in defining the panedge. The roles of influencers and targets should also be clearly described in an interaction modification. Furthermore, one vertex can appear multiple times in a panedge. We define panedge's \textbf{order} $o(e)$ (as well as incoming $o^{\tin}(e)$, and outgoing $o^{\out}(e)$ order) that counts each vertex once, and a \textbf{weighted order} $o_w(e)$ (as well as weighted incoming $o^{\tin}_w(e)$, and weighted outgoing $o^{\out}_w(e)$ order) \textbf{of panedge} $e\in E_\mP$ that counts all occurrences of vertices.

In order to give a formal definition of the order of a given $e\in E_\mP$ of $D(e)$-depth, we recursively define the families of sets  
$\mathcal{E}^\textrm{x}_j(e), \mathcal{E}_j(e)$ and $\mathcal{V}^\textrm{x}_j(e),\mathcal{V}_j(e)$, $j=0,\ldots,$ $D(e)$, $\textrm{x}=\tin,\out$: 
\begin{eqnarray}\nonumber
\mathcal{E}_j^\textrm{x}(e)&:=&\bigcup_{e_i\in \mathcal{E}_{j-1}\setminus V}e_i^\textrm{x},\qquad \mathcal{E}_j(e):=\bigcup_{s=\tin,\out}\mathcal{E}_j^s(e),\qquad \mathcal{E}^\textrm{x}_0(e)=\mathcal{E}_0(e)=\{e\},\\ \label{eq:v}
\mathcal{V}^\textrm{x}_j(e)&:=&\mathcal{E}^\textrm{x}_j(e)\cap V,\qquad \mathcal{V}_j(e):=\mathcal{E}_j(e)\cap V,\qquad \mathcal{V}^\textrm{x}_0(e)=\mathcal{V}_0(e)=\emptyset.
\end{eqnarray}

The order and weighted order of a panedge $e$ are given respectively by
\begin{equation}\label{eq:o^x}
o^\textrm{y}(e):=\left|\bigcup_{j=1}^{D(e)} \mathcal{V}^\textrm{y}_j(e)\right|,\qquad o^\textrm{y}_w(e)=\sum_{j=1}^{D(e)} |\mathcal{V}_j^\textrm{y}(e)|,\qquad \text{for}\,\,\textrm{y}=\tin,\out,\emptyset.
\end{equation}
Note that incoming and outgoing order counts only the fundamental vertices that are used to be a head/tail at each step of an algorithm. Consequently, it is likely that the following relations hold
\begin{equation}\label{eq:o_neq}
o^{\tin}(e)+o^{\out}(e)\neq o(e),\quad  \textrm{whereas} \quad o_w^{\tin}(e)+o_w^{\out}(e)= o(e).
\end{equation}

For later use, we also define the set of all vertices appearing in the heads/tails of a panedge's subcomponents:
\begin{equation}\label{def:all_tails_heads}
   \mathcal{V}^{\textrm{y}}(e)=\bigcup_{j=1}^{D(e)} \mathcal{V}^\textrm{y}_j(e), \quad \textrm{for y}=\tin, \out,\emptyset.
\end{equation}

Let us return to pangraph $\mP'$ in Example \ref{exam:pangr}. We note that the edge $e_5$ has depth-2. Indeed, it follows 
 \begin{equation}
 D(e_5)=\max\left(D(v_1),D(v_4),D(e_2)\right)+1=D(e_2)+1=2.
 \end{equation}
Based on \eqref{eq:o^x}, one can calculate the order and weighted order of this edge. Let us start with the outgoing order.
\begin{equation*}
\mathcal{E}_0^{\textrm{out}}=\{e_5\},\quad \mathcal{E}_1^{\textrm{out}}=\{v_4,e_2\},\quad \mathcal{E}_2^{\textrm{out}}=\{v_1\}; \qquad \mathcal{V}_0^{\textrm{out}}=\emptyset, \quad \mathcal{V}_1^{\textrm{out}}=\{v_4\},\quad \mathcal{V}_2^{\textrm{out}}=\{v_1\}.
\end{equation*}
Consequently, $\mathcal{V}^{\textrm{out}}(e_5)=\{v_1,v_4\}$ and $o^{\textrm{out}}(e_5)=2$. Repeating this calculation for incoming order and order, one obtains
\begin{equation*}
\mathcal{V}^{\textrm{in}}(e_5)=\{v_1, v_3 \},\quad  \mathcal{V}(e_5)=\{v_1,v_3, v_4\}\qquad \textrm{and}\qquad o^{\textrm{in}}(e_5)=2,\quad o(e_5)=3. 
\end{equation*}
We note that the first condition in \eqref{eq:o_neq} holds for $e_5$. 

The principal novelty in the concept of pangraph is that the panedges that become heads
or tails of other panedges begin to play a role akin to vertices. They influence other objects
or vice versa. At the same time, panedges not included as a head
or tail are carriers of such influences.  This leads us to define the set of \textbf{generalized vertices} being a subset of $V\cup E_{\mP}$. Namely,
\begin{eqnarray}\label{eq:U_ver}
V_{\mP}&:=&V\cup E_{\mP}\setminus \left\{v\in E_{\mP}\;|\; v\notin e^{\mathrm{out}}\cup e^{\mathrm{in}}, \quad \forall e=(e^{\tin},e^{\out})\in E_{\mP} \right\}\\ \nonumber
&=&\{v_i\;|\; i\in I'\},\qquad |I'|=n'.
\end{eqnarray}

Obviously, $V\subseteq V_{\mP}\subseteq V\cup E_{\mathcal{P}}$. For a pangraph that is a dihypergraph, $\mP=\mH$, the sets of fundamental vertices and generalized vertices coincide $V=V_{\mP}$. 

\begin{definition}\label{df:pan_izo} We say that two pangraphs $\mP^A=(V^A,\EP^A)$ and $\mP^B=(V^B,\EP^B)$ are \textbf{isomorphic} if there exists an isomorphism $f:V_{\mP}^{A} \to V_{\mP}^{B}$, for $V_{\mP}^{A}, V_{\mP}^{B}$ defined in \eqref{eq:U_ver}, such that 
\begin{equation*}
(f(v),f(w))\in \EP^B,\,\, \text{for every }(v,w)\in \EP^A.
\end{equation*}
\end{definition}

We can now define a weighted pangraph.

\begin{definition}
\textbf{A weighted $k_0$-depth pangraph}, for $k_0\geq 1$, is a 4-tuple $\mP= (V,\EP, \incip,\incop)$ where $ (V,\EP)$ is an unweighted $k_0$-depth pangraph and 
\begin{equation}
\incip=(({\incip})_{ij})_{i\in K,j\in I'},\quad \incop=(({\incop})_{ij})_{i\in K,j\in I'}\in \mathbb{M}_{m\times n'}([0,\infty))
\end{equation}
are respectively incoming and outgoing incidence matrices. 
\end{definition}
The incidence matrices of a pangraph generalise the ones in Definition \ref{def:WeightedDigraph}, by changing the set of vertices $V$ enumerated by $I$ into the set of generalized vertices $V_{\mP}$ enumerated by $I'$. Each panedge has exactly $|e_i^{\tin}\cup e_i^{\out}|$ weights, similar to the dihypergraph case. Introducing weights through incidence matrices will become more natural after we introduce a Levi digraph representation of a pangraph (Subsection \ref{sec:Levi_pangraph}).

We define the adjacency matrices of a pangraph in analogy to dihypergraphs (see Eq.~\eqref{eq:adj_and_inc}),
\begin{equation}\label{eq:adj_and_inc3}
\mathcal{A}_{\mP}^{\tin}=(\overline{\mathcal{I}}^{\textrm{out}}_{\mP})^T \inci_{\mP} ,\qquad \mathcal{A}_{\mP}^{\out}=(\overline{\mathcal{I}}^{\textrm{in}}_{\mP})^T \inco_{\mP}.
\end{equation}

Pangraph adjacency matrix does not contain complete information about its structure, just as in the case of dihypergraphs. In further consideration, we omit the subscript in the notation of the adjacency matrices of a pangraph when it does not cause ambiguity.

\subsection{Incidence multilayer digraph representation of a weighted pangraph}\label{sec:Levi_pangraph}

 The above considerations show that the concept of a pangraph is a natural generalisation of a dihypergraph. In this section, we show that a pangraph can be represented through its incidence (Levi~\protect\cite{Levi}) digraph, just like graphs and hypergraphs. In order to visualise the transformation of a pangraph into its incidence digraph, let us imagine that we add a vertex in the middle of every panedge to represent it.
  
\begin{definition}\label{def:levi-graph}
    A \textbf{digraph incidence representation (or a Levi digraph)} of a $k_0$-depth pangraph $\mathcal{P}=(V, \EP,\incip,\incop )$ is a multilayer digraph $\mL(\mathcal{P})=(V_{\mL}, L, E_{\mathcal{L}},\inci_{\mathcal{L}}, \inco_{\mathcal{L}})$, where
    \begin{enumerate}
    \item $V_{\mL}=V\cup \EP=\{v_i\;|\; i\in I_{\mL}\}$,
        \item $L = \{L_j\;|\;j\in J_{\mL}\}$, $J_{\mL}=\{1,\ldots,k_0+1\}$, is a partition of the set $V \cup E_\mathcal{P}$ such that  
        \begin{equation}
            L_j = \{v \in V \cup E_\mathcal{P}: D(v) = j-1\}, \quad j\in J_{\mL}\label{eq:wartswyL}
        \end{equation}
        called layers,
        \item the set of edges is given by 
        \begin{equation}
        E_{\mathcal{L}}=\{e_k \; | \; k \in K_\mL\}=\{(\{v_i\},\{v_j\})\in V_\mL \times V_\mL |\ v_i \in v_j^{\mathrm{in}} \text{or}\,\, v_j \in v_i^{\mathrm{out}},\,i,j\in I_\mL\},
        \end{equation}
        \item $\inci_{\mL}=({\inci_\mL}_{ij})_{i\in K_\mL, j \in I_\mL}, \ \  \inco_{\mL}=({\inco_{\mL}}_{ij})_{i\in K_\mL, j \in I_\mL}$ are weighted in- and out- incidence matrices satisfying conditions: 
        \begin{eqnarray*} 
        (\mathcal{I}_{\mL}^\tin)_{ki}=(\mathcal{I}_{\mL}^\out)_{kj}&:=&(\mathcal{I}_{\mP}^\tin)_{ji}, \quad  e_k=(\{v_i\},\{v_j\})\in E_{\mathcal{L}}\quad , v_{i} \in v_{j}^{\tin},\\
        (\mathcal{I}_{\mL}^\tin)_{ki}=(\mathcal{I}_{\mL}^\out)_{kj}&:=&(\mathcal{I}_{\mP}^{\out})_{ij}, \quad e_k=(\{v_i\},\{v_j\})\in E_{\mathcal{L}}\quad , v_{j} \in v_{i}^{\out}.
        \end{eqnarray*}
    \end{enumerate} 
\end{definition}

The incidence representation of a pangraph is well-defined. Namely, for every pangraph, there exists a unique associated Levi digraph. Furthermore, considering $\mP=\mH$, we obtain the classical definition of Levi representation for a dihypergraph, see~\protect\cite{Levi}. 

Note that for a dihypergraph $\mH= (V,\EH,\inci,\inco)$, its Levi representation is a bipartite graph such that the existence of an edge $e=({v_i},{e_j})$  (resp. $e=({e_j},{v_i})$), $v_i\in V$, $e_i\in \EH$ in a Levi graph, signifies being included in an edge $v_i\in e_j^{\tin}$ (resp. $v_i\in e_j^{\out}$) in the original dihypergraph. However, Levi representation of a $k_0$-depth pangraph has $k_0+1$ layers, and it is bipartite only in the case $k_0=1$. 

The classical interpretation of the directions of the Levi graph's edges is modified, and the information being included in an edge in the original pangraph also depends on the layer of its Levi digraph. The directions of edges in Levi representation presented in Definition \ref{def:levi-graph} allow us to conserve the direction from cause to effect in the original pangraph, which is crucial for further applications.

\begin{figure}
    \centering
    \includegraphics[width=0.7\linewidth]{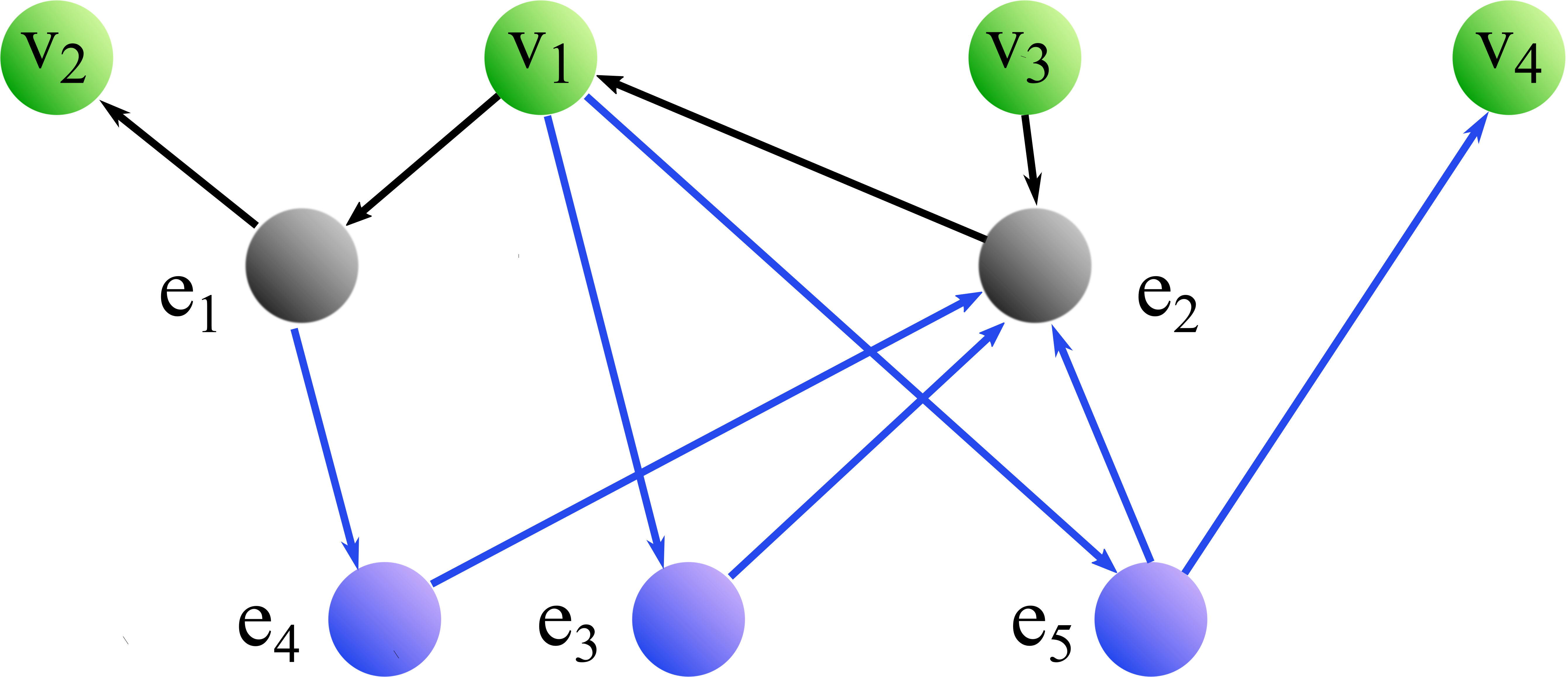}
    \caption{A Levi graph of the pangraph from Example 1. Fundamental vertices (layer $L_1$) are shown in green, edges in layer $L_2$ in grey, and edges in layer $L_3$ in blue.}\label{fig:Levi_example_1}
\end{figure}

\begin{exam}
    Consider a Levi graph of the 2-depth unweighted pangraph defined in Example \ref{exam:pangr}. Vertices are given by $V_{\mathcal{L}}=\{v_1,\ldots,v_4,e_1,\ldots,e_5\}$ while layers $L=\{L_1,L_2,L_3\}$ such that $L_1=\{v_1,\ldots, v_4\},$ $L_2=\{e_1,e_2\},$ $L_3=\{e_3,e_4, e_5\}$. Finally, edges are given by
\begin{eqnarray*}
E_{\mathcal{L}}&=&\left\{(\{v_1\},\{e_1\}), (\{e_1\},\{v_2\}), (\{v_3\},\{e_2\}),(\{e_2\},\{v_1\}), (\{v_1\},\{e_3\}), (\{e_3\},\{e_2\}),\right.\\
&&\left.(\{e_1\},\{e_4\}),(\{e_4\},\{e_2\}) , (\{v_1\},\{e_5\}), (\{e_5\},\{e_2\}), (\{e_5\},\{v_4\}) \right\}.
\end{eqnarray*}
This multilayer digraph is shown in Fig.~\ref{fig:Levi_example_1}.
\end{exam}

What are the properties of the Levi digraph of a pangraph? Which multilayer digraphs can be Levi digraphs of some pangraphs? To answer the last question, let us represent the adjacency matrix $\mathcal{A}^{\tin}=(\mathcal{A}^{\tin}_{ij})_{i,j\in I}$ of a weighted multilayer digraph satisfying \eqref{eq:multi-weights}
in the block-matrix form: 
\begin{equation}\label{eq:adj_block}
\mathcal{A}^{\tin}=(\mathbb{A}_{ij})_{i,j\in J_{\mL}},
\end{equation}
where each block $\mathbb{A}_{ij}$, for some $i,j\in J_{\mL}$, represents the weights of edges having heads in the layer $L_i$ and tails in $L_j$.  

\begin{thm}\label{thm:Levi_vs_pan}
A weighted multilayer digraph $\mG_L$ without multiple edges, see Definitions \ref{def:WeightedDigraph}--\ref{def:multi_layer_dirgaph}, satisfying \eqref{eq:multi-weights}, is an incidence representation of a $k_0$-depth pangraph if and only if there exists a permutation of layers such that the adjacency matrix $\mathcal{A}^{\tin}=(\mathcal{A}^{\tin}_{ij})_{ij\in I}$ of digraph $\mG_L$, presented in a block form in \eqref{eq:adj_block}, satisfies the following conditions:
\begin{itemize}

\item[i)] $\mathbb{A}_{ii}$ is a zero matrix for any $i\in J_{\mL}$;
\item[ii)] for every $v_j\in L_l$, $l\in L_{\mL}\setminus \{1\}$, there exists a vertex $v_i\in L_{l'}$, for $l'\in L_{\mL}$, $l'<l$ such that $\mathcal{A}_{ij}^{\tin}\neq 0;$
\item[iii)] for every $v_j\in L_l$, $l\in L_{\mL}\setminus \{1\}$, there exists a vertex $v_i\in L_{l'}$, for $l'\in L_{\mL}$ $l'<l$ such that $\mathcal{A}_{ji}^{\tin}\neq 0.$
\end{itemize}  
\end{thm}

If for a multilayer digraph $\mathcal{G}_L$ there exist a pangraph $\mP$ stated in Theorem \eqref{thm:Levi_vs_pan}, then we call $\mP$ a \textbf{pangraph realisation} of $\mathcal{G}_L$.
\begin{rem}\label{rem:uniq_pan}
The multilayer digraph $\mathcal{G}_L$ has a unique pangraph realisation if there exists exactly one ordering of layers satisfying conditions ii) - iii) in Theorem \ref{thm:Levi_vs_pan}.
\end{rem}

The introduction of layers into the Levi graph is crucial to obtain the uniqueness. In the following example, by changing the order of layers, one can obtain a different pangraph. 

\begin{exam}\label{exam:Levi_to_pan}
Let us consider two unweighted multilayer digraphs $$\mathcal{G}_{\mathcal{L}_x}=\left(\{v_1,v_2,v_3,V_4,V_5\}, \{L^x_1,L^x_2\}, E_{\mathcal{L}}\right),\qquad x=A,B;$$
such that all edges' weights are equal to one and 
\begin{eqnarray*}
&L^A_1=L^B_2=\{v_1,v_2,v_3\},\qquad L^A_2=L^B_1=\{V_4,V_5\}\\
&E_{\mathcal{L}}=\{(v_i,V_4),(V_4,v_i),(v_j,V_5),(V_5,v_j):\,\, i=1,2;\, j=2,3\}.
\end{eqnarray*}

Using the construction of a pangraph presented in the proof of Theorem \ref{thm:Levi_vs_pan}, we note that  $\mathcal{G}_{\mathcal{L}_A}$ is a Levi representation of pangraph $\mP_A=(V, \EP,\incip,\incop) $ such that
$$V^A=\{v_1,v_2,v_3\}, \quad \EP^A=\{(\{v_1,v_2\},\{v_1,v_2\}), (\{v_2,v_3\},\{v_2,v_3\})\},$$
whereas  $\mathcal{G}_{\mathcal{L}_B}$ is a Levi representation of pangraph $\mP_B=(V^B, \EP^B,\incip,\incop)$ such that 
$$V^B=\{V_4,V_5\}, \quad \EP^B=\{(\{V_4\},\{V_4\}), (\{V_5\},\{V_5\}), (\{V_4,V_5\},\{V_4,V_5\})\}.$$
\end{exam}

Two pangraphs obtained in Example \ref{exam:Levi_to_pan} are entirely different objects. However, if we fix the first layer in Levi representation, which is reflected in the set of fundamental vertices in the pangraph, then we have a uniqueness of the representation up to a pangraph isomorphism, see Definition \ref{df:pan_izo}. 

\begin{rem}
Consider two weighted multilayer digraphs, satisfying (2), such that one can be transformed into another by relabeling the layers. If their first layers, concerning the reordering defined in Theorem \ref{thm:Levi_vs_pan}, are equal, then their pangraph realisations are homomorphic.
\end{rem}

\section{Dynamics on pangraphs}\label{sec:pan_dynamics}

Let us present our first thoughts on how to use a pangraph to receive additional information on the dynamics of a complex system. The interpretation of panedges that we propose in this paper is a generalisation of existing results in a chemical reaction and ecosystem modelling, presented in detail in Section \ref{sec:other_structures}. In these networks, the transfer of the information described by the flow between fundamental vertices is modified by the deeper interactions represented in our approach by panedges. The deeper interactions are only the modifications of other interactions but do not result in any transfer of matter themselves. On the contrary, the flow between fundamental vertices is a physically observable quantity. In designing the pangraph framework, we aim to enable the modeller to quantify the strengths of interaction modifications and their observable impact. 

In this section, we present the mathematical models based on statistical network analysis, ordinary and partial differential equations, and the concept of their pangraph interpretation. Incorporation of a new structure into a well-established theory depends strictly on the type of a panedge. In
particular, we are interested in the statistical network methods in this study. Therefore, we focus on the case of a vertex influencing an edge. Other two types of panedges, namely an edge influencing a vertex and an edge influencing another edge, can be applied in the case of models based on differential equations. In this case, we show basic concepts based on linear dynamics representing a mass flow. This concept needs further study.

\subsection{A vertex influencing an edge, a statistical approach}\label{sec:v->e}
Centrality measures form the basis of the statistical network analysis. They rely on combining edges and weights and studying walks. In order to properly generalise various graph measures into the pangraph case, one needs to interpret the process of traversing the pangraph and consequently choose an appropriate definition of a walk. 

\begin{definition}\label{def:walk}
    A \textbf{causal walk} $W_c$ of length $l_{W_c}$ in an unweighted pangraph $\mP=(V,\EP)$ is a sequence $W_c=(w_{0},\ldots, w_{l_{W_c}})$ such that
    
\begin{enumerate}
    \item $w_{0},w_{l_{W_c}} \in V$, and $\{w_{1},\ldots, w_{l_{W_c}-1}\} \in V\cup\EP$;
    \item if $w_i\in V$, for any $i=0,\ldots, l_{W_c}-1$; then $w_{i+1}\in \EP$ and $w_i \in w_{i+1}^{\tin};$
    \item if $w_i\in \EP$, $i=1,\ldots, l_{W_c}-1$ we have $w_i \in w_{i+1}^{\tin}$ or $w_{i+1} \in w_{i}^{\out}$.
\end{enumerate}
We say that a causal walk is a \textbf{transport walk} $W_t=(w_1,\ldots, w_{l_{W_c}})$ if additionally the condition
$$D(w_i)\leq 1,\qquad \text{for all}\,\,i=1,\ldots, l_{W_t}-1$$
holds.
\end{definition}

The notions of walks in an unweighted pangraph are both well-defined since, in the case $\mP=\mG$, both walks simplify to the classical definition of a walk on a digraph, see Subsection \ref{sec:m_graph}. 

Depending on the application, a modeller may prefer one definition of a walk over another. We note that a causal walk is, in fact, a walk in the Levi representation of a pangraph. On the other hand, causal walks treat interactions described by 1-depth panedges and deeper panedges as the same type of influence. However, the deeper panedges represent modifiers of main interactions given by 1-depth panedges. We will show how weights of 1-depth panedges can be modified to include interactions' modifications.

In this study, we consider quantitative modifications of interactions defined in relative terms so their effect scales with the size of the modified interaction in the first approximation. Such is the case, e.g., of the rates of chemical reactions (see Sec.~\ref{sec:other_structures}). Two proposed ecological models~\cite{Arditi_2005_non_trophic_rheagogies, Goudard_nontrophic_2008} have also used this principle as their basis. However, real-world interaction modifications may combine in both non-linear and non-multiplicative ways~\cite{Golubski_2011_non_trophic_combining}, so the formula for combining the modifications will always depend on the particular applied model.  

In order to formally introduce the modification effect on the network parameters, let us define recursively a new digraph representation of pangraphs by projecting the interactions described by deeper panedges into 1-depth panedeges. Let $ \mP=(V,\EP, \inci_{\mP},\inco_{\mP})$ be a $k_0$-depth pangraph and $L=\{L_j\;|\; j=1,\ldots, k_0 +1\}$ be a set of layers introduced in Definition \ref{def:levi-graph}. Consider a family of pangraphs 
 such that for any $s=1,\ldots, k_0$, we have
 \begin{equation}\label{eq:E_Ge}
 E_{\mG_s} = \bigcup_{i=2}^{k_0-s+2} L_i.
 \end{equation}
 For any panedge $e_i\in E_{\mP}$, let us define respectively a set $A_i$ and a vector $\inci_{\cdot \, i}$ by 
 $$A_i:=\{p=1,\ldots, m\;|\; e_i\in e_p^{\out}\},\quad \inci_{\cdot \, i}=(\inci_{pi}\;|\;p\in A_i)\in \RR^{|A_i|}.$$
We define the matrices $\inci_{\mG_s},\inco_{\mG_s}$ recursively in the following way. For $\mathcal{I}^{\text{x}}_{\mG_1}:=\mathcal{I}^{\text{x}}_\mP, x=\tin,\out$; and for $s=2,\ldots,k_0$, $v_i\in V,$ and $e_i\in E_{\mG_s}\subset  E_{\mG_{s-1}}$, we have
\begin{subequations} \label{eq:I_Ge}
\begin{flalign}
&(\inci_{\mG_s})_{ij}:=f_i^s(\inci_{\cdot \, i})(\inci_{\mG_{s-1}})_{ij}, \quad f_i:\RR^{|A_i|}\to \RR;\\[.2cm]
&(\inco_{\mG_s})_{ij}:=(\inco_{\mG_{s-1}})_{ij}.
\end{flalign}
\end{subequations}

We say that $f_i^s$ ( $i=1,\ldots,m$) is an aggregation function for a panedge $e_i\in E_{\mP}$ of depth $s$. We can interpret this parameter as the strength of the interaction modification at depth $s$. In a classical example, a weight represents the transition probability, and a natural aggregation would be to multiply weights, hence, one has 
\begin{equation}\label{eq:agreg_f}
f_i(x_1,\ldots,x_{\|A_{i}\|})=\prod_{l=1}^{|A_i|}x_l.
\end{equation}
\begin{definition} \label{def:eff_flow}We say that $\mG_{\mathcal{E}}(\mP)$ is a graph with \textbf{effective flow} of a $k_0$-depth pangraph $\mP$ if $\mG_{\mathcal{E}}(\mP)=\mG_{k_0}(\mP)$ defined in the recursive algorithm \eqref{eq:E_Ge}--\eqref{eq:I_Ge}.
\end{definition}
The graph of effective flow $\mG_{\mathcal{E}}(\mP)$ allows us to calculate network parameters based on the observable impact of all deeper interactions, whereas weights of panedges allow us to preserve the data related to the strength of interaction modifications. 

In Section \ref{sec:measures}, we compare graph measures calculated based on Levi digraph representation and digraph of effective flow to understand the strengths and weaknesses of both approaches.

\subsection{ An edge influencing a vertex, ODE approach}\label{sec:e->v}

In this section, we show how a panedge can influence the network dynamics of ODE systems. Let us first consider a digraph $\mG$ and some quantities $u(t)=(u_i(t))_{i\in I}$ stored in vertices. Mass conservation implies that change is achieved by the difference between
mass flowing in and out. These quantities are specified by weighted adjacency matrix $\mathcal{A}^{\tin}$ and weighted degree matrix $\mathcal{D}^{\out}$, defined respectively in \eqref{eq:adj_and_inc} and \eqref{eq:degree}. This amounts to an initial value problem:
\begin{equation}\label{eq:mass_acts}
\frac{d}{dt}u(t)=\left(\mathcal{A}_{\mG}^{\tin}-\mathcal{D}_{\mG}^{\out}\right)u(t),\qquad u(0)=\mathring{u},
\end{equation}
where $u(t)=(u_i(t))_{i\in I}$ is a continuously differentiable vector function $u:[0,\infty)\to \mathbb{R}^n.$  This  system of equations could be also interpreted as laws of mass action on a network. Two edges with the same endpoints would then describe two separate processes with the same output rather than one process with two inputs and one output.

Now, let us introduce a quantitative interaction modification by adding a $2$-depth panedge $e_k=(e_k^{\tin},e_k^{\out})\in E_{\mP}$. The edge $e_j\in e_k^{\tin}$ modifies the dynamics in vertices $v_i\in e_k^{\out}$ by a multiplicative factor (as in Section \ref{sec:v->e}). The modified system for the $i$-th coordinate reads
\begin{eqnarray*}
\frac{d}{dt}u_i(t)&=&(\mathcal{I}^{\out}_{\mG})_{ki}\left(\sum_{s\in I}(\mathcal{A}_{\mG}^{\tin})_{is}u_s(t)-\kappa_{\mG}^{\text{out}}(v_i)u_i(t)\right),\qquad \text{if}\,\,v_i\in e_k^{\out};\\
\frac{d}{dt}u_i(t)&=&\left(\sum_{s\in I}(\mathcal{A}^{\tin}_{\mG})_{is}u_s(t)-\kappa_{\mG}^{\text{out}}(v_i)u_i(t)\right),\qquad \text{if}\,\,v_i\notin e_k^{\out},
\end{eqnarray*}
with an initial condition similar to \eqref{eq:mass_acts}. A recursive procedure, similar to the one presented in \eqref{eq:E_Ge}--\eqref{eq:I_Ge}, allows us to generalise this approach to a dynamical system on a $k$-depth pangraph. 

\subsection{Panedges in PDE approach}
Let us consider three kinds of panedges: a vertex influencing an edge, an edge influencing a vertex, and one edge influencing another. Metric graphs facilitate their description.

\begin{definition}
We say that a pair $\mG_M=(\mG,d)$ is a metric graph where $\mG$ is a digraph satisfying Def. \eqref{eq:multi-weights} and $d:E_{\mG}\to \mathcal{B}(\RR)$, $\mathcal{B}(\RR)$ being a Borel $\sigma$-algebra on $\RR$, is a mapping that associates each edge with a finite interval. 
\end{definition}

For simplicity, let us consider a metric graph $\mathcal{G}_M$ such that $d:E_{\mathcal{G}}\to [0,1]$ and define a process of advection with flux conservation in every graph vertex $k$, namely 
\begin{equation}\label{eq:F}
\begin{array}{rcll}
\frac{\partial}{\partial t}\, u_k(x,t) &=& -c_k\,\frac{\partial}{\partial x}\, u_k(x,t)
,&t >0,\ x\in(0,1), k\in K,\\[.1cm]
c_k\inci_{ki} u_{k}(0,t)&=&\sum_{e_j\in E_{\mG}}c_j\inco_{ji}u_{j}(1,t),& t >0,\,k\in K,\,  i\in I,
\\[.1cm]
u_k(0,x)&=& \mathring{u}_k(x), &x\in\left[0,1\right],\,\,k\in K;
\end{array}
\end{equation}
where $u(x,t)=(u_k(x,t))_{k\in K}$ is a density function on graph's edges and $c=(c_k)_{k\in K}\in \mathbb{R}_+^m$ is the real-valued transport velocity along the edge, which can depend on some of the variables. Again, adding a $2$-depth panedge $e_k=(e_k^{\tin},e_k^{\out})\in E_{\mP}$ to $\mathcal{G}_M$, we obtain a pangraph $\mP$. Depending on the type of the panedge, we consider three cases.
\begin{enumerate}
\item  A vertex influencing an edge. Assume that there exists $v_i\in V_{\mP}$ and $e_j\in E_{\mP}$ such that 
\begin{equation*}
v_i\in e_k^{\textrm{in}}, \qquad e_j\in e_k^{\textrm{out}}, 
\end{equation*}
and consider the flux $F_i:C([0,\infty)])^m\times [0,\infty)\to \mathbb{R}$, 
$$F_i:(u,t)\mapsto \sum_{e_k\in E_{\mP}}\, c_k\inci_{ki} u_{k}(0,t)$$ through the vertex $v_i$, which accelerates or decelerates the dynamics on edge $e_j$ at rate $\inco_{kj}$ by changing the velocity $c_j$. This changes the system of equations into 
\begin{equation*}
\begin{array}{rcll}
\frac{\partial}{\partial t}\, u_j(x,t) &=& - \inco_{kj}\, c_j\left(F_i(u,t)\right)\,\frac{\partial}{\partial x}\, u_j(x,t)
,&t >0,\ x\in(0,1),\\[.1cm]
\frac{\partial}{\partial t}\, u_k(x,t) &=& -v_k\,\frac{\partial}{\partial x}\, u_k(x,t)
,&t >0,\ x\in(0,1), k\in K\setminus\{j\},\\[.1cm]
c_k\inci_{ki} u_{k}(0,t)&=&\sum_{e_j\in E_{\mP}}c_j\inco_{ji}u_{j}(1,t),& t >0,\,k\in K,
\\[.1cm]
u_k(0,x)&=& \mathring{u}_k(x), &x\in\left[0,1\right],\,\,k\in K.
\end{array}
\end{equation*}
In this system, $c_j:\mathbb{R}\to \mathbb{R}_+$ is a velocity function, while $c_k\in \mathbb{R}_+$ for $k\neq j$.
\item An edge influencing a vertex. Consider $e_j\in E_{\mP}$ and $v_i\in V_{\mP}$ such that 
\begin{equation*}
e_j\in e_k^{\textrm{in}}, \qquad v_i \in e_k^{\textrm{out}},
\end{equation*}
and denote by $M_j:C([0,\infty))\times [0,\infty) \to [0,\infty)$,
\begin{equation}\label{eq:edge_mass}
M_j:(u_j,t)\mapsto \int_{0}^1u_j(x,t)dx,
\end{equation}
the total mass on the edge $e_j$ that influences a flow through the vertex $v_i$ at rate $\inco_{ki}$.
The system of equations in such a situation reads
\begin{equation*}
\begin{array}{rcll}
\frac{\partial}{\partial t}\, u_k(x,t) &=& -c_k\,\frac{\partial}{\partial x}\, u_k(x,t)
,&t >0,\ x\in(0,1), k\in K,\\[.1cm]
c_k\inci_{ki} u_{k}(0,t)&=&\inco_{ki}M_j(u_j,t)\; \sum_{e_j\in E_{\mP}}v_j\inco_{ji}u_{j}(1,t),& t >0, k\in \{s\in K\;|\;v_i\in e_s^{in}\}
\\[.1cm]
v_k\inci_{ki} u_{k}(0,t)&=&\sum_{j\in K}v_j\inco_{ji}u_{j}(1,t),& t >0,\,k\in K\setminus \{s\in K\;|\;v_i\in e_s^{in}\},
\\[.1cm]
u_k(0,x)&=& \mathring{u}_k(x), &x\in\left[0,1\right],\,\,k\in K.
\end{array}
\end{equation*}

\item One edge influencing another edge. Flow on one edge can influence the velocity of flow on another. Let us consider edges $e_j,e_l\in E_{\mP}$ connected by such a higher-order influence 
\begin{equation*}
e_j\in e_k^{\textrm{in}} , \qquad e_l \in e_k^{\textrm{out}}.
\end{equation*}
Let the total mass gathered at $e_j$, defined in \eqref{eq:edge_mass}, influence the velocity on the edge $e_l$ at rate $\inco_{kl}$, 
\begin{equation*}
\begin{array}{rcll}
\frac{\partial}{\partial t}\, u_l(x,t) &=& - \inco_{kl}\, c_l\left(M_j(u_j,t)\right)\,\frac{\partial}{\partial x}\, u_l(x,t)
,&t >0,\ x\in(0,1),\\[.1cm]
\frac{\partial}{\partial t}\, u_k(x,t) &=& -c_k\,\frac{\partial}{\partial x}\, u_k(x,t)
,&t >0,\ x\in(0,1), k\in K\setminus\{l\},\\[.1cm]
c_k\inci_{ki} u_{k}(0,t)&=&\sum_{e_j\in E_{\mP}}c_j\inco_{ji}u_{j}(1,t),& t >0,\,k\in K,
\\[.1cm]
u_k(0,x)&=& \mathring{u}_k(x), &x\in\left[0,1\right],\,\,k\in K.
\end{array}
\end{equation*}
\end{enumerate}
Up to authors' best knowledge, no models of that kind have been considered in the literature. The most similar existing results to the open problem presented in the second case are related to transport with McKendric boundary conditions, see \cite{BGS2011}.

\section{Related structures}
\label{sec:other_structures}
Network science has seen a proliferation of structures aiming to represent higher-order interactions. Some are tailor-made for specific applications, e.g., in chemistry, whereas others adopt the general theory by adding additional assumptions about the network, e.g., food webs. In this section, we offer a short review of mathematical methods of description of higher-order interactions and show their relation to pangraph theory. It turns out that many methods fit the pangraph framework, which paves the way to generalisations of results obtained in particular fields.

\subsection{Mutualistic networks}\label{subsec:mutualistic_nets}

Modifications to the physical mass
flow in ecosystems emerge from various interactions other than feeding. One used to call them non-trophic interactions. They are studied under the broad term of mutualistic networks, even though the represented phenomena go beyond just mutualism.

We show that mutualistic~\cite{Bascompte_Mutualistic_networks} interaction networks in ecosystems and conceptually related causal graphs~\cite{Pearl_2009, Morgan_Winship_2014} are particular cases of pangraphs. An example of a coffee agroecosystem~\cite{GOLUBSKI2016344}, discussed in detail in~\ref{sec:coffee_agroeco}, belongs to this category.

The simplest mutualistic networks are undirected graphs, where an edge signifies the existence of a relationship. This framework allows us to describe, for instance, plant-animal interactions such as pollination of flowers or dispersal of seeds. More generally, non-tropic interactions can be classified according to their impact on the participants into three categories: positive-positive (e.g., mutualism), positive-negative (e.g., parasitism), and negative-negative (e.g., competition). Although typically studied separately, they can be combined to generalise of signed graphs~\cite{signed_graph_1936, signed_graph_1955}. In such a case, the distinction between the types of interaction is kept in the signs given to the edges.

From a quantitative perspective, these networks capture relations between variables that constitute a dynamical system, e.g., species populations represented as vertices, see Subsection \ref{sec:v->e}. The dynamics of observed phenomena are rarely linear. In order to fit the model into a linear framework offered by network structure, one can linearise the system since there is a correspondence between these two mathematical objects. In consequence, the edges are related to the entries of the system's Jacobian matrix~\cite{HIGASHI_1995_ecological_interaction_networks, NAKAJIMA_1995_ecological_interaction_networks}. Bilateral relationships - edges $e_{ij}$ and $e_{ji}$ - are frequently represented through one edge with two signs at its ends. Such is the case of the coffee agroecosystem~\cite{GOLUBSKI2016344}, with edges carrying the signs of the Jacobian terms, $(+,-), (-,-), (0,+)$, etc. We denote the lack of a sign/weight as $0$. The real-valued incidence matrices, see Definition~\ref{def:WeightedDigraph}, can conveniently accommodate these signed edge ends. We formalize the rough notion found in ecological literature and apply definitions formulated in Section~\ref{sec:graph_theo}.

\begin{definition}\label{def:mutualistic_graph}
    A \textbf{mutualistic graph} is a weighted digraph $\mathcal{G}_M=(V,\EG, \inci_\mG,\inco_\mG)$ with incidence matrices $\inci_\mG$ and $\inco_\mG$ being sign pattern matrices, namely matrices having signs $+,-,0$ instead of classical entries.
\end{definition}

\begin{figure}
    \centering
    \includegraphics[width=0.9\linewidth]{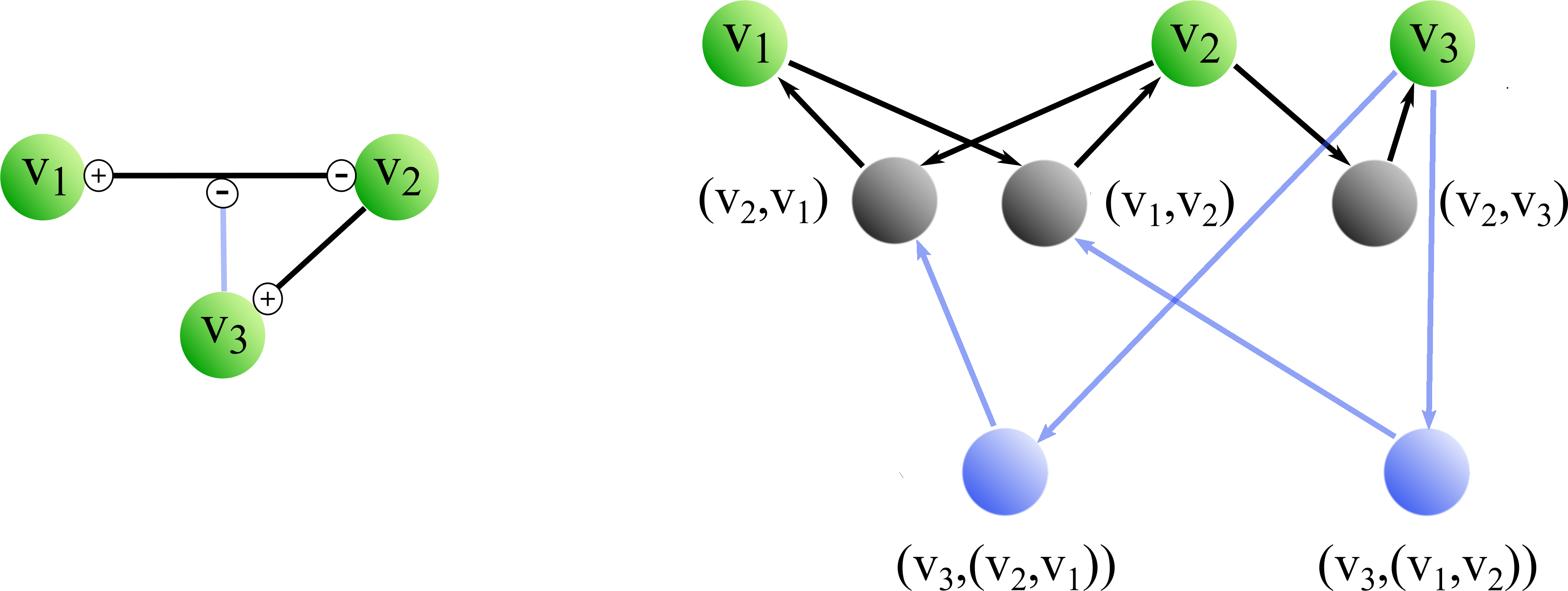}
    \caption{Left: examples of mutualistic network interactions~\cite{GOLUBSKI2016344}. A bilateral interaction, a unilateral influence, and an interaction modification. Right: the Levi graph of their pangraph representation. Each signed edge end in a mutualistic network is a separate influence, mapped to a distinct pangraph edge.}\label{fig:minus_minus_rep}
\end{figure}

Let us consider a bijection $\psi:\{+,-,0\} \to \{1,-1,0\}$ defined as
\begin{equation}\label{eq:psi}
\psi(x)=\left\{\begin{array}{ll} \pm 1, &\textrm{if }x=\pm,\\
0,&\textrm{if }x=0.
\end{array}\right.
\end{equation}
We denote incidence matrices with signs transformed into integers as $\psi(\mathcal{I}^x_{\mG})=\left((\psi(\mathcal{I}^x_{\mG})_{ij})\right)_{i\in K,j\in I}$, $x=\textrm{in},\textrm{out}$. Then, the mutualistic network $\mG_M=(V,\EG, \inci_\mG,\inco_\mG)$ is isomorphic to the weighted digraph $\mG=(V,\EG, \psi(\inci_\mG),\psi(\inco_\mG)).$ In further considerations, we omit $\psi$ with no ambiguity.

It is a widespread practice to draw the network in the form consistent with Definition~\ref{def:mutualistic_graph}, but quantitative analyses neglect the signs, see instance in~\cite{GOLUBSKI2016344}. However, a qualitative difference exists between a $(+,-)$ interaction and $(+,0)$ one. In the latter, one of the variables is unaffected by the other, while the first depicts a bilateral influence. A proper translation into unweighted graphs requires splitting the bilateral influences as in~\cite{HIGASHI_1995_ecological_interaction_networks}. Even though they might depict one biological interaction, the corresponding Jacobian entries are generally independent.

The process of translating a mutualistic graph into its pangraph counterpart is straightforward. An undirected edge $e_k=\{v_i, v_j\}$ is replaced by a pair of directed edges, from the cause to the effect (identified through the signed endpoint) $e_{k'}=(v_i, v_j)$, $e_{k''}=(v_j, v_i)$. Incidence matrix entries change accordingly, with $$\inci_{k'i}=1, \quad \inci_{k''j}=1, \quad \inco_{k'i}=\inci_{ki}, \quad \inco_{k''j}=\inco_{kj}.$$ If an edge carries just one sign, it corresponds to just one influence, thus one edge. This is automatically represented by the zero entry of the incidence matrix that signifies no impact. Fig.~\ref{fig:minus_minus_rep} depicts this procedure using the Levi graph representation of the resultant digraph. It enables the qualitative distinction between bilateral interactions and unilateral influences to be kept even when weights are neglected.

\subsection{Multilayer ecological networks}

Multilayer ecological networks~\cite{Pilosof_2015, Hutchinson2019} supplement food webs with layers representing other types of interactions, such as parasitism, mutualism, or competition. Each layer is a digraph with the same set of vertices. The first such model of a real-world ecosystem portrayed Chilean rocky shores~\cite{Kefi_multilayer_Chilean} and was qualitative. It indicated the existence of a particular trophic (feeding) or non-trophic interaction. 

The physical law of mass conservation means that most non-trophic interactions impact biomass flows rather than directly vertices. From a general perspective, every population change corresponds to a biomass flow. So, influencing a population change (having a non-zero Jacobian entry indicated by a mutualistic network edge) means influencing a biomass flow. In a more specific approach, an ecological interaction may influence a biomass flow - e.g., predation rate, birth processes, and flow of nutrients. A comprehensive quantitative ecosystem model demanded by the multilayer network proponents~\cite{Kefi_multilayer_Chilean} would have to be able to connect the non-trophic interactions to biomass flows.

Panedges can represent precisely this type of causal coupling. A non-trophic influence of $v_1$ on $v_2$ represented by a multilayer ecological network edge $(v_1,v_2)$ means in most cases that some physical process involving $v_2$ is modified by $v_1$. In network studies, such a process would be represented by an edge, say $e_1=(v_2,v_3)$. Using the pangraph notation, the correct formulation of the considered process would be a 2-depth panedge $(v_1, e_1)$ rather than a 1-depth panedge $(v_1,v_2)$.

Multilayer ecological networks fit into the pangraph framework, and this theory can be substantially enhanced using pangraphs. The 0-depth panedges, namely the fundamental vertices, may contain information about population size, quantified most commonly by the population density or biomass, which changes in time. Meanwhile, 1-depth panedges represent flows/predation or other direct interactions, while deeper panedges represent modifications to other interactions or the size of the respective population.
 
\subsection{Petri nets}\label{subsec:petri_nets}

The weighted Levi graph of a pangraph is also a generalisation of a mathematical object known as a Petri net. Petri nets~\cite{Petri_thesis, Petri_Peterson_book, Baez_open_petri_2017} are bipartite digraphs (see Section \ref{sec:m_graph}) with sets of vertices representing objects (elements of a set $S\subset \mathbb{N}$) and transitions between them (elements of a set $T\subset \mathbb{N}$). An edge $(\{v_i\},\{v_j\})$ signifies that an object participates in a process as either an input ($v_i\in S$ is an input to the process $v_j\in T$) or an output ($v_j\in S$ is an output of a process $v_i\in T$). Even though Petri nets can be defined using classical graph theory, their community applies their notions.

Petri nets usually describe chemical reaction networks or ecological interactions. Consequently, edge weights (called multiplicities and denoted by $W$) represent stoichiometric indices and are thus constrained to natural numbers. Markings, analogous to stocks in stock and flow networks, represent the number of molecules/individuals of a given substance/species and can be considered as weights of vertices being objects $S$. Consequently, the definition of Petri net reads:
\begin{definition}\label{def:petr}
    A \textbf{Petri net} is a pair $\mathcal{G}_{\textrm{P}}=(\mathcal{G}_L,M)$ such that $\mathcal{G}_L=(V,L,E_{\mathcal{G}_L}, \inci_{\mG_L},\inco_{\mG_L})$ is a connected weighted multilayer digraph satisfying the conditions:
    \begin{enumerate}
  \item  The set of vertices $V=S\cup T$ consists of places (species, objects) $S$ and transitions $T$, hence a set of layers reads $L=\{S,T\}$;
    \item  Elements of $E_{\mG_L}$ are called arcs and every arc joins two vertices from different layers, namely for any $e_{k}=(\{v_i\},\{v_j\})\in E_{\mG_L}$, we have
    \begin{equation}\label{eq:bipart_ST}
    v_p\in S \quad \textrm{and}\quad v_q\in T,\qquad \textrm{for }\,\, p\in \{i,j\},\,\, q\in \{i,j\}\setminus\{p\};
    \end{equation}
    \item  Weights of arcs are given by the function $W:E_{\mG_L}\to \mathbb{N}$ are called multiplicities, and have the following relation with incidence matrices  $\inci_{\mG_L},\inco_{\mG_L}\in \mathbb{M}_{m\times n}(\mathbb{N})$
    \begin{equation*}
W(e_k)=\left(\inci_{\mG_L}\right)_{ki}=\left(\inco_{\mG_L}\right)_{kj},\qquad \textrm{ if }e_k=(\{v_i\},\{v_j\})\in E_{\mG_L}.
    \end{equation*}
    \end{enumerate}
     The function $M: S \rightarrow \mathbb{Z}^{+}$ is a places' weight function called markings.
\end{definition}

We say that $\mG_{P}=(\mG_L,M)$ is an \textrm{unweighted Petri net} if $\mG_L$ is an unweighted multilayer network and function $M$ is not specified. Consequently, unweighted Petri net can associated with $\mG_L=(V,L,E_{\mG_L})$.

In the description of a Petri net one can add quite a natural assumption: 
\begin{equation}\label{eq:in_out_PN}
\textrm{each process in } T \textrm{ has at least one input and at least one output.} 
\end{equation}
In such a case, by Subsection \ref{sec:Levi_pangraph}, we can prove that any Petri net can be associated with a pangraph.

\begin{prop}\label{prop:Petri}

For any Petri net $\mG_P=((V,\{S,T\},E_{\mathcal{G}_L}, \inci_{\mG_L},\inco_{\mG_L}),M)$ satisfying \eqref{eq:in_out_PN} there exists exactly one $1$-depth pangraph $\mP=(V_{\mP},E_{\mP}, \inci_{\mP},\inco_{\mP})$ such that $\mG_P$ is a Levi digraph of pangraph $\mP$ with
\begin{equation}\label{eq:S_fixed}
V_{\mP}=S,\quad E_{\mP}=T.
\end{equation}
\end{prop}
\begin{proof}
The existence of a pangraph $\mP$ follows directly from Theorem \ref{thm:Levi_vs_pan}, since condition i) holds due to \eqref{eq:bipart_ST}, while ii) and iii) by \eqref{eq:in_out_PN}. The uniqueness follows from Remark \ref{rem:uniq_pan}, and the condition \ref{eq:S_fixed} which fixes the layer order.   
\end{proof}

A generalisation of a Petri net considers an additional category of vertices (species) and catalysts~\cite{Baez_2019_catalysts} that represent chemical catalysis. Catalysts can influence a reaction, but the reaction does not change the amount of a catalyst. Classically, catalysts are modelled as a subset of vertices $C \subset S$ with equal incoming and outgoing weights with every transition $T$ they connect to. Let us formalise the definition presented in ~\cite{Baez_2019_catalysts}.

\begin{definition}[The classical definition of a Petri net with catalyst]\label{def:Petr&catal}
    A \textbf{ Petri net with catalyst} is a pair $\mathcal{G}_{\textrm{Pc}}=(\mathcal{G}_L,M)$ such that $\mathcal{G}_L=(V,L,E_{\mathcal{G}_L}, \inci_{\mG_L},\inco_{\mG_L})$ is a connected weighted multilayer digraph satisfying the conditions:
    \begin{enumerate}
  \item  The set of vertices $V=S\cup T \cup C$ consists of places (species) $S$, transitions $T$ and catalysts $C$, and a set of layers reads $L=\{S\cup C,T\}$;
    \item  Elements of $E_{\mG_L}$ are called arcs and every arc $e_{k}=(\{v_i\},\{v_j\})\in E_{\mG_L}$ satisfies
    \begin{equation}\label{eq:bipart2}
    v_p\in S\cup C \quad \textrm{and}\quad v_q\in T,\qquad \textrm{for }\,\, p\in \{i,j\},\,\, q\in \{i,j\}\setminus\{p\};
    \end{equation}
    \item  Weights of arcs are given by function $W:E_{\mG_L}\to \mathbb{N}$ are called multiplicities, and have the following relation with incidence matrices  $\inci_{\mG_L},\inco_{\mG_L}\in \mathbb{M}_{m\times n}(\mathbb{N})$
    \begin{equation*}
W(e_k)=\left(\inci_{\mG_L}\right)_{ki}=\left(\inco_{\mG_L}\right)_{kj},\qquad \textrm{ if }e_k=(\{v_i\},\{v_j\})\in E_{\mG_L};
    \end{equation*}
    \item If $e_k=(\{v_i\},\{v_j\})\in E_{\mG_L}$, then there exists $e_{k'}=(\{v_j\},\{v_i\})\in E_{\mG_L}$ and
    \begin{equation}
(\inci_{\mG_L})_{ki}=(\inco_{\mG_L})_{kj}=(\inci_{\mG_L})_{k'j}=(\inco_{\mG_L})_{k'i}.
    \end{equation}
    \end{enumerate}
     The function $M: S \cup C\rightarrow \mathbb{Z}^{+}$ is a weight function called markings.
\end{definition}

$\mG_{Pc}=(\mG_L,M)$ is an \textrm{unweighted Petri net with catalyst} if $\mG_L$ is an unweighted multilayer network and function $M$ is not specified, and again we associate it with $\mG_L=(V,L,E_{\mG_L})$. Let us now construct a pangraph associated with a Petri net with a catalyst, in analogy to Proposition~\ref{prop:Petri}.

\begin{exam}\label{exam:2petri}
Let us consider an unweighted Petri net with catalyst $\mG_{Pc}=(S\cup C\cup T, \{S\cup C, T\}, E_{P_c})$ such that 
\begin{equation*}
S=\{s_1,s_2\},\quad T=\{t_1\}\quad \text{and}\quad C=\{c_1\};
\end{equation*}
and edges are given by
\begin{equation*}
E_{Pc}=\{(\{s_1\},\{t_1\}), (\{t_1\},\{s_2\}), (\{t_1\},\{c_1\}), (\{c_1\},\{t_1\})\}.
\end{equation*}

\noindent  The pangraph realisation of this Petri net is a pangraph $\mP=(V,E_{\mP})$ with vertices $V=\{s_1,s_2,c_1\}$ and panedge $E_{\mP}=\{t_1\}$ such that $t_1=(\{s_1,c_1\},\{s_2,c_1\})$ is 1-depth panedge.

We propose an alternative description of this process by a pangraph $\mP'=(V',E_{\mP}')$ with three vertices that represent substances $V'=\{s_1,s_2,c_1\}$ and two edges describing processes $e_{\mP}'=\{t_1,t_c\}$ where $t_1=(\{s_1\},\{s_2\})$ is a 1-depth representation of the reaction and $t_c=(\{c_1\},\{t_1\})$ is a 2-depth representation of the catalysis. The Levi graph representation of $\mP'$ is given by $\mG_{Pc}=(\{S' \cup T' \cup C'\},\{S',T',C'\},E_{Pc}')$ where 
\begin{equation*}
S'=\{s_1,s_2,c_1\},\quad T'=\{t_1\}\quad \text{and}\quad C'=\{t_c\};
\end{equation*}
and edges are given by
\begin{equation*}
   E_{Pc}'=\{(\{s_1\},\{t_1\}), (\{t_1\},\{s_2\}), (\{c_1\},\{t_c\}), (\{t_c\},\{t_1\})\}.
\end{equation*}
\end{exam}

Example \ref{exam:2petri} shows that the classical definition of a Petri net with catalysis (Definition \ref{def:Petr&catal}) leads to a pangraph in which transitions involving catalysts are hyperedges connecting them and other species involved in the transition in question. The hyperedge indicates the qualitative requirement of the catalyst's presence for this reaction to occur but does not affect transition rates. In the case of reactions in which the catalyst is not required but can enhance the reaction rate, the above-described structure suffices, but this would require the presence of two transitions - one with the catalyst and one without the catalyst.

However, for a clear structural representation and the possibility of capturing the transition rates, the higher-order impact of a catalyst on the reaction (transition) seems more appropriate. Given the foregoing, we propose an alternative definition of Petri net with catalysts. 
\begin{definition}[An alternative definition of a Petri net with catalyst]\label{def:Petr&catal2}
    A \textbf{ Petri net with catalyst} is a pair $\mathcal{G}_{\textrm{Pc}}'=(\mathcal{G}_L',M')$ such that $\mathcal{G}_L'=(V',L',E'_{\mathcal{G}_L'}, {\inci_{\mG_L}}',{\inco_{\mG_L}}')$ is a connected weighted multilayer digraph satisfying the conditions:
    \begin{enumerate}
  \item The set of vertices $V=S\cup C\cup T \cup T_C$ consists of places (species) $S$, catalysts $C$, transitions $T$ and process of catalysis $T_C$, and a set of layers reads $L=\{S\cup C,T,T_C\}$;
    \item  Elements of $E_{\mG_L}$ are called arcs and every arc $e_{k}=(\{v_i\},\{v_j\})\in E_{\mG_L}$ satisfies either \ref{eq:bipart2} or 
     \begin{equation}\label{eq:bipart3}
    (v_i\in C \quad \textrm{and}\quad v_j\in T_C),\qquad \textrm{or}\qquad (v_i\in T_C \quad \textrm{and}\quad v_j\in T);
    \end{equation}
    
    \item  Weights of arcs are given by the function $W:E_{\mG_L}\to \mathbb{N}$ and called multiplicities, and have the following relation with incidence matrices  $\inci_{\mG_L},\inco_{\mG_L}\in \mathbb{M}_{m\times n}(\mathbb{N})$
    \begin{equation*}
W(e_k)=\left(\inci_{\mG_L}\right)_{ki}=\left(\inco_{\mG_L}\right)_{kj},\qquad \textrm{ if }e_k=(\{v_i\},\{v_j\})\in E_{\mG_L}.
    \end{equation*}
    \item for every $v_k\in T_C$, there is exactly one outgoing edge and at least one incoming, namely  
    \begin{equation*}
   \kappa_{\mG}(v_k)=1,\qquad \kappa_{\mG}(v_k)\geq 1.
    \end{equation*}
    \end{enumerate}
     The function $M: S \cup C\rightarrow \mathbb{Z}^{+}$ is a weight function called markings.
\end{definition}

\begin{prop}\label{prop:Petri&catal}
Let us consider a classical Petri net with catalysts $\mG_{Pc}$, see Definition \ref{def:Petr&catal}, and a Petri net with catalysts $\mG_{Pc}'$ defined in an alternative way, see Definition \ref{def:Petr&catal2}. For both $\mG_{Pc}$ and $\mG_{Pc}'$, there exists exactly one pangraph such that Petri net is its Levi digraph. Furthermore,
\begin{enumerate}
\item For $\mathcal{G}_{\textrm{Pc}}=((V,L,E_{\mathcal{G}_L}, \inci_{\mG_L},\inco_{\mG_L}),M)$, we have 
\begin{equation}\label{eq:pan_petri1}
V_{\mP}=S\cup C,\quad \{e\in E_{\mP}\,\,|\,\,D(e)=1\}=T,
\end{equation}
and for every catalysts $c\in V_\mP$, there exists $e\in E_{\mP}$ such that $e$ is a hyperloop, see the definition in Subsec. \ref{sec:H_graph}, namely $c\in e^{\textrm{in}}\cap e^{\textrm{out}}$;
\item For $\mG_{Pc}'=((V',\{S',T'\},E'_{\mathcal{G}_L'},  {\inci_{\mG_L}}',{\inco_{\mG_L}}'),M')$, we have 
\begin{equation}\label{eq:pan_petri2}
V_{\mP}=S\cup C,\quad \{e\in E_{\mP}\,\,|\,\,D(e)=1\}=T,\quad \{e\in E_{\mP}\,\,|\,\,D(e)=2\}=T_c
\end{equation}
and for every catalysts $c\in V_\mP$, there exists $e\in T_c$ such that $c\in e^{\textrm{in}}$.
\end{enumerate}
\end{prop}

\begin{proof} The existence of a pangraph that satisfies \eqref{eq:pan_petri1} and \eqref{eq:pan_petri2} is analogous to the proof of Proposition \ref{prop:Petri}. Let us show only the properties in 1 and 2. 
\begin{enumerate}
\item By connectedness of Petri net for every $c\in C$, there exists $t\in T$ such that $(\{c\},\{t\})\in E_{P_c}$ (or $(\{t\},\{c\})\in E_{P_c}$). By the property 4 in Definition \ref{def:Petr&catal}, we have also $(\{t\},\{c\})\in E_{P_c}$ (resp. $(\{c\},\{t\})\in E_{P_c}$). By the construction of a Levi representation of a pangraph, there exists $e\in E_{\mP}$ such that $c\in e^{\textrm{in}}\cap e^{\textrm{out}}$. 
\item The proof follows analogously to 1.
\end{enumerate}
\end{proof}

As the final conclusion of this subsection, let us note that pangraphs offer a consistent generalisation of Petri nets with catalysts to arbitrarily complex interaction modifications.

\subsection{Metabolic graphs}
Analogously, pangraphs generalise the notion of metabolic graphs defined in~\cite{Metabolic_graphs_McQuade}. A metabolic graph is a weighted dihypergraph endowed with additional signed objects called uberedges that can be associated with 2-depth panedges in a pangraph, connecting nodes to dihyperedges. We propose a strict mathematical definition of this object that agrees with considerations in~\cite{Metabolic_graphs_McQuade}.

\begin{definition}
    A \textbf{metabolic graph} is a 4-tuple $\mathcal{H}_{M}=(\mathcal{H},\mathcal{U}, \inci_M,\inco_M)$, such that $\mathcal{H}=(V,\EH,\inci_\mH,\inco_\mH)$ is a weighted dihypergraph, see Definition \ref{def:dihypergraph}, that satisfies \eqref{eq:multi-weights}, and
    \begin{enumerate}
  \item  $\mathcal{U}=\{u_k\,\,|\,\,k\in K'\}\subset \{(\{v_i\},\{e_j\})\;\;|\;\;v_i\in V,\,\,e_j\in E_{\mH}\}$ is a set of additional panedges that join vertices with dihyperedges;
  
    \item  $\inci_M,\inco_M\in \mathbb{M}_{|K'|\times |I|}(\{+,-\})$ are sign incidence matrices.
        \end{enumerate}
\end{definition}

Using the formula for $\psi$ defined in \eqref{eq:psi}, we conclude the next result. 
\begin{prop}
    A metabolic graph $\mathcal{H}_{M}=(\mathcal{H},U, \inci_M,\inco_M)$ is a $2$-depth pangraph $\mP$.
\end{prop}

\begin{proof} 
Let us denote by $\mH= (V_{\mH},\EH, \inci_{\mH},\inco_{\mH})$ a weighted dihypergraph from the definition of metabolic graph $\mG_{M}$. The demanded pangraph $\mP= (V_{\mP},\EP, \incip,\incop)$ is given by
\begin{equation*}
    V_{\mP}=V_{\mH},\quad E_{\mP}=E_{\mH}\cup U, 
\end{equation*}
while pangraph incidence matrices $\inci_{\mP},\inco_{\mP}\in \mathbb{M}_{|K\cup K'|\times |I|}$ are defined based on $\inci_M,\inco_M\in \mathbb{M}_{|K'|\times |I|}$ and $\inci_{\mH},\inco_{\mH}\in \mathbb{M}_{|K|\times |I|}$ as
\begin{equation}
(\mathcal{I}^{\textrm{x}}_{\mathcal{P}})_{ij}=\left\{\begin{array}{ll}
(\mathcal{I}^{\textrm{x}}_{\mathcal{H}})_{ij}&\textrm{for }e_i\in E_{\mH}\\ [.2cm]
(\mathcal{I}^{\textrm{x}}_{M})_{i-m j}&\textrm{for }e_i\in \mathcal{U},
\end{array}
\right.\quad x=\textrm{in, out},\quad  i\in K\cup K',\,\,j\in I,
\end{equation}
where $m=|K|$.
\end{proof}

Finally, it is worth mentioning that adding a dynamics into metabolic graphs in \cite{Metabolic_graphs_McQuade}, authors apply effective weights that in this paper are generalized to pangraphs in Definition~\ref{def:eff_flow} (compare Eq. (1.8) in~\cite{Metabolic_graphs_McQuade}).

\section{Network measures for pangraphs}\label{sec:measures}

In this section, we generalise commonly used graph measures for use with pangraphs. We focus on the duality offered by considering either the pangraphs themselves or their Levi graphs. A pangraph defines the system; consequently, it is the first choice when one needs to define a process properly without losing information about the roles of edges. On the other hand, the application of a Levi graph facilitates computations. 

In Section \ref{sec:real_exam}, we present the interpretation of each measure in the context of higher-order interactions in food webs.

\subsection{Degree Centrality}\label{subsec:deg_central}

The degree centrality of a vertex measures the connectivity of that vertex with other
vertices in the network. Classically, \textbf{degree centrality} of a vertex $v$ in a graph is defined as the sum of entries of the corresponding row of incidence or adjacency matrix (i.e. the number of edges that are incident to $v$ or, equivalently, the number of vertices that are adjacent to $v$). As stated in Subsection \ref{sec:m_graph}, these notions can be used interchangeably in the case of graphs, and consequently, the degree centrality of a vertex $v$ can be defined by either using its incidence or adjacency matrix, see \eqref{eq:degree}. 

One panedge can contain an arbitrary number of adjacent vertices and two adjacent vertices can be contained in more than one panedge. Then, similarly to the dihypergraph case, the degree centrality of a vertex in a pangraph can be defined in two separate ways: as incidence degree centrality or adjacency degree centrality. 

What should be the domain for which we define the degree centrality? Section \ref{sec:pangraph} specifies three candidates: fundamental vertices $V$, generalized vertices $V_{\mP}$, and elements of $V\cup E_{\mP}$. Vertices in $V_{\mP}$ turn out to offer a coherent generalisation from hypergraph degree centralities to the pangraph case. 
\begin{definition}[Incidence degree centrality in a weighted pangraph]\label{def:inci_deg}
    Let $\mP= (V,\EP, \incip,\incop)$ be a weighted pangraph and let $V_{\mP}$ be the set of vertices defined in \eqref{eq:U_ver}. \textbf{Incidence in-degree (resp. out-degree) centrality} $ \kappa_{\mP,I}^{\tin}(v_i)$ (resp. $\kappa_{\mP,I}^{\out}(v_i)$) of a vertex $v_i \in V_{\mP}$ is the sum of weights of the directed panedges $e_j=(e_j^{\mathrm{\tin}}, e_j^{\mathrm{\out}})$ such that $v_i \in e_j^{\tin}$ (resp. $v_i \in e_j^{\out}$). Namely,
    
\begin{equation} \label{eq:incidence_deg}
    \kappa_{\mP,I}^{\mathrm{x}}(v_i):=\sum_{e_j\in E_{\mP}}\mathcal{I}^{\textrm{x}}_{ji},\qquad x=\textrm{in},\textrm{out}.
\end{equation}
\end{definition}

Note that the formula \eqref{eq:incidence_deg} can be applied also to $v_i\in E_{\mP}\setminus V_{\mP}$. However, since $v_i$ is not a head nor a tail of any other panedge, then $\kappa_{\mP,I}^{\mathrm{in}}(v_i)=\kappa_{\mP,I}^{\mathrm{out}}(v_i)=0$. Thus, it does not bear any valuable information. On the other hand, if we decide to define incidence degree centrality for all $v\in V\cup E_{\mP}$, then for a digraph, hence $1$-depth pangraph, it simplifies to a set $V\cup E_{\mG}$ and this does not agree with the standard definition of degree centrality. On the contrary, $V_{\mP}=V$ for a digraph, and hence Definition \ref{def:inci_deg} can be considered as a generalisation of the classical notion of vertex degree centrality.

On the other hand, the adjacency degree centrality of $v_i$ (which we denote by $\kappa_{\mP,A} (v_i)$) measures the weighted number of vertices $v_i$ interacts with. Each interaction that contains $v_i$ and its adjacent vertices is counted separately and added. To be more precise let us consider an example.

\begin{exam}\label{exam:I_A_degree}
Consider an unweighted $\mP=(V,\EP)$ such that $V=\{ v_1,v_2,v_3\}$ is a set of fundamental vertices, 
\begin{eqnarray*}
E_{\mP}=\left\{ e_1 = (\{v_1\},\{v_2\}), e_2 = (\{v_1\},\{v_2,v_3, e_1\} ), e_3 =(\{v_1,e_2\}, \{e_1\})\right\}.
\end{eqnarray*} 
Under the assumption that all panedges are of the same weight equal to $1$ we calculate vertex dergees.

Adjacency in-degree centrality of $v_1$ should take into account five vertices: $v=v_2$ via edge $e_1$, $v=v_2,v_3,e_1$ via edge $e_2$ and $v=e_1$ via edge $e_3$. Hence $$\kappa_{\mP,A}^{\tin}(v_1)=5\quad \textrm{while}\quad \kappa^{\tin}_{\mP,I}(v_1)=3.$$ 
Analogously, adjacency out-degree centrality of $v=e_1$ should take into account two vertices $v=v_1,e_2$ via $e_3$, and $v=v_1$ via $e_2$ consequently $$\kappa_{\mP,A}^{\out}(e_1)=3\quad \textrm{while} \quad \kappa^{\out}_{\mP,I}(e_1)=2.$$
\end{exam}

The adjacency degree centrality captures how many other influencing/influenced vertices a vertex is connected to.

\begin{definition}[Adjacency degree centrality in a weighted pangraph]\label{def:adj_deg}
     Let $\mP= (V,\EP, \incip,\incop)$ be a weighted pangraph. \textbf{Adjacency in-degree centrality} (resp. \textbf{out-degree centrality}) $ \kappa_{\mP,A}^{\tin}(v_i)$ (resp. $ \kappa_{\mP,A}^{\out}(v_i)$) of a vertex $v_i \in V_{\mP}$ is defined as the sum of weights of incident panedges including the multiplicity of their heads (resp. tails) namely for $x\in\{\textrm{in},\textrm{out}\}$ and $y\in\{\textrm{in},\textrm{out}\}\setminus\{\textrm{x}\}$
     \begin{eqnarray*}
        \kappa_{\mP,A}^{\textrm{x}}(v_i):=\sum_{v_k\in V_{\mP}}(\mathcal{A}^{\textrm{x}}_{\mP})_{ki}=\sum_{e_j\in E_{\mP}} \mathcal{I}^{\textrm{x}}_{ji}\left(\sum_{v_k\in V_{\mP}}  \overline{\mathcal{I}}^{\textrm{y}}_{jk}\right) .
    \end{eqnarray*}
\end{definition}

In order to better understand the relation between the pangraph and its Levi graph representation let us compare their incidence degree centralities for elements $v_i\in V_{\mP}\subset V_{\mathcal{L}}$, hence elements for which both indices are defined. 

Let us denote by  $\kappa_{\mP,I}$, $\kappa_{\mL,I}$ the incidence degree centrality of pangraph $\mP$ and its Levi graph $\mL(\mathcal{P})$; and by $\kappa_{\mP,A}$, $\kappa_{\mL,A}$ adjacency degree centrality of pangraph $\mP$ and its Levi graph $\mL(\mathcal{P})$.

\begin{prop}
Let $\mP= (V,\EP, \incip,\incop)$ and  $\mL(\mathcal{P})=(V_{\mathcal{L}},L,E_{\mathcal{L}}, \mathcal{I}^{\textrm{in}}_{\mathcal{L}},\mathcal{I}^{\textrm{out}}_{\mathcal{L}})$ be a pangraph and its Levi graph, respectively. If $\mathrm{x}=\textrm{in},\textrm{out}$ and $i\in K$, then the following relations hold for the incidence and adjacency degree centrality in pangraph and its Levi graph:
\begin{enumerate}
\item $\kappa_{\mL,I}^{\mathrm{x}}(v_i)=\kappa_{\mL,A}^{\mathrm{x}}(v_i)
$ for $v_i \in  V_{\mathcal{L}}$;
\item $\kappa_{\mP,I}^{\mathrm{x}}(v_i)\leq \kappa_{\mP,A}^{\mathrm{x}}(v_i)$ for $v_i\in V_{\mP}$;

\item for $y\in \{\textrm{in},\textrm{out}\}\setminus\{\textrm{x}\}$
\begin{equation*}
\kappa_{\mL, I}^{\mathrm{x}}(v_i)=\begin{cases}
\kappa_{\mP,I}^{\mathrm{x}}(v_i)&\textrm{for }v_i\in V\subset V_{\mathcal{L}};\\
\kappa_{\mP,I}^{\mathrm{x}}(v_i)+ \sum_{\{e_j\in E_{\mP}\,\,|\,\,D(e_j^{\textrm{x}}),D(e_j^{\textrm{y}})<D(v_i)\}}(\mathcal{I}^{\textrm{x}}_{\mP})_{ji}& \text{for }v_i\in V_{\mP}\setminus V\subset V_{\mathcal{L}}.
\end{cases}
\end{equation*}
\end{enumerate}
\end{prop}

\begin{proof}
The equalities follow from the definitions of incidence and adjacency degree centrality, Definitions \ref{def:inci_deg} -- \ref{def:adj_deg}, as well as the definition of Levi digraph, Definition \ref{def:levi-graph} .
\end{proof}

The above result indicates that a subtle difference in the incidence and adjacency degree centralities of a vertex $v_i\in V_{\mP}$ is lost while approximating a pangraph by its Levi graph. In the Levi graph case, both incidence and adjacency degree centralities are always equal. Furthermore, a comparison of the degree centrality of fundamental vertices ($v\in V$) and other vertices $v\in V_{\mP}\setminus V$ using Levi representation cannot give a reliable result. An example below shows that an error can not only change the centrality results quantitatively but also qualitatively by changing the order of the most central vertices. 

\begin{exam}
Let us return to the Example \ref{exam:I_A_degree}. The only nonzero incidence in-degrees in the pangraph $\mP$  are given by
\begin{equation*}
\kappa_{\mP,A}^{\mathrm{in}}(v_1)=5,\quad \textrm{and}\quad \kappa_{\mP,A}^{\mathrm{in}}(e_2)=1;
\end{equation*}
whereas in its Levi graph $\mG_{\mathcal{L}}(\mP)$ they read
\begin{equation*}
\kappa_{\mL,A}^{\mathrm{in}}(v_1)=3,\quad \kappa_{\mL,A}^{\mathrm{in}}(e_2)=4,\quad \kappa_{\mL,A}^{\mathrm{in}}(e_3)=1.
\end{equation*}
Consequently, the order of centrality in pangraph and Levi graph differ and is given respectively by
\begin{equation*}
\kappa_{\mP,A}^{\mathrm{in}}(v_1)>\kappa_{\mP,A}^{\mathrm{in}}(e_2);\quad \kappa_{\mL,A}^{\mathrm{in}}(e_2)>\kappa_{\mL,A}^{\mathrm{in}}(v_1)>\kappa_{\mL,A}^{\mathrm{in}}(e_3).
\end{equation*}

\end{exam}

Consequently, degree centralities for pangraphs cannot be derived from the degree centralities of its Levi representation.

\subsection{Katz centrality}\label{sec:Katz_centrality}
A classic example of a recursive measure is the \textbf{Katz centrality} that defines vertex importance as being connected to (many) important vertices. The contribution from a~neighbour is multiplied by the weight of the connecting edge and a universal damping factor $\alpha$. In directed graphs \textbf{Katz in-centrality} means being influenced by many important vertices, while \textbf{Katz out-centrality} signifies influencing many important vertices. 

Formally, for any vertex $v_i\in V$ in a digraph $\mathcal{G}=(V,E_{\mG}, \inci_{\mG},\inco_{\mG})$ one can define Katz in-- and out-- centrality, respectively $c^{\tin}(v_i), c^{\out}(v_i)$, using adjacency matrices in the following way:
\begin{eqnarray}\label{eq:c_recursive}
   c_{\mG}^{\textrm{in}}(v_i)&=&\alpha \sum_{v_j\in V}  (\mathcal{A}^{\textrm{in}}_{\mG})_{ij}  c_{\mG}^{\textrm{in}}(v_j) + \beta_i,\\
   c_{\mG}^{\textrm{out}}(v_i)&=&\alpha \sum_{v_j\in V}  (\mathcal{A}^{\textrm{out}}_{\mG})_{ij}  c_{\mG}^{\textrm{out}}(v_j) + \beta_i,
\end{eqnarray}

\noindent where $\mathcal{A}_{\mG}^{\tin},\mathcal{A}_{\mG}^{\out}$ are in- and out-adjacency matrices of a digraph $\mathcal{G}$ defined in \eqref{eq:adjacency-in} -- \eqref{eq:adjacency-out}, and $\alpha, \beta_i\in \mathbb{R}$, $i\in I$, are fixed constants. A Levi representation of a pangraph $\mP$, $\mL(\mP)$ is a digraph, hence one can define a Katz centrality measure for $\mL(\mP)$. We denote it by $c_{\mL}^{\textrm{x}}$, $\textrm{x}=\tin, \out$. Katz centrality vector $c_{\mL}=(c_{\mL}(v_i))_{i\in I}$ is well defined, according to formula \eqref{eq:c_recursive}, if $\|\alpha\mathcal{A}_{\mL}^{\textrm{x}}\|<1$ and it is given for Levi graph by concise formula, with $\beta=(\beta_i)_{i\in I}$,
\begin{equation}\label{eq:c_concise}
c_{\mL}^{\textrm{x}}:=(\mathcal{I}-\alpha\mathcal{A}_{\mL}^{\textrm{x}})^{-1}\beta=\sum_{n=0}^{\infty}(\alpha\mathcal{A}_{\mL}^{\textrm{x}})^n\beta,\qquad \textrm{where }x\in \{\tin,\out\}.
\end{equation}

If we note that the $n$-th power of the adjacency matrix informs about the weight of all walks between fixed vertices, one can interpret Katz centrality as the weight of all walks starting/terminating at $v_i$, reaching all possible vertices, with different weights $\beta$ depending on the vertex at the end/beginning of a walk.

Pangraph walks can lead over panedges, see Def.~\ref{def:walk}, which guides us to define Katz centrality using adjacency matrix giving the relation between elements of $V_{\mP}$, see Def.~\ref{eq:adj_and_inc3}. This domain contains the panedges that play the role akin to vertices. Exactly those contained in the head or tail of another vertex allow walks to proceed panedge after panedge. Namely,
\begin{equation}\label{katz_pangraph_adjacency}
c_{\mP}^{\textrm{x}}:=(\mathcal{I}-\alpha\mathcal{A}_{\mP}^{\textrm{x}})^{-1}\beta=\sum_{n=0}^{\infty}(\alpha\mathcal{A}_{\mP}^{\textrm{x}})^n\beta,\qquad \textrm{where }x\in \{\tin,\out\}.
\end{equation}

Similarly to the considerations in Subsection \ref{subsec:deg_central}, we show that calculations of Katz centrality for a pangraph and its Levi representation give qualitatively different results.

\begin{exam}
Let us return to the Example \ref{exam:I_A_degree}. Since for the pangraph $\mP$ and its Levi representation $\mL(\mP)$ all weights are equal $1$, $$\left\|\mathcal{A}^{\tin}_{\mP}\right\|=\max_{j\in I'}\sum_{i\in I'}(\mathcal{A}^{\tin}_{\mP})_{ij}=4,\quad  \textrm{and} \quad \left\|\mathcal{A}^{\tin}_{\mL}\right\|=\max_{j\in I'}\sum_{i\in I'}(\mathcal{A}^{\tin}_{\mL})_{ij}=3,$$
then we choose $\alpha=0.2$. For $\beta=\mathbf{1}^T$ we obtain
\begin{equation*} c_{\mP}=\left[\begin{array}{ccccc}1&1.4&1.2&1.4&1\end{array}\right]^T\qquad \textrm{and}\qquad  c_{\mL}= \left[\begin{array}{cccccc}1&1.3744&1.16&1.072&0.8&1.36\end{array}\right]^T.
\end{equation*}
Consequently, the rank of centralities differs between pangraph and its Levi digraph, e.g. comparing $e_1$ and $v_3$ in terms of Katz centrality. Their ranks are given by \begin{eqnarray*}
&c_{\mP}^{\tin}(v_1)=c_{\mP}^{\tin} (e_2)<\underline{c_{\mP}^{\tin} (v_3)}<\underline{c_{\mP}^{\tin} (e_1)}=c_{\mP}^{\tin} (v_2), \\[0.4cm]
&c_{\mL}^{\tin} (e_2)<c_{\mL}^{\tin} (v_1)<\underline{c_{\mL}^{\tin} (e_1)}<\underline{c_{\mL}^{\tin} (v_3)}<c_{\mL}^{\tin} (e_3) < c_{\mL}^{\tin} (v_2).
\end{eqnarray*}
\end{exam}

This has general consequences for any recursive graph measure. To know its value for a given vertex we have to compute it for another, effectively walking over the network. The measure's proper generalization to pangraphs can walk only over elements of $V_{\mP}$. Its digraph version for the Levi graph will walk also over all the remaining panedges that do not belong to $V_{\mP}$.

\subsection{Generalized pangraph Katz centrality}\label{sec:gen_pan_Katz}

The classical Katz centrality treats relations differently based on whether they are modified and belong to $V_{\mP}$, or not. This might also treat processes of identical physical nature differently. In this subsection, we propose a centrality measure based on Katz centrality that treats vertices and all interactions equally and assigns centralities to all of them.

 We say that $\tilde{\mathcal{A}}_{\mP}^{\mathrm{x}} \in \mathbb{M}_{(n+m) \times (n+m)}([0,\infty)])$, $\mathrm{x}=\tin, \out$ is a \textbf{generalized in- and out- adjacency matrix} of a pangraph $\tilde{\mathcal{A}}_{\mP}^{\mathrm{x}}=(\tilde{\mathcal{A}}_{\mP}^{\mathrm{x}})_{ij})_{i,j=1,\ldots n+m}$ if

\begin{equation}\label{eq:def_gen_adj}
(\tilde{\mathcal{A}}_{\mP}^{\mathrm{x}})_{ij}=\left\{\begin{array}{cl}
(\mathcal{A}_{\mL}^{\mathrm{x}})_{ij}&\textrm{for } i\geq j\\
1&\textrm{for } i< j\textrm{ and } (\mathcal{A}_{\mL}^{\mathrm{x}})_{ij}\neq 0\\
0&\textrm{otherwise}
\end{array}\right.;\qquad
\textrm{x}=\tin,\out.
\end{equation}

Note that in the case when all pangraph edges are equal to one, we have $\tilde{\mathcal{A}}_{\mP}^{\mathrm{x}}=\mathcal{A}_{\mL}^{\mathrm{x}}$, $\textrm{x}=\tin,\out$.

We define \textbf{generalized} \textbf{Katz centrality} in pangraphs for any vertex or a panedge $v_i\in V\cup \EP$ in the following way. The generalized Katz in-/out-centrality means being influenced by/influencing other central panedges or fundamental vertices. This retains the recursive logic of the classical digraph Katz centrality.

\begin{align}\label{eq:Katz_from_Levi}
    \tilde{c}^{\,\tin}=\tilde{\alpha}^{\,\tin} \, \tilde{\mathcal{A}}^{\,\tin}_{\mP}\, \tilde{c}^{\,\tin} + \tilde{\beta}^{\,\tin}, \qquad \textrm{and}\qquad
    \tilde{c}^{\,\out}=\tilde{\alpha}^{\,\out}  \,\tilde{\mathcal{A}}^{\,\out}_{\mP}\, \tilde{c}^{\,\out} + \tilde{\beta}^{\,\out}.
\end{align}

We note that the generalized pangraph Katz centrality $\tilde{c}$ does not \emph{simply} reduce to classical Katz centrality when the pangraph is a digraph or a dihypergraph, as also shown in subsection \ref{sec:Katz_centrality}. However, one can find values of $\tilde{\alpha}, \tilde{\beta}$ that would make the generalized Katz centrality vector $\tilde{c}$ equal to that of Katz centrality $c_{\mP}$ for another choice of $\alpha, \beta$. In general, $\beta$ would have to be a nontrivial vector. Let us split the generalized Katz centrality vector $\tilde{c}$ and vector $\tilde{\beta}$ into two subvectors associated with centralities of vertices and edges of a hypergraph 
\begin{equation}\label{eq:c_split}
\tilde{c}^{\,\textrm{x}}_{\mH, (\tilde{\alpha},\tilde{\beta})}=[\tilde{c}^{\,\textrm{x}}_{\mH, (\tilde{\alpha},\tilde{\beta})}(v), \tilde{c}^{\,\textrm{x}}_{\mH, (\tilde{\alpha},\tilde{\beta})}(e)]^T,\qquad \tilde{\beta}^{\,\textrm{x}}=[\tilde{\beta}^{\,\textrm{x}}(v),\tilde{\beta}^{\,\textrm{x}}(e)]^T.
\end{equation}

\begin{thm}\label{thm:alpha-beta}
For any hypergraph $\mH=(V, \EH,\inci_\mH,\inco_\mH)$ there exists two pairs of parameters $(\alpha^{\,\textrm{x}}, \beta^{\,\textrm{x}}),$ $ (\tilde{\alpha}^{\,\textrm{x}}, \tilde{\beta}^{\,\textrm{x}})\in [0,\infty)^{n+m+1}$, such that 
\begin{itemize}
\item[i)] generalized in- and out- Katz centrality measures $\tilde{c}_{\mH,(\tilde{\alpha},\tilde{\beta})}^{\,\textrm{x}}$, $\textrm{x}=\tin, \out$ calculated for fundamental vertices $v\in V$ with coefficients $(\tilde{\alpha}^{\,\textrm{x}}, \tilde{\beta}^{\,\textrm{x}})$, ; and in- and out- Katz centarility measures $c_{\mH,(\alpha,\beta)}^{\,\textrm{x}}$, $\textrm{x}=\tin, \out$, for corresponding fundamental vertices calculated with coefficients $(\alpha^{\,\textrm{x}}, \beta^{\,\textrm{x}})$, are well-defined;
\item[ii)] if for $\textrm{x}\in\{\tin,\out\}$ and $\textrm{y}\in\{\tin,\out\}\setminus\{\textrm{x}\}$  
\begin{equation}\label{eq:as_tw2}
\alpha^{\textrm{x}}=(\tilde{\alpha}^{\textrm{x}})^2,\qquad \beta^{\textrm{x}}=\tilde{\alpha}^{\textrm{x}}(\overline{\mathcal{I}}_{\mH}^{\,\textrm{y}})^T \tilde{\beta}^{\textrm{x}}(e)+\tilde{\beta}^{\textrm{x}}(v),
\end{equation}
then the following equality holds
\begin{equation}\label{eq:K_cent_equal}
\tilde{c}^{\,\textrm{x}}_{\mH,(\tilde{\alpha},\tilde{\beta})}=c^{\,\textrm{x}}_{\mH,(\alpha,\beta)},\qquad \textrm{x}=\tin,\out.
\end{equation}
\end{itemize}
\end{thm}
\begin{proof}
Let us represent a generalized adjacency matrix of a hypergraph using the block-matrix representation \eqref{eq:adj_block}. Levi digraph of hypergraph is a bipartile digraph, see \eqref{partition}. Furthermore, by the Theorem \ref{thm:Levi_vs_pan} the diagonal blocks are zero matrices. Using the notation $0_{p\times q}\in \mathbb{M}_{p\times q}(\{0\})$, for any $p,q\in \mathbb{N}$, a generalized adjacency matrix of dihypergraph reads
\begin{equation}\label{eq:block-adj}
\tilde{\mathcal{A}}_{\mH}^{\mathrm{x}}=\left[\begin{array}{cc}
0_{n\times n}&(\overline{\mathcal{I}}_{\mH}^{\,\textrm{y}})^T\\
\mathcal{I}_{\mH}^{\,\textrm{x}}&0_{m\times m}
\end{array}\right],\qquad \textrm{x}\in\{\tin,\out\},\,\, \textrm{y}\in\{\tin,\out\}\setminus\{\textrm{x}\}.
\end{equation}
Using the representations \eqref{eq:Katz_from_Levi}, \eqref{eq:block-adj} and \eqref{eq:c_split} one can calculate the generalized Katz centralities for dihypergraph vertices
\begin{eqnarray}
\tilde{c}^{\,\textrm{x}}_{\mH, (\tilde{\alpha},\tilde{\beta})}(v)=\tilde{\alpha}\, (\overline{\mathcal{I}}_{\mH}^{\,\textrm{y}})^T \,\left(\tilde{\alpha}\mathcal{I}_{\mH}^{\,\textrm{x}} \, \tilde{c}^{\,\textrm{x}}_{\mH, (\tilde{\alpha},\tilde{\beta})}(v)+\tilde{\beta}(e)\right)+\tilde{\beta}(v).
\end{eqnarray}
Using the assumption \eqref{eq:as_tw2} and the definitions of adjacency matrices for hypergrphs \eqref{eq:adj_and_inc2}, we note that centralities $\tilde{c}^{\,\textrm{x}}_{\mH, (\tilde{\alpha},\tilde{\beta})}(v)$ satisfy condition
\begin{eqnarray}
\tilde{c}^{\,\textrm{x}}_{\mH, (\tilde{\alpha}^{\,\textrm{x}},\tilde{\beta}^{\,\textrm{x}})}(v)=\alpha\,  \mathcal{A}^{\textrm{x}}_{\mH}\tilde{c}^{\,\textrm{x}}_{\mH, (\tilde{\alpha}^{\,\textrm{x}},\tilde{\beta}^{\,\textrm{x}})}(v)+\beta^{\,\textrm{x}}.
\end{eqnarray}
Hence, condition \eqref{eq:K_cent_equal} is satisfied. Finally, we choose parameters $(\tilde{\alpha}^{\,\textrm{x}}_{\mH})^{\star}, (\alpha^{\,\textrm{x}}_{\mH})^{\star}<1$ such that the generalized Katz centralities and Katz centralities are well defined. Define \begin{equation}\label{eq:alpha}
\tilde{\alpha}^{\,\textrm{x}}_{\mH}:=\min((\tilde{\alpha}^{\,\textrm{x}}_{\mH})^{\star}, (\alpha^{\,\textrm{x}}_{\mH})^{\star})<1
\end{equation}
and consequently $\alpha^{\,\textrm{x}}_{\mH}<(\alpha^{\,\textrm{x}}_{\mH})^{\star}$ and both centrality measures with parameters satisfying \eqref{eq:alpha} and \eqref{eq:K_cent_equal} are well-defined.
\end{proof}
Theorem~\ref{thm:alpha-beta} indicates that for hypergraphs, generalized Katz centrality should be indeed considered as the generalization of classical Katz centrality, provided that the appropriate parameters satisfy conditions \eqref{eq:K_cent_equal}. The graph measure defined in this way agrees numerically with the classical Katz centrality measure but allows us to determine additionally the centrality of interaction given by dihypergraph edges. 

Note that in the case of pangraphs a similar comparison is not true.

\section{Comparison of different graph models of higher-order phenomena}\label{sec:real_exam}

The existing literature proposed mapping interaction modifications to hyperedges~\cite{GOLUBSKI2016344, BATTISTON20201}. We show that it leads to a loss of critical information about the roles of vertices in such an interaction. We also compare Katz centralities of vertices in both hypergraph and pangraph realization of the same real-world system~\cite{GOLUBSKI2016344}, analytically and numerically. We show that the choice of representation leads to significantly different conclusions about vertex importance, also changing vertex order in terms of centrality.

\subsection{Dihypergraph and pangraph representations of interaction modifications}\label{subsec:hyp_vs_pan}

Let us start by formalizing the correspondence between a dihypergraph model of a system with higher-order interactions as described in~\cite{GOLUBSKI2016344} and a pangraph representation of the same system. In the first approach, each interaction modification is mapped to a dihyperedge connecting all vertices involved in the modified relation. In the pangraph case, the modification appears as a panedge of depth larger than $1$. 

We map a causal walk $W_c$ from $v_0$ to $v_l$, $(v_0,v_l\in V)$ over panedges $e_1, \ldots, e_{l-1}$, see Definition \ref{def:walk}, to a directed hyperedge whose set of tails consists of all tails of edges $e_1, \ldots, e_{l-1}$ that are vertices. Its set of heads equals the set of all heads of $e_1, \ldots, e_{l-1}$ that are vertices. In this representation, we consider walks that do not contain a cycle, i.e., $e_i\neq e_j$, for any $i=1,\ldots,l-1$, $i\neq j$. Equation~\eqref{def:all_tails_heads} allows us to define this dihyperedge formally.

\begin{definition}\label{def:related_pangraph}
    Given an unweighted pangraph $\mP = (V, \EP)$, a related unweighted hypergraph $\mH(\mP) =(V, \EH)$ satisfies condition
    \begin{equation*}
    \EH=\left\{\left(\left\{v_0,\bigcup_{m=1}^{l-1}\mathcal{V}^{\textrm{in}}(e_m)\right\},\left\{v_l,\bigcup_{m=1}^{l-1}\mathcal{V}^{\textrm{out}}(e_m)\right\}\right)\,\,|\,\, (v_0, e_1, ... , e_{l-1},v_l)\,-\textrm{causal walk}\right\},
    \end{equation*}
    where the sets $\mathcal{V}^{\tin}$, $\mathcal{V}^{\out}$ are defined recursively in procedure \eqref{eq:v}, \eqref{def:all_tails_heads}.
\end{definition}

\begin{figure}[ht]
    \centering\includegraphics[width=0.9\linewidth]{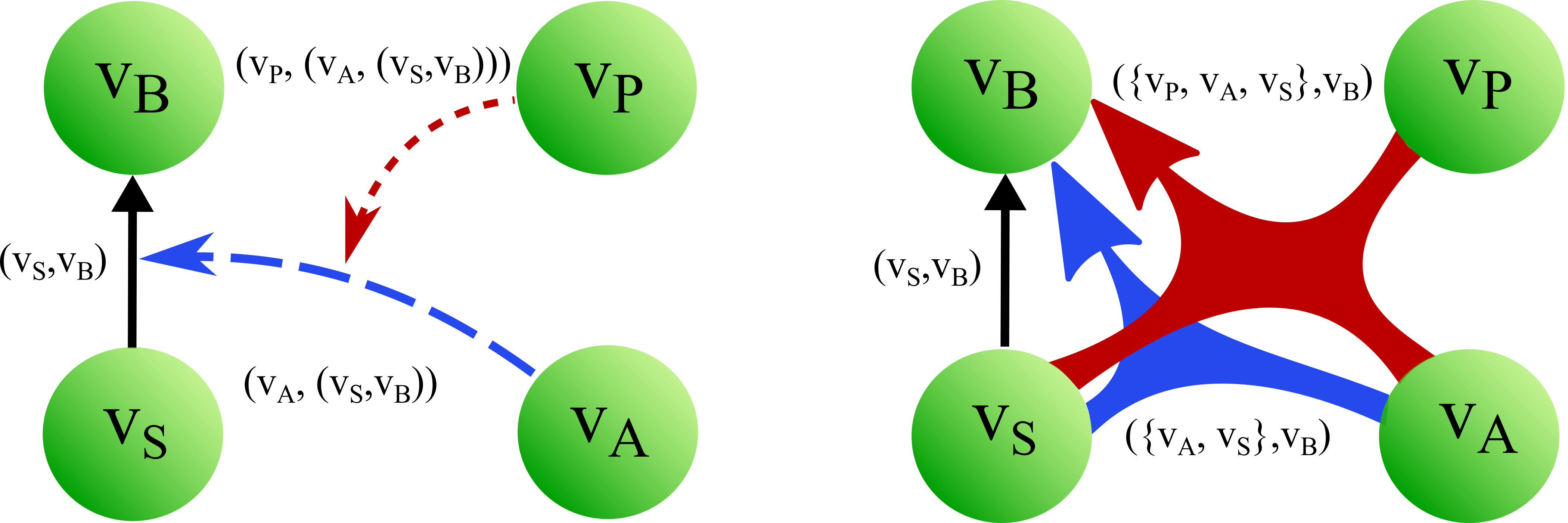}
    \caption{Pangraph walks that do not pass through intermediate fundamental vertices are mapped to hyperedges. Left: a subgraph of the coffee agroecosystem model. Right: its hypergraph representation as proposed in~\cite{GOLUBSKI2016344}. We skipped brackets around one-element sets.}\label{fig:pangraph_hypergraph_mapping}
\end{figure}

This correspondence allows us to compare graph measures calculated for a hypergraph with a pangraph representation of the same system. Let us trace the relationships between the Katz centralities of the vertices $v_0$ and $v_l$ in both representations. As our main example~\cite{GOLUBSKI2016344} and the reason for the hypergraph-pangraph comparison used only unweighted graphs, we constrain ourselves to this setting. We focus on the relationship between the centralities of vertices connected by the hyperedges in the hypergraph representation. 

Let us consider the pangraph Katz out-centrality $c_{\mP}^{\out}$ and hypergraph Katz out-centrality $c_{\mH}^{\out}$. The reasoning for in-centralities would be analogous, just backward from $v_l$ to $v_0$. We denote by $\textrm{Foll}_{W_c}(s)$, $s\in \{v_0,e_1,\ldots, e_{l-1}\}$ the following entity on the walk $W_c$. Vertices $v_0, v_l$ and edges $e_1, \ldots, e_{l-1}$ may have more connections than those on the walk $W_c$, and we denote their additional contributions to the centrality $c_{\mP}^{\out}(v_i)$ of a generalized vertex $v_i\in V\cup E_{\mP}$ as $c_{\
\mP}^{\out}(R_{v_i})$, namely
\begin{equation*}
c_{\mP}^{\out}(R_{v_i})=\alpha_{\mP}\sum_{v_j\in V\cup E_{\mP}\setminus \{\textrm{Foll}_{W_c}(v_i)\}}(\mathcal{A}_{\mP}^{\out})_{ij} c_{\mP}^{\out}(v_j).
\end{equation*}
 We have
\begin{eqnarray*}
    (c_{\mP}^{\out})_{v_0}&=&\alpha_{\mP} (c_{\mP}^{\out})_{e_1} + (c_{\mP}^{\out})_{R_{v_0}} + \beta_{\mP},\\
\end{eqnarray*}
 Then,
\begin{equation}
    (c_{\mP}^{\out})_{v_0}=\alpha_{\mP}^2  (c_{\mP}^{\out})_{e_2}  + \alpha_{\mP}  \left(  (c_{\mP}^{\out})_{R_{e_1}}+ \beta_{\mP} \right) + (c_{\mP}^{\out})_{R_{v_0}} + \beta_{\mP} .
\end{equation}

Applying the recursive definition of Katz centrality further along the walk $w$, we finally obtain
\begin{equation}\label{eq:pangraph_Katz_walk}
    (c_{\mP}^{\out})_{v_0}=\alpha_{\mP}^l (c_{\mP}^{\out})_{v_l}+ \sum_{m=1}^{l-1} \alpha_{\mP} ^{m-1} \left(  (c_{\mP}^{\out})_{R_{e_m}} + \beta_{\mP} \right) + (c_{\mP}^{\out})_{R_{v_0}} + \beta_{\mP} .
\end{equation}

In the hypergraph case, the walk and all its subwalks have been substituted by hyperedges. All tails of each edge $e_m$ have as many direct connections to $v_l$ as their distance to the $v_0$, so
\begin{equation}\label{eq:hypergraph_Katz_walk}
    (c_{\mH}^{\out})_{v_0}=\alpha_{\mH}  \left(\sum_{m=1}^{l-1} m \sum_{v \in \mathcal{V}_{D(e_m)^{\out}}} (c_{\mH}^{\out})_v\right) + (c_{\mH}^{\out})_{R_{v_0}}+ \beta_{\mH} .
\end{equation}
The impact of $v_0$ on $v_l$ is thus reduced by a factor $\alpha_{\mP}^l$ in a pangraph and by $\alpha_{\mH}$ in a hypergraph. A contribution from a vertex connected to $e_m$ is multiplied by $\alpha_{\mP}^{m-1}$, while in a hypergraph by $\alpha_{\mH}m$. 

In summary, the dihypergraph representation creates direct connections between vertices whereas in pangraph this structure is represented by the walk. It also gives vertices connected to intermediate edges additional connections to the final vertex, one for each edge earlier than themselves.

\subsection{Coffee agroecosystem}\label{sec:coffee_agroeco}

An empirical network model of a coffee agroecosystem~\cite{Vandermeer_2010, Perfecto_coffe_agroeco, GOLUBSKI2016344} describes direct pairwise interactions between species, as well as influences strengthening or weakening other interactions. The authors of~\cite{GOLUBSKI2016344} mapped these interactions to an unweighted undirected hypergraph. They postulated such a representation of systems with interaction modifications, which was further reiterated by~\cite{BATTISTON20201}. In this paper, we prove that the described system explicitly portrays a 3-depth pangraph and discuss the advantages of pangraph approach. 

The notion of a pangraph substantially simplifies the model for at least two reasons. Representing a complex ecosystem as an undirected hypergraph introduces ambiguity about the role of a vertex in an interaction. The test of the soundness of the representation, conducted in ~\cite{GOLUBSKI2016344} by vertex removal and edge addition, clearly showed that unweighted undirected hypergraphs might incorrectly inflate centrality measures by considered higher-order interactions which in reality weaken the interaction. This problem can be easily resolved by considering the directed case that clearly distinguishes between being influenced and influencing. 

The second challenge indicated by the authors of ~\cite{GOLUBSKI2016344} is the need to use different weights for each vertex on a hyperedge, remarking that assigning and combining weights would be a non-trivial task. In pangraph approach, the same goal can be achieved with standard graph weights assigned to the whole panedge in which each vertex's role is clear.

\subsubsection{Dihypergraph and pangraph representations}
We extend the representation of the coffee agroecosystem~\cite{GOLUBSKI2016344} to a dihypergraph and construct its pangraph representation. They represent interactions between species mapped to (fundamental) vertices $V$. The original coffee agroecosystem hypergraph, which can be found in the supplementary material of~\cite{GOLUBSKI2016344}, is undirected, and neglects the difference between bidirectional and unidirectional mutualistic interactions, see Sec.~\ref{subsec:mutualistic_nets}. We added information about directions using Fig.~1 of~\cite{GOLUBSKI2016344}. We also applied it to the pangraph representation. Each edge ending with signs at both ends has been replaced by two panedges/hyperedges with appropriate orientation, as described in Sects.~\ref{subsec:mutualistic_nets} and~\ref{subsec:hyp_vs_pan}. The one-sided edges have been kept in an appropriate direction, pointing to the sign.

The weights (1 and -1) from Fig.~1 of ~\cite{GOLUBSKI2016344} were omitted in the corresponding matrix representations and numerical analysis. We set the weights to $1$ in order to consistently compare the representations. We also derive the Levi graphs of the dihypergraph and the pangraph representations as well as present visualizations of the Levi graphs in Appendix~\ref{sec:visualisations}.

\subsubsection{Comparison of centralities}

In this section, we compare dihypergraph and pangraph representations of the coffee agroecosystem using Katz centrality. The analysis of the generalized pangraph Katz centralities, as a measure which for the purpose of the comparison is supplementary to the classic Katz centrality, (defined in Section~\ref{sec:gen_pan_Katz}) can be found in Appendix~\ref{sec:generalized-centrality-analysis}.
A vector of Katz centralities is calculated based on the adjacency relations of each network representation, as encoded by respective adjacency matrices (see Eq.~\ref{katz_pangraph_adjacency}). We computed Katz centralities with $\beta = 1$, $\alpha=\frac{0.9}{\lambda}$, where $\lambda$ stands for the largest eigenvalue of the respectful in- or out-adjacency matrix. We obtained 20 centrality values for the fundamental vertices $v_i \in V$ that we compare between dihypergraph and pangraph representation. The pangraph representation also allows us to estimate centralities of the 46 panedges $e_j \in V_P$.

\begin{figure}[ht]
    \centering
    \includegraphics[width=1\linewidth]{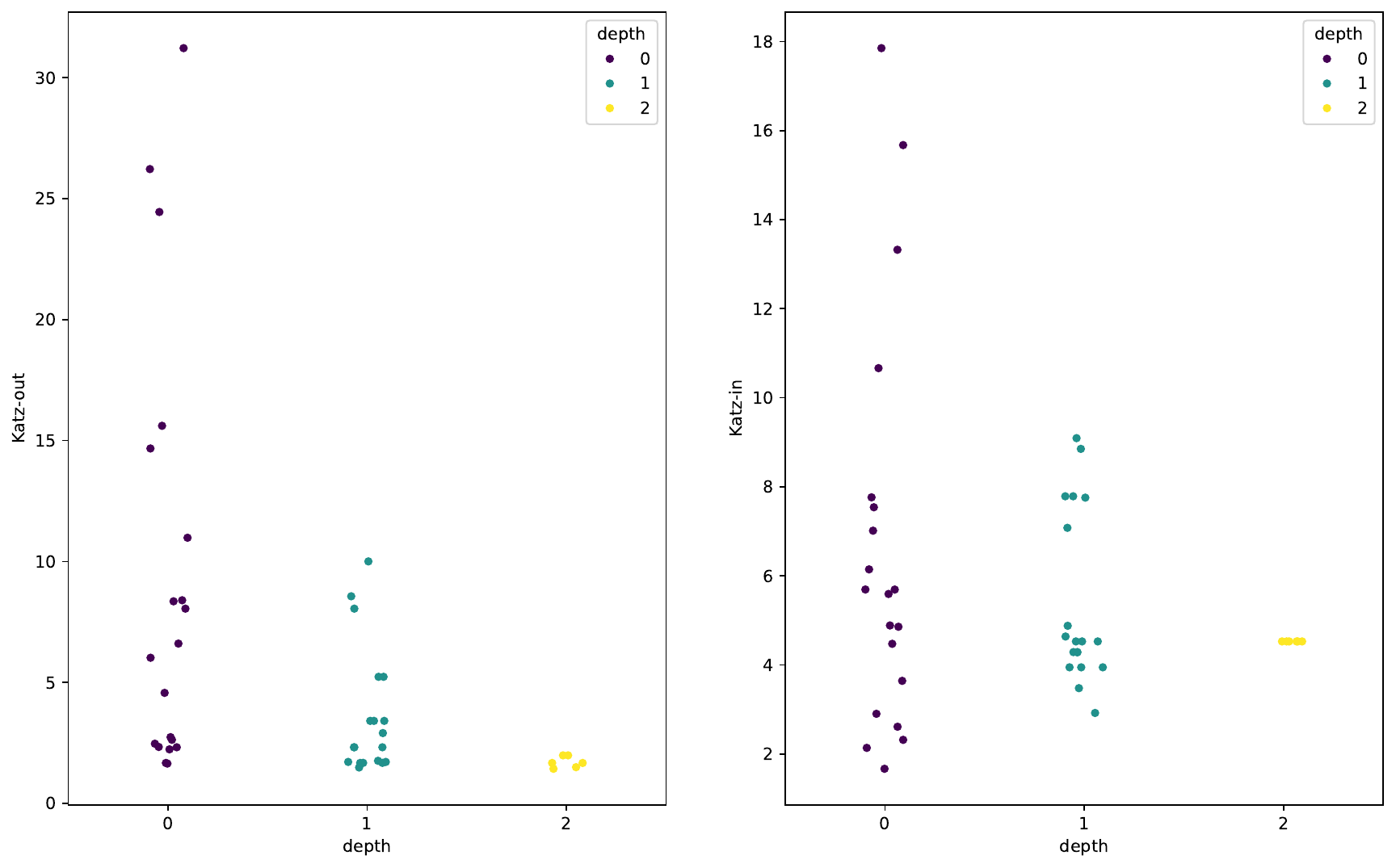}
    \caption{Katz centrality scores for pangraph generalized vertices}
    \label{fig:katz-stripplots-updated}
\end{figure}

The first observation is that Katz centralities of generalized vertices decrease with depth, see Fig.~\ref{fig:katz-stripplots-updated}. Linear trends estimated for Katz in- $c_i^{\tin}$ and out-centrality $c_i^{\out}$ of a pangraph as functions of depth $D(v_i)$ are given, for any $i\in I$ by:
\begin{eqnarray*}
    c_{\mP}^{\tin}(v_i) &=& -1.1 \; D(v_i) + 6.58, \\
    c_{\mP}^{\out}(v_i) &=& -4.22 \; D(v_i) + 8.73.
\end{eqnarray*}

The result confirms the intuition that panedges adjacent to fundamental vertices with high Katz centrality scores should have higher centrality than some other fundamental vertices.  

\begin{figure}[h!]
    \centering\includegraphics[width=\linewidth]{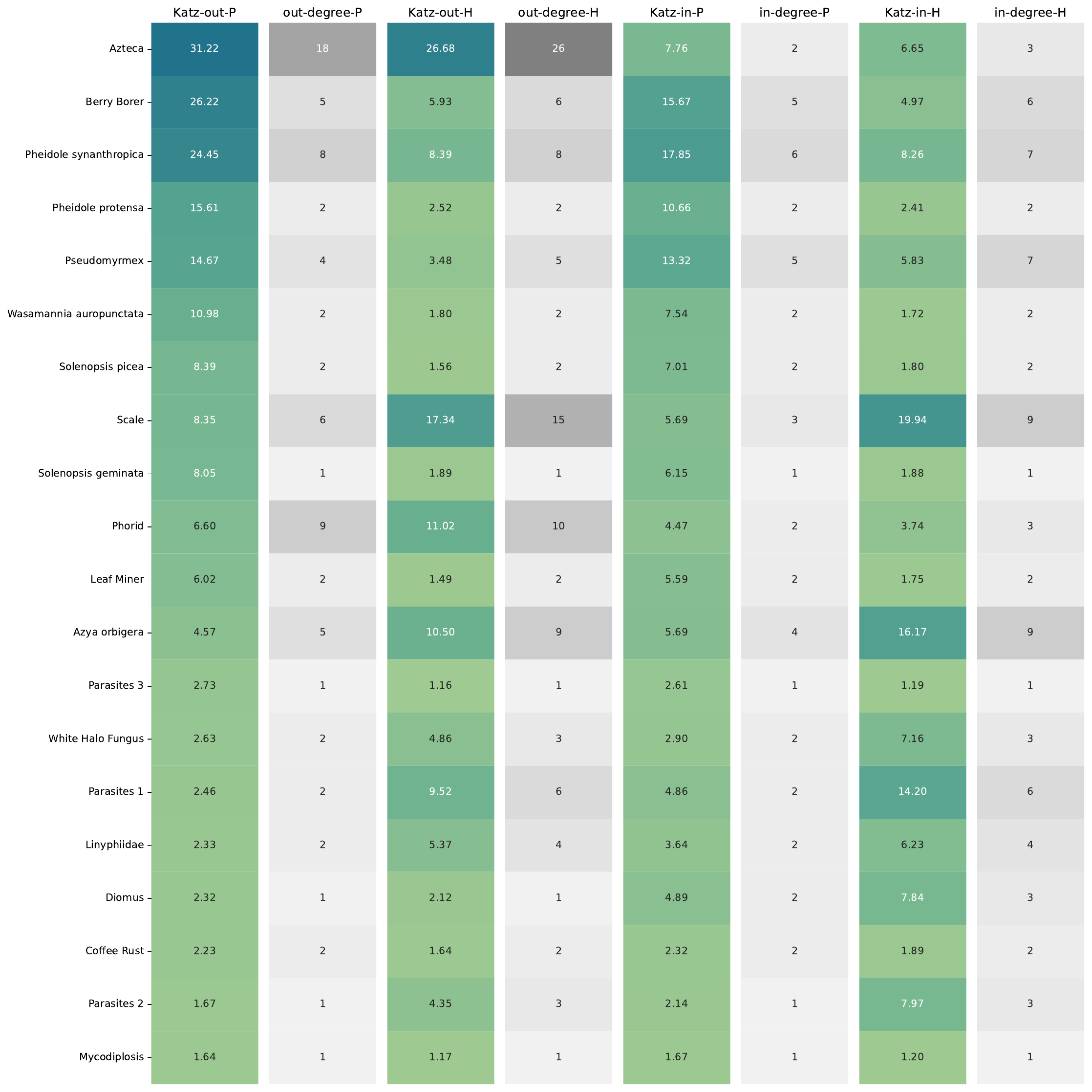}
    \caption{Katz in- and out-centralities for directed hypergraph and pangraph (fundamental vertices)
representations. In-centralities signify being a focal node receiving influences, whereas a high out-centrality means being an important influencer.}
    \label{fig:centralities-heatmap-updated}
\end{figure}

We compare the hypergraph and pangraph Katz centralities of fundamental vertices. In Fig.~\ref{fig:centralities-heatmap-updated} we present the numerical values of Katz out- and Katz in-centralities as well as in- and out-degree values of the fundamental vertices in pangraph and dihipergraph representations as a heatmap. In both coffee agroecosystem representations, it is the Katz out-centrality, and therefore the information about how the vertex influences its neighbors, that has a bigger range of values than Katz in-centrality. The versatile interactions of the Azteca ants~\cite{Vandermeer_2010, Perfecto_coffe_agroeco} consistently make them the most important influencers.

Vertex degree centralities show how HOIs, translated into additional hyperedges, inflate Katz centralities. The differences in dihypergraph and pangraph degree centralities are significant for vertices such as Scale ($\kappa_{\mH,A}^{out}(\textrm{Scale})=15$, $\kappa_{\mP,A}^{out}(\textrm{Scale})=6$), Azya orbigera ($\kappa_{\mH,A}^{out}(\textrm{Azya orbigera})=9$, $\kappa_{\mP,A}^{out}(\textrm{Azya orbigera})=5$) and Parasites 1 ($\kappa_{\mH,A}^{out}(\textrm{Parasites 1})=6$, $\kappa_{\mP,A}^{out}(\textrm{Parasites 1})=2$). This illustrates the consequences of inflating the degree centralities of participants of modified interactions, e.g. Scale and Azya orbigera, by the hypergraph representation, see Fig.~\ref{fig:pangraph_hypergraph_mapping}. 

The most central vertices increase the centralities of their neighbors, but this feature is sensitive to edge direction. The dominant Azteca out-centrality propagates to Berry Borer and Pheidole synantropica, but not to Parasites 1 which is not a direct neighbor, or Phorid (wrong edge direction). This is strengthened by the reciprocal nature of the Azteca - Berry Borer relationship, which creates a two-step cycle.

Another difference between the dihypergraph and pangraph approach can be observed for the vertices that take part in numerous HOIs: Azteca and Phorid. In pangraph representation, Phorid is no longer influencing Azteca directly as in dihypergraph representation. The walk from Phorid to Azteca goes through two panedges. Consequently, Phorid out-centrality decreases in pangraph representation even though its degree does not change much. As we observed in Fig.~\ref{fig:katz-stripplots-updated}, deeper edges generally have lower Katz centralities. They also contribute less to their source centrality than a direct connection to another vertex (see Eq.~\ref{eq:pangraph_Katz_walk},~\ref{eq:hypergraph_Katz_walk}). A similar situation occurs for Azya Orbigera, which does not affect Azteca directly in pangraph representation and its out-centrality is much lower than in dihypergraph representation.

Pangraph in-centralities differ from dihypergraph ones, also changing the ranks of vertices. Apart from Azteca, the vertices of Scale, Azya orbigera, and Parasites 1 (see Fig.~\ref{fig:centralities-heatmap}) are the most in-central and perhaps the most sensitive in the dihypergraph. They lose importance in the pangraph in favor of Berry Borer, Pheidole synanthropica, and Pseudomyrmex. Scale, Azya Orbigera, and Parasites 1 also lose outgoing connections.

\begin{figure}[ht]
    \centering\includegraphics[width=1\linewidth]{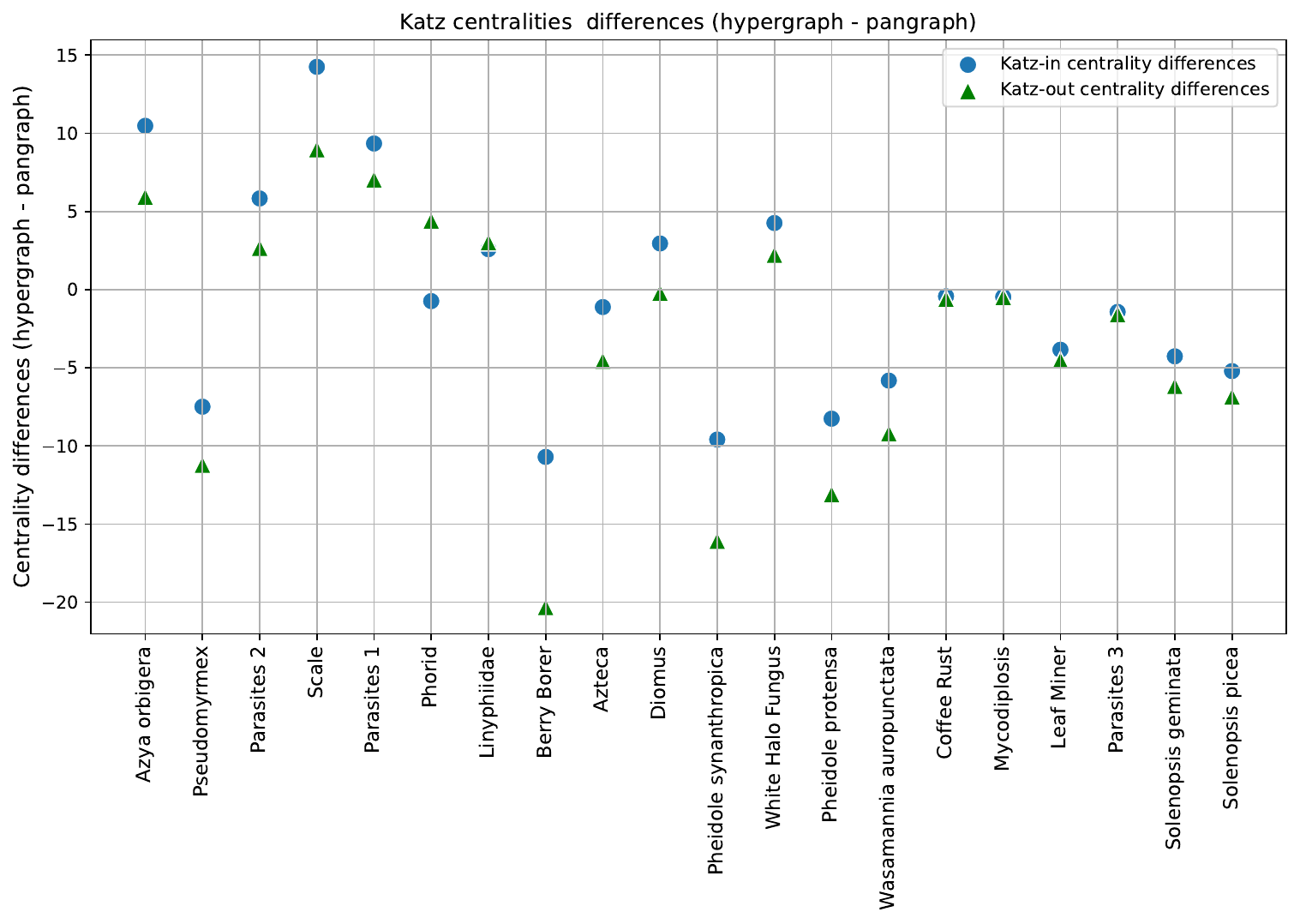}
    \caption{Differences between hypergraph and pangraph values of generalized Katz centrality for fundamental vertices}
    \label{fig:katz-diff-updated}
\end{figure}

Are there any general patterns in the differences between pangraph and hypergraph centrality scores? Fig.~\ref{fig:katz-diff-updated} shows a plot of the differences between in- and out-centralities for hypergraph and pangraph. It is possible to distinguish a group of vertices, such as Coffee Rust, Parasites 3, or Mycodiplosis, where the differences between hypergraph and pangraph representations are close to zero. These vertices are neither directly adjacent to vertices of high centrality nor involved in numerous HOIs. Vertices involved in numerous HOIs or adjacent to such vertices have more diverse centrality scores depending on the chosen graph representation.  

This analysis reveals several key differences between hypergraph and pangraph representations of interaction modifications. Hypergraph walks tend to be shorter than pangraph ones, which impacts not only Katz centralities but all measures dependent on walks. The vertex ranks concerning Katz in- and out-centralities also change depending on representation choice, impacting conclusions about the real system. The changes in centrality ranks result from additional edge ends present in the hypergraph approach, with a visible impact on vertex degrees. In addition, panedge centrality systematically decreases with depth, an observation enabled by the explicit assignment of centralities to interactions. Many differences in the centralities of particular vertices can be explained by the existence of direct connections to highly central vertices, such as Azteca. Katz centrality values and ranks vary with greater discrepancies for vertices involved in multiple HOIs.

\section{Discussion}\label{sec:discussion}

Higher-order interactions play a substantial role in complex system modeling as they can determine system stability~\cite{Grilli2017} and are key to addressing real-world problems with significant impacts. Understanding biological pest control~\cite{Vandermeer_2010, Perfecto_coffe_agroeco, GOLUBSKI2016344}, antibiotic resistance~\cite{bac_34_reading1977clavulanic, bac_31_kelsic2015counteraction, bac_32_perlin2009protection, bac_33_abrudan2015socially}, climate interdependencies~\cite{40_climate_Boers_2019, 41_climate_Su_2021} requires a proper representation of HOIs.

We propose the pangraph as a universal structure that can accurately map arbitrarily complex interaction modifications. It resolves important shortcomings of the hypergraph approach, including those found in its special case of simplicial complexes. We apply our framework to a real-world coffee agroecosystem. Pangraphs not only map vertex roles accurately, but they also alter conclusions regarding vertex importance when compared to hypergraphs. We described it analytically and evidenced by changes in Katz centrality rankings in the numerical example. In each approach, different species belonged to the group of the most central.

The introduction of a new structure always raises the question of whether the same information can be captured using existing, and preferably simpler, terminology. While it might seem that a Levi graph representation of pangraphs could replace them with a conventional digraph structure, the consistent generalization of recursive graph measures uses the domain of generalized vertices $V_{\mP}$, rather than Levi graph vertices $V\cup E_{\mP}$, see Section~\ref{sec:Katz_centrality}.

A persistent challenge, common to hypergraph and pangraph approaches, is determining which interactions can be decomposed into pairwise edges and which should remain as true, non-divisible hyperedges or panedges with multiple tails/heads. In the hypergraph context, this issue was studied in~\cite{letten_stouffer_non_additivity}. Our intuition points to the criterion of the effects of vertex removal. If removing any of the panedge elements causes interactions' cessation in the real system, it should be represented as a single edge. If the interactions are independent, a representation through several processes might be more appropriate.

\section{Acknowledgements}

We thank Mwawi Nyirenda for her involvement in the initial discussions and Marcin Pilipczuk for his remarks on Levi graphs.

This article was made in a collaboration founded at a workshop of COST Action CA18232, supported by COST (European Cooperation in Science and Technology).   G\"okhan Mutlu has received a Short
Term Scientific Mission (STSM) Grant funded by COST Action CA18232 entitled ”Ubergraphs and ecological dynamics”.

\bibliographystyle{naturemag} 
\bibliography{hyp_dyn}	

\begin{thebibliography}{10}
\expandafter\ifx\csname url\endcsname\relax
  \def\url#1{\texttt{#1}}\fi
\expandafter\ifx\csname urlprefix\endcsname\relax\def\urlprefix{URL }\fi
\providecommand{\bibinfo}[2]{#2}
\providecommand{\eprint}[2][]{\url{#2}}

\bibitem{GOLUBSKI2016344}
\bibinfo{author}{Golubski, A.~J.}, \bibinfo{author}{Westlund, E.~E.}, \bibinfo{author}{Vandermeer, J.} \& \bibinfo{author}{Pascual, M.}
\newblock \bibinfo{title}{Ecological networks over the edge: Hypergraph trait-mediated indirect interaction (tmii) structure}.
\newblock \emph{\bibinfo{journal}{Trends in Ecology \& Evolution}} \textbf{\bibinfo{volume}{31}}, \bibinfo{pages}{344--354} (\bibinfo{year}{2016}).
\newblock \urlprefix\url{https://www.sciencedirect.com/science/article/pii/S0169534716000513}.

\bibitem{Moleon_scavenging}
\bibinfo{author}{Moleón, M.}, \bibinfo{author}{Sánchez-Zapata, J.~A.}, \bibinfo{author}{Selva, N.}, \bibinfo{author}{Donázar, J.~A.} \& \bibinfo{author}{Owen-Smith, N.}
\newblock \bibinfo{title}{Inter-specific interactions linking predation and scavenging in terrestrial vertebrate assemblages}.
\newblock \emph{\bibinfo{journal}{Biological Reviews}} \textbf{\bibinfo{volume}{89}}, \bibinfo{pages}{1042--1054} (\bibinfo{year}{2014}).
\newblock \urlprefix\url{https://doi.org/10.1111/brv.12097}.

\bibitem{Levine_2017}
\bibinfo{author}{Levine, J.~M.} \emph{et~al.}
\newblock \bibinfo{title}{Beyond pairwise mechanisms of species coexistence in complex communities}.
\newblock \emph{\bibinfo{journal}{Nature}}  (\bibinfo{year}{2017}).

\bibitem{Mayfield2017}
\bibinfo{author}{Mayfield, M.~M.} \& \bibinfo{author}{Stouffer, D.~B.}
\newblock \bibinfo{title}{Higher-order interactions capture unexplained complexity in diverse communities}.
\newblock \emph{\bibinfo{journal}{Nature Ecology and Evolution}} \textbf{\bibinfo{volume}{1}} (\bibinfo{year}{2017}).

\bibitem{Grilli2017}
\bibinfo{author}{Grilli, J.}, \bibinfo{author}{Barabás, G.}, \bibinfo{author}{Michalska-Smith, M.~J.} \& \bibinfo{author}{Allesina, S.}
\newblock \bibinfo{title}{Higher-order interactions stabilize dynamics in competitive network models}.
\newblock \emph{\bibinfo{journal}{Nature}} \textbf{\bibinfo{volume}{548}} (\bibinfo{year}{2017}).
\newblock \urlprefix\url{https://doi.org/10.1038/nature23273}.

\bibitem{BATTISTON20201}
\bibinfo{author}{Battiston, F.} \emph{et~al.}
\newblock \bibinfo{title}{Networks beyond pairwise interactions: Structure and dynamics}.
\newblock \emph{\bibinfo{journal}{Physics Reports}} \textbf{\bibinfo{volume}{874}}, \bibinfo{pages}{1--92} (\bibinfo{year}{2020}).
\newblock \urlprefix\url{https://www.sciencedirect.com/science/article/pii/S0370157320302489}.

\bibitem{hoi_book}
\bibinfo{author}{Battiston, F.} \& \bibinfo{author}{Petri, G.}
\newblock \emph{\bibinfo{title}{Higher-Order Systems}} (\bibinfo{publisher}{Springer Cham}, \bibinfo{year}{2022}).
\newblock \urlprefix\url{https://doi.org/10.1007/978-3-030-91374-8}.

\bibitem{levine2017}
\bibinfo{author}{Levine, J.}, \bibinfo{author}{Bascompte, J.}, \bibinfo{author}{Adler, P.} \& \bibinfo{author}{Allesina, S.}
\newblock \bibinfo{title}{Beyond pairwise mechanisms of species coexistence in complex communities}.
\newblock \emph{\bibinfo{journal}{Nature}} \textbf{\bibinfo{volume}{546}}, \bibinfo{pages}{56--64} (\bibinfo{year}{2017}).

\bibitem{letten_stouffer_non_additivity}
\bibinfo{author}{Letten, A.~D.} \& \bibinfo{author}{Stouffer, D.~B.}
\newblock \bibinfo{title}{The mechanistic basis for higher-order interactions and non-additivity in competitive communities}.
\newblock \emph{\bibinfo{journal}{Ecology Letters}} \textbf{\bibinfo{volume}{22}}, \bibinfo{pages}{423--436} (\bibinfo{year}{2019}).
\newblock \urlprefix\url{https://onlinelibrary.wiley.com/doi/abs/10.1111/ele.13211}.
\newblock \eprint{https://onlinelibrary.wiley.com/doi/pdf/10.1111/ele.13211}.

\bibitem{lichen_duran_nebreda}
\bibinfo{author}{Duran-Nebreda, S.} \& \bibinfo{author}{Valverde, S.}
\newblock \bibinfo{title}{Composition, structure and robustness of lichen guilds}.
\newblock \emph{\bibinfo{journal}{Scientific Reports}} \textbf{\bibinfo{volume}{13}} (\bibinfo{year}{2023}).

\bibitem{39_chemistry_Jost_2019}
\bibinfo{author}{Jost, J.}, \bibinfo{author}{Jost, J.}, \bibinfo{author}{Mulas, R.} \& \bibinfo{author}{Mulas, R.}
\newblock \bibinfo{title}{Hypergraph laplace operators for chemical reaction networks}.
\newblock \emph{\bibinfo{journal}{Advances in Mathematics}}  (\bibinfo{year}{2019}).

\bibitem{37_brain_Giusti_2016}
\bibinfo{author}{Giusti, C.} \emph{et~al.}
\newblock \bibinfo{title}{Two's company, three (or more) is a simplex}.
\newblock \emph{\bibinfo{journal}{Journal of Computational Neuroscience}}  (\bibinfo{year}{2016}).

\bibitem{38_brain_Faskowitz_2021}
\bibinfo{author}{Faskowitz, J.} \emph{et~al.}
\newblock \bibinfo{title}{Edges in brain networks: Contributions to models of structure and function.}
\newblock \emph{\bibinfo{journal}{arXiv: Neurons and Cognition}}  (\bibinfo{year}{2021}).

\bibitem{40_climate_Boers_2019}
\bibinfo{author}{Boers, N.} \emph{et~al.}
\newblock \bibinfo{title}{Complex networks reveal global pattern of extreme-rainfall teleconnections}.
\newblock \emph{\bibinfo{journal}{Nature}}  (\bibinfo{year}{2019}).

\bibitem{41_climate_Su_2021}
\bibinfo{author}{Su, Z.}, \bibinfo{author}{Meyerhenke, H.} \& \bibinfo{author}{Kurths, J.}
\newblock \bibinfo{title}{The climatic interdependence of extreme-rainfall events around the globe}.
\newblock \emph{\bibinfo{journal}{Chaos}}  (\bibinfo{year}{2021}).

\bibitem{53_Alvarez-Rodriguez_2020}
\bibinfo{author}{Alvarez-Rodriguez, U.} \emph{et~al.}
\newblock \bibinfo{title}{Evolutionary dynamics of higher-order interactions in social networks}.
\newblock \emph{\bibinfo{journal}{Nature Human Behaviour}}  (\bibinfo{year}{2020}).

\bibitem{Holling_1959}
\bibinfo{author}{Holling, C.~S.}
\newblock \bibinfo{title}{The components of predation as revealed by a study of small-mammal predation of the european pine sawfly}.
\newblock \emph{\bibinfo{journal}{The Canadian Entomologist}} \textbf{\bibinfo{volume}{91}}, \bibinfo{pages}{293–320} (\bibinfo{year}{1959}).

\bibitem{Getz_2011}
\bibinfo{author}{Getz, W.~M.}
\newblock \bibinfo{title}{Biomass transformation webs provide a unified approach to consumer-resource modelling}.
\newblock \emph{\bibinfo{journal}{Ecology Letters}} \textbf{\bibinfo{volume}{14}}, \bibinfo{pages}{113--124} (\bibinfo{year}{2011}).
\newblock \urlprefix\url{https://doi.org/10.1111/j.1461-0248.2010.01566.x}.

\bibitem{42_Millán_2019}
\bibinfo{author}{Millán, A.~P.} \emph{et~al.}
\newblock \bibinfo{title}{Explosive higher-order kuramoto dynamics on simplicial complexes}.
\newblock \emph{\bibinfo{journal}{arXiv: Adaptation and Self-Organizing Systems}}  (\bibinfo{year}{2019}).

\bibitem{43_Skardal_2019}
\bibinfo{author}{Skardal, P.~S.} \& \bibinfo{author}{Arenas, A.}
\newblock \bibinfo{title}{Abrupt desynchronization and extensive multistability in globally coupled oscillator simplices}.
\newblock \emph{\bibinfo{journal}{arXiv: Adaptation and Self-Organizing Systems}}  (\bibinfo{year}{2019}).

\bibitem{44_Zhang_2021}
\bibinfo{author}{Zhang, Y.} \emph{et~al.}
\newblock \bibinfo{title}{Unified treatment of synchronization patterns in generalized networks with higher-order, multilayer, and temporal interactions}.
\newblock \emph{\bibinfo{journal}{Communications in Physics}}  (\bibinfo{year}{2021}).

\bibitem{45_Mulas_2020}
\bibinfo{author}{Mulas, R.} \emph{et~al.}
\newblock \bibinfo{title}{Coupled dynamics on hypergraphs: Master stability of steady states and synchronization.}
\newblock \emph{\bibinfo{journal}{arXiv: Dynamical Systems}}  (\bibinfo{year}{2020}).

\bibitem{47_St-Onge_2021}
\bibinfo{author}{St-Onge, G.} \emph{et~al.}
\newblock \bibinfo{title}{Universal nonlinear infection kernel from heterogeneous exposure on higher-order networks.}
\newblock \emph{\bibinfo{journal}{Physical Review Letters}}  (\bibinfo{year}{2021}).

\bibitem{48_Arruda_2020}
\bibinfo{author}{de~Arruda, G.~F.} \emph{et~al.}
\newblock \bibinfo{title}{Social contagion models on hypergraphs}.
\newblock \emph{\bibinfo{journal}{Physical Review Research}}  (\bibinfo{year}{2020}).

\bibitem{49_Iacopini_2018}
\bibinfo{author}{Iacopini, I.} \emph{et~al.}
\newblock \bibinfo{title}{Simplicial models of social contagion}.
\newblock \emph{\bibinfo{journal}{arXiv: Physics and Society}}  (\bibinfo{year}{2018}).

\bibitem{50_Arruda_2020}
\bibinfo{author}{de~Arruda, G.~F.} \emph{et~al.}
\newblock \bibinfo{title}{Phase transitions and stability of dynamical processes on hypergraphs}.
\newblock \emph{\bibinfo{journal}{arXiv: Physics and Society}}  (\bibinfo{year}{2020}).

\bibitem{51_Sun_2021}
\bibinfo{author}{Sun, H.} \& \bibinfo{author}{Bianconi, G.}
\newblock \bibinfo{title}{Higher-order percolation processes on multiplex hypergraphs}.
\newblock \emph{\bibinfo{journal}{Physical Review E}}  (\bibinfo{year}{2021}).

\bibitem{52_Taylor_2015}
\bibinfo{author}{Taylor, D.} \emph{et~al.}
\newblock \bibinfo{title}{Topological data analysis of contagion maps for examining spreading processes on networks}.
\newblock \emph{\bibinfo{journal}{Nature Communications}}  (\bibinfo{year}{2015}).

\bibitem{46_random_walk_Carletti_2020}
\bibinfo{author}{Carletti, T.} \emph{et~al.}
\newblock \bibinfo{title}{Random walks on hypergraphs.}
\newblock \emph{\bibinfo{journal}{Physical Review E}}  (\bibinfo{year}{2020}).

\bibitem{Rnd_walk_edge_vertex_weights}
\bibinfo{author}{Chitra, U.} \& \bibinfo{author}{Raphael, B.}
\newblock \bibinfo{title}{Random walks on hypergraphs with edge-dependent vertex weights}.
\newblock In \bibinfo{editor}{Chaudhuri, K.} \& \bibinfo{editor}{Salakhutdinov, R.} (eds.) \emph{\bibinfo{booktitle}{Proceedings of the 36th International Conference on Machine Learning}}, vol.~\bibinfo{volume}{97} of \emph{\bibinfo{series}{Proceedings of Machine Learning Research}}, \bibinfo{pages}{1172--1181} (\bibinfo{publisher}{PMLR}, \bibinfo{year}{2019}).
\newblock \urlprefix\url{https://proceedings.mlr.press/v97/chitra19a.html}.

\bibitem{Landi2018}
\bibinfo{author}{Landi, P.}, \bibinfo{author}{Minoarivelo, H.~O.}, \bibinfo{author}{Åke Brännström}, \bibinfo{author}{Hui, C.} \& \bibinfo{author}{Dieckmann, U.}
\newblock \bibinfo{title}{Complexity and stability of ecological networks: a review of the theory}.
\newblock \emph{\bibinfo{journal}{Population Ecology}} \textbf{\bibinfo{volume}{0}}, \bibinfo{pages}{1--27} (\bibinfo{year}{2018}).
\newblock \urlprefix\url{http://dx.doi.org/10.1007/s10144-018-0628-3}.

\bibitem{May1972}
\bibinfo{author}{May, R.~M.}
\newblock \bibinfo{title}{Will a large complex system be stable?}
\newblock \emph{\bibinfo{journal}{Nature}} \textbf{\bibinfo{volume}{238}}, \bibinfo{pages}{413--414} (\bibinfo{year}{1972}).

\bibitem{Vandermeer_2010}
\bibinfo{author}{Vandermeer, J.}, \bibinfo{author}{Perfecto, I.} \& \bibinfo{author}{Philpott, S.}
\newblock \bibinfo{title}{{Ecological Complexity and Pest Control in Organic Coffee Production: Uncovering an Autonomous Ecosystem Service}}.
\newblock \emph{\bibinfo{journal}{BioScience}} \textbf{\bibinfo{volume}{60}}, \bibinfo{pages}{527 -- 537} (\bibinfo{year}{2010}).
\newblock \urlprefix\url{https://doi.org/10.1525/bio.2010.60.7.8}.

\bibitem{Perfecto_coffe_agroeco}
\bibinfo{author}{Perfecto, I.}, \bibinfo{author}{Vandermeer, J.} \& \bibinfo{author}{Philpott, S.~M.}
\newblock \bibinfo{title}{Complex ecological interactions in the coffee agroecosystem}.
\newblock \emph{\bibinfo{journal}{Annual Review of Ecology, Evolution, and Systematics}} \textbf{\bibinfo{volume}{45}}, \bibinfo{pages}{137--158} (\bibinfo{year}{2014}).
\newblock \urlprefix\url{https://www.annualreviews.org/content/journals/10.1146/annurev-ecolsys-120213-091923}.

\bibitem{preisser_intimidation}
\bibinfo{author}{Preisser, E.~L.}, \bibinfo{author}{Bolnick, D.~I.} \& \bibinfo{author}{Benard, M.~F.}
\newblock \bibinfo{title}{Scared to death? the effects of intimidation and consumption in predator–prey interactions}.
\newblock \emph{\bibinfo{journal}{Ecology}} \textbf{\bibinfo{volume}{86}}, \bibinfo{pages}{501--509} (\bibinfo{year}{2005}).
\newblock \urlprefix\url{https://esajournals.onlinelibrary.wiley.com/doi/abs/10.1890/04-0719}.

\bibitem{Mickalide_microbial}
\bibinfo{author}{Mickalide, H.} \& \bibinfo{author}{Kuehn, S.}
\newblock \bibinfo{title}{Higher-order interaction between species inhibits bacterial invasion of a phototroph-predator microbial community}.
\newblock \emph{\bibinfo{journal}{Cell Systems}} \textbf{\bibinfo{volume}{9}}, \bibinfo{pages}{521--533.e10} (\bibinfo{year}{2019}).
\newblock \urlprefix\url{https://www.sciencedirect.com/science/article/pii/S2405471219303904}.

\bibitem{bac_31_kelsic2015counteraction}
\bibinfo{author}{Kelsic, E.~D.}, \bibinfo{author}{Zhao, J.}, \bibinfo{author}{Vetsigian, K.} \& \bibinfo{author}{Kishony, R.}
\newblock \bibinfo{title}{Counteraction of antibiotic production and degradation stabilizes microbial communities}.
\newblock \emph{\bibinfo{journal}{Nature}} \textbf{\bibinfo{volume}{521}}, \bibinfo{pages}{516--519} (\bibinfo{year}{2015}).

\bibitem{bac_32_perlin2009protection}
\bibinfo{author}{Perlin, M.~H.} \emph{et~al.}
\newblock \bibinfo{title}{Protection of salmonella by ampicillin-resistant escherichia coli in the presence of otherwise lethal drug concentrations}.
\newblock \emph{\bibinfo{journal}{Proceedings of the Royal Society B: Biological Sciences}} \textbf{\bibinfo{volume}{276}}, \bibinfo{pages}{3759--3768} (\bibinfo{year}{2009}).

\bibitem{bac_33_abrudan2015socially}
\bibinfo{author}{Abrudan, M.~I.} \emph{et~al.}
\newblock \bibinfo{title}{Socially mediated induction and suppression of antibiosis during bacterial coexistence}.
\newblock \emph{\bibinfo{journal}{Proceedings of the National Academy of Sciences}} \textbf{\bibinfo{volume}{112}}, \bibinfo{pages}{11054--11059} (\bibinfo{year}{2015}).

\bibitem{bac_34_reading1977clavulanic}
\bibinfo{author}{Reading, C.} \& \bibinfo{author}{Cole, M.}
\newblock \bibinfo{title}{Clavulanic acid: a beta-lactamase-inhibiting beta-lactam from streptomyces clavuligerus}.
\newblock \emph{\bibinfo{journal}{Antimicrobial agents and chemotherapy}} \textbf{\bibinfo{volume}{11}}, \bibinfo{pages}{852--857} (\bibinfo{year}{1977}).

\bibitem{Sun_2023_triadic_Bianconi}
\bibinfo{author}{Sun, H.}, \bibinfo{author}{Radicchi, F.}, \bibinfo{author}{Kurths, J.} \& \bibinfo{author}{Bianconi, G.}
\newblock \bibinfo{title}{The dynamic nature of percolation on networks with triadic interactions}.
\newblock \emph{\bibinfo{journal}{Nature Communications}}  (\bibinfo{year}{2023}).

\bibitem{Joslyn2017_ubergraph}
\bibinfo{author}{Joslyn, C.} \& \bibinfo{author}{Nowak, K.~E.}
\newblock \bibinfo{title}{Ubergraphs: A definition of a recursive hypergraph structure}.
\newblock \emph{\bibinfo{journal}{ArXiv}} \textbf{\bibinfo{volume}{abs/1704.05547}} (\bibinfo{year}{2017}).

\bibitem{food_webs_pimm}
\bibinfo{author}{Pimm, S.~L.}
\newblock \emph{\bibinfo{title}{Food Webs}} (\bibinfo{publisher}{Springer Netherlands}, \bibinfo{year}{1982}).

\bibitem{Bascompte_Mutualistic_networks}
\bibinfo{author}{Bascompte, J.} \& \bibinfo{author}{Jordano, P.}
\newblock \emph{\bibinfo{title}{Mutualistic networks}} (\bibinfo{publisher}{Princeton University Press}, \bibinfo{year}{2013}).

\bibitem{Pilosof_2015}
\bibinfo{author}{Pilosof, S.}, \bibinfo{author}{Porter, M.~A.}, \bibinfo{author}{Pascual, M.} \& \bibinfo{author}{Kéfi, S.}
\newblock \bibinfo{title}{The multilayer nature of ecological networks}.
\newblock \emph{\bibinfo{journal}{Nat. Ecol. Evol.}} \textbf{\bibinfo{volume}{1}}, \bibinfo{pages}{1--9} (\bibinfo{year}{2015}).
\newblock \urlprefix\url{http://dx.doi.org/10.1038/s41559-017-0101}.

\bibitem{Lurgi_2020}
\bibinfo{author}{Lurgi, M.} \emph{et~al.}
\newblock \bibinfo{title}{Geographical variation of multiplex ecological networks in marine intertidal communities}.
\newblock \emph{\bibinfo{journal}{Ecology}} \textbf{\bibinfo{volume}{101}} (\bibinfo{year}{2020}).
\newblock \urlprefix\url{http://dx.doi.org/10.1002/ecy.3165}.

\bibitem{Hutchinson2019}
\bibinfo{author}{Hutchinson, M.~C.} \emph{et~al.}
\newblock \bibinfo{title}{Seeing the forest for the trees: Putting multilayer networks to work for community ecology}.
\newblock \emph{\bibinfo{journal}{Functional Ecology}} \textbf{\bibinfo{volume}{33}}, \bibinfo{pages}{206--217} (\bibinfo{year}{2019}).
\newblock \urlprefix\url{http://dx.doi.org/10.1111/1365-2435.13237}.

\bibitem{Petri_thesis}
\bibinfo{author}{Petri, C.}
\newblock \bibinfo{title}{Communication with automata}.
\newblock \bibinfo{type}{Tech. Rep.} \bibinfo{number}{AD0630125}, \bibinfo{institution}{DTIC Research Report} (\bibinfo{year}{1966}).

\bibitem{Petri_Peterson_book}
\bibinfo{author}{Peterson, J.~L.}
\newblock \emph{\bibinfo{title}{Petri Net Theory and the Modeling of Systems}} (\bibinfo{publisher}{Prentice–Hall}, \bibinfo{address}{New Jersey}, \bibinfo{year}{1981}).

\bibitem{Baez_open_petri_2017}
\bibinfo{author}{Baez, J.~C.} \& \bibinfo{author}{Pollard, B.~S.}
\newblock \bibinfo{title}{A compositional framework for reaction networks}.
\newblock \emph{\bibinfo{journal}{Reviews in Mathematical Physics}} \textbf{\bibinfo{volume}{29}}, \bibinfo{pages}{1750028} (\bibinfo{year}{2017}).
\newblock \urlprefix\url{https://doi.org/10.1142/S0129055X17500283}.
\newblock \eprint{https://doi.org/10.1142/S0129055X17500283}.

\bibitem{KivArena2014}
\bibinfo{author}{Kivelä, M.} \emph{et~al.}
\newblock \bibinfo{title}{{Multilayer networks}}.
\newblock \emph{\bibinfo{journal}{Journal of Complex Networks}} \textbf{\bibinfo{volume}{2}}, \bibinfo{pages}{203--271} (\bibinfo{year}{2014}).
\newblock \urlprefix\url{https://doi.org/10.1093/comnet/cnu016}.
\newblock \eprint{https://academic.oup.com/comnet/article-pdf/2/3/203/9130906/cnu016.pdf}.

\bibitem{Mugnolo2013}
\bibinfo{author}{Mugnolo, D.}
\newblock \emph{\bibinfo{title}{Semigroup Methods for Evolution Equations on Networks}} (\bibinfo{publisher}{Springer Cham}, \bibinfo{year}{2013}).

\bibitem{Bretto2013}
\bibinfo{author}{Bretto, A.}
\newblock \bibinfo{title}{Hypergraph theory: An introduction}.
\newblock \emph{\bibinfo{journal}{Mathematical Engineering}} \textbf{\bibinfo{volume}{11}} (\bibinfo{year}{2013}).

\bibitem{Michoel2012}
\bibinfo{author}{Michoel, T.} \& \bibinfo{author}{Nachtergaele, B.}
\newblock \bibinfo{title}{Alignment and integration of complex networks by hypergraph-based spectral clustering}.
\newblock \emph{\bibinfo{journal}{Physical Review E}} \textbf{\bibinfo{volume}{86(5)}}, \bibinfo{pages}{056111} (\bibinfo{year}{2012}).

\bibitem{COOPER2012}
\bibinfo{author}{Cooper, J.} \& \bibinfo{author}{Dutle, A.}
\newblock \bibinfo{title}{Spectra of uniform hypergraphs}.
\newblock \emph{\bibinfo{journal}{Linear Algebra and its Applications}} \textbf{\bibinfo{volume}{436}}, \bibinfo{pages}{3268--3292} (\bibinfo{year}{2012}).
\newblock \urlprefix\url{https://www.sciencedirect.com/science/article/pii/S0024379511007610}.

\bibitem{Hu2013}
\bibinfo{author}{Hu, S.}
\newblock \bibinfo{title}{Spectral hypergraph theory}.
\newblock \emph{\bibinfo{journal}{PhD thesis, The Hong Kong Polytechnic University}}  (\bibinfo{year}{2013}).
\newblock \urlprefix\url{https://theses.lib.polyu.edu.hk/handle/200/7238}.

\bibitem{Levi}
\bibinfo{author}{Levi, F.~W.}
\newblock \emph{\bibinfo{title}{Finite Geometrical Systems}} (\bibinfo{publisher}{University of Calcutta}, \bibinfo{address}{Calcutta}, \bibinfo{year}{1942}).

\bibitem{Arditi_2005_non_trophic_rheagogies}
\bibinfo{author}{Arditi, R.} \emph{et~al.}
\newblock \bibinfo{title}{Rheagogies: Modelling non-trophic effects in food webs}.
\newblock \emph{\bibinfo{journal}{Ecological Complexity}}  (\bibinfo{year}{2005}).

\bibitem{Goudard_nontrophic_2008}
\bibinfo{author}{Goudard, A.} \& \bibinfo{author}{Loreau, M.}
\newblock \bibinfo{title}{Nontrophic interactions, biodiversity, and ecosystem functioning: An interaction web model.}
\newblock \emph{\bibinfo{journal}{The American Naturalist}} \textbf{\bibinfo{volume}{171}}, \bibinfo{pages}{91--106} (\bibinfo{year}{2008}).
\newblock \urlprefix\url{https://doi.org/10.1086/523945}.
\newblock \bibinfo{note}{PMID: 18171154}, \eprint{https://doi.org/10.1086/523945}.

\bibitem{Golubski_2011_non_trophic_combining}
\bibinfo{author}{Golubski, A.~J.}, \bibinfo{author}{Golubski, A.~J.}, \bibinfo{author}{Abrams, P.~A.} \& \bibinfo{author}{Abrams, P.~A.}
\newblock \bibinfo{title}{Modifying modifiers: what happens when interspecific interactions interact?}
\newblock \emph{\bibinfo{journal}{Journal of Animal Ecology}}  (\bibinfo{year}{2011}).

\bibitem{BGS2011}
\bibinfo{author}{Banasiak, J.}, \bibinfo{author}{Goswami, A.} \& \bibinfo{author}{Shindin, S.}
\newblock \bibinfo{title}{Aggregation in age and space structured population models: an asymptotic analysis approach}.
\newblock \emph{\bibinfo{journal}{Journal of Evolution Equations}} \textbf{\bibinfo{volume}{11}}, \bibinfo{pages}{121--154} (\bibinfo{year}{2011}).

\bibitem{Pearl_2009}
\bibinfo{author}{Pearl, J.}
\newblock \emph{\bibinfo{title}{Causality}} (\bibinfo{publisher}{Cambridge University Press}, \bibinfo{year}{2009}), \bibinfo{edition}{2} edn.

\bibitem{Morgan_Winship_2014}
\bibinfo{author}{Morgan, S.~L.} \& \bibinfo{author}{Winship, C.}
\newblock \bibinfo{title}{Causal graphs}.
\newblock In \emph{\bibinfo{booktitle}{Counterfactuals and Causal Inference: Methods and Principles for Social Research}}, Analytical Methods for Social Research, \bibinfo{pages}{77–102} (\bibinfo{publisher}{Cambridge University Press}, \bibinfo{year}{2014}).

\bibitem{signed_graph_1936}
\bibinfo{author}{K\H{o}nig, D.}
\newblock \emph{\bibinfo{title}{Theorie der endlichen und unendlichen Graphen}} (\bibinfo{publisher}{Akademische Verlagsgesellschaft}, \bibinfo{year}{1936}).

\bibitem{signed_graph_1955}
\bibinfo{author}{Harary, F.}
\newblock \bibinfo{title}{{On the notion of balance of a signed graph.}}
\newblock \emph{\bibinfo{journal}{Michigan Mathematical Journal}} \textbf{\bibinfo{volume}{2}}, \bibinfo{pages}{143 -- 146} (\bibinfo{year}{1953}).
\newblock \urlprefix\url{https://doi.org/10.1307/mmj/1028989917}.

\bibitem{HIGASHI_1995_ecological_interaction_networks}
\bibinfo{author}{Higashi, M.} \& \bibinfo{author}{Nakajima, H.}
\newblock \bibinfo{title}{Indirect effects in ecological interaction networks i. the chain rule approach}.
\newblock \emph{\bibinfo{journal}{Mathematical Biosciences}} \textbf{\bibinfo{volume}{130}}, \bibinfo{pages}{99--128} (\bibinfo{year}{1995}).
\newblock \urlprefix\url{https://www.sciencedirect.com/science/article/pii/0025556494001197}.

\bibitem{NAKAJIMA_1995_ecological_interaction_networks}
\bibinfo{author}{Nakajima, H.} \& \bibinfo{author}{Higashi, M.}
\newblock \bibinfo{title}{Indirect effects in ecological interaction networks ii. the conjugate variable approach}.
\newblock \emph{\bibinfo{journal}{Mathematical Biosciences}} \textbf{\bibinfo{volume}{130}}, \bibinfo{pages}{129--150} (\bibinfo{year}{1995}).
\newblock \urlprefix\url{https://www.sciencedirect.com/science/article/pii/0025556494001161}.

\bibitem{Kefi_multilayer_Chilean}
\bibinfo{author}{Kéfi, S.} \emph{et~al.}
\newblock \bibinfo{title}{Network structure beyond food webs: mapping non-trophic and trophic interactions on chilean rocky shores}.
\newblock \emph{\bibinfo{journal}{Ecology}} \textbf{\bibinfo{volume}{96}}, \bibinfo{pages}{291--303} (\bibinfo{year}{2015}).
\newblock \urlprefix\url{https://esajournals.onlinelibrary.wiley.com/doi/abs/10.1890/13-1424.1}.
\newblock \eprint{https://esajournals.onlinelibrary.wiley.com/doi/pdf/10.1890/13-1424.1}.

\bibitem{Baez_2019_catalysts}
\bibinfo{author}{Baez, J.~C.}, \bibinfo{author}{Foley, J.} \& \bibinfo{author}{Moeller, J.}
\newblock \bibinfo{title}{Network models from petri nets with catalysts}.
\newblock \emph{\bibinfo{journal}{Compositionality}} \textbf{\bibinfo{volume}{1}}, \bibinfo{pages}{4} (\bibinfo{year}{2019}).
\newblock \urlprefix\url{https://doi.org/10.32408%2Fcompositionality-1-4}.

\bibitem{Metabolic_graphs_McQuade}
\bibinfo{author}{McQuade, S.~T.}, \bibinfo{author}{Merrill, N.~J.} \& \bibinfo{author}{Piccoli, B.}
\newblock \bibinfo{title}{Metabolic graphs, life method and the modeling of drug action on mycobacterium tuberculosis}.
\newblock In \emph{\bibinfo{booktitle}{Advances in Nonlinear Biological Systems: Modeling and Optimal Control}} (\bibinfo{publisher}{American Institute of Mathematical Sciences}, \bibinfo{year}{2020}).
\newblock \urlprefix\url{http://arxiv.org/abs/2003.12400}.

\end{thebibliography}
\newpage

\appendix

\section{Comparison of generalized pangraph Katz centralities in an empirical ecosystem}
\label{sec:generalized-centrality-analysis}

Katz centrality of a generalized vertex in a pangraph is equal to its digraph Katz centrality in the corresponding Levi graph (Eq.~\ref{eq:Katz_from_Levi}). The underlying reason is that the Katz centrality of a generalized vertex is a linear combination of Katz centralities of its heads and tails.

As in the case of classic Katz centrality, we computed generalized Katz centralities with the same parameters $\beta = 1,\;  \alpha = \frac{0.9}{\lambda}.$ After obtaining 20 centrality values for the fundamental vertices $v_i \in P_0$, we compare them between dihypergraph and pangraph representation. The pangraph representation also allows us to estimate centralities of the 76 panedges $e_j \in P_{3}$.

\begin{figure}[h!]
    \centering
    \includegraphics[width=1\linewidth]{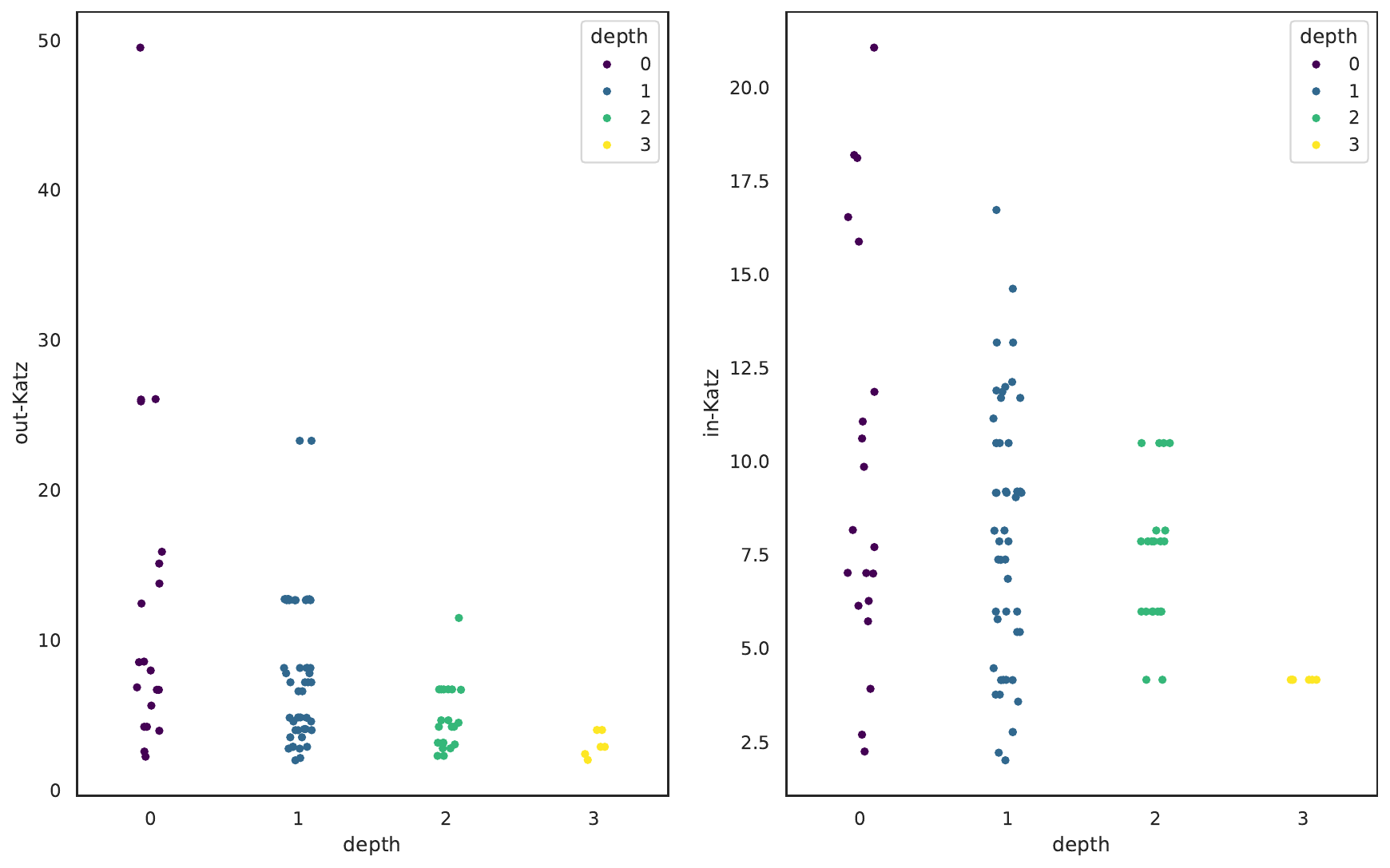}
    \caption{Generalized Katz centrality scores for fundamental vertices and panedges.}
    \label{fig:katz-stripplots}
\end{figure}

Just as in the case of the classic Katz centrality, generalized Katz centralities of panedges decrease with depth. Fig.~\ref{fig:katz-stripplots} presents generalized Katz centrality scores for all panedges, grouped by depth. Linear trends estimated for the in- and out-centrality of a pangraph are determined by:
\begin{eqnarray*}
    c_i^{in} &=& -1.52 k_i + 9.75, \\
    c_i^{out} &=& -3.47 k_i + 11.97.
\end{eqnarray*}
 
\begin{figure}[h!]
    \centering\includegraphics[width=\linewidth]{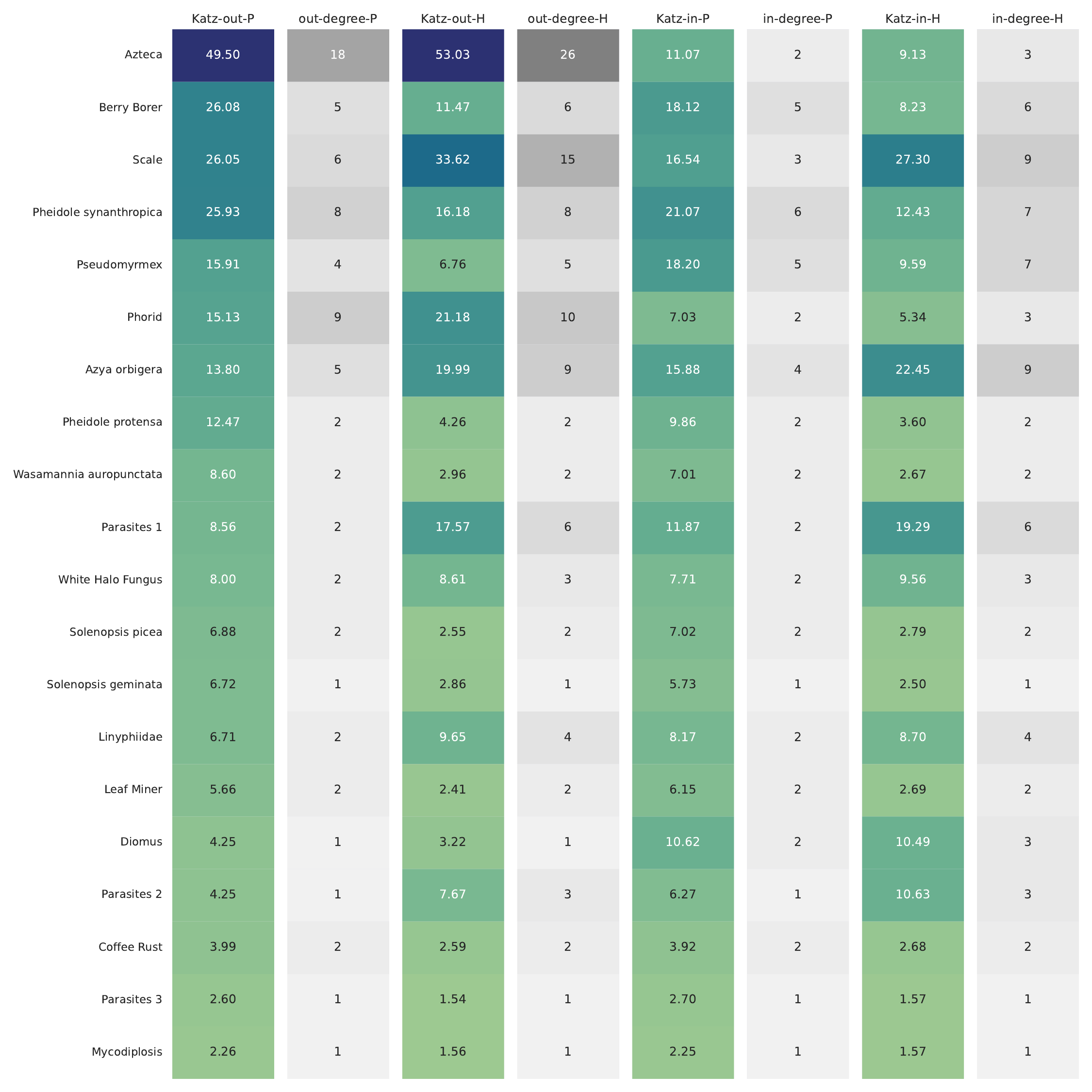}
    \caption{Generalized Katz in- and out-centralities for directed hypergraph and pangraph (fundamental vertices)
representations. In-centralities signify being a node receiving influences, whereas a high out-centrality means being an important influencer.}
    \label{fig:centralities-heatmap}
\end{figure}

We compare the hypergraph and pangraph generalized Katz centralities of fundamental vertices.  In Fig.~\ref{fig:centralities-heatmap} we present the numerical values of generalized Katz in-, and out-centralities as well as in- and out-degree centralities of the fundamental vertices in pangraph and dihipergraph representations as a heatmap. In both coffee agroecosystem representations, it is the generalized Katz out-centrality, and therefore the information about how the vertex influences its neighbors, that has a bigger range of values than generalized Katz in-centrality. The versatile interactions of the Azteca ants~\cite{Vandermeer_2010, Perfecto_coffe_agroeco} consistently make them the most important influencers.

Pangraph centralities significantly differ from dihypergraph ones, also changing the ranks of vertices. Scale, Azya orbigera, and Parasites 1 (see Fig.~\ref{fig:centralities-heatmap}) are the most central and perhaps sensitive in the dihypergraph. They lose importance in the pangraph in favor of Pheidole synanthropica, Pseudomyrmex, and Berry Borer.

Scale, Azya Orbigera, and Parasites 1 lose outgoing connections too. This time 
Scale centrality rank remains higher than that of Pseudomyrmex and Pheidole Synanthropica. 

Vertex degree centralities show how HOIs translated into additional hyperedges inflate generalized Katz centralities. The difference in dihypergraph and pangraph degree centralities of Scale, Azya orbigera, and Parasites 1 are much larger than those of the abovementioned three other vertices. This illustrates the consequences of inflating the degree centralities of participants of modified interactions, e.g. Scale and Azya orbigera, by the hypergraph representation, see Fig.~\ref{fig:pangraph_hypergraph_mapping}. Very large Azteca out-centrality contributes to Berry Borer and Scale out-centralities. In contrast, Azya Orbigera has a connection with Azteca in the other direction, which explains its much lower generalized Katz out-centrality.

However, vertices whose degree centralities do not change much may still have significantly different generalized Katz centrality values and ranks, e.g. Phorid. Its outgoing connections consist of three $3$-depth edges, two $2$-depth edges, and one 1-depth edge. As we observed in Fig.~\ref{fig:katz-stripplots}, deeper edges have generally lower generalized Katz centralities. They also contribute less to their source centrality than a direct connection to another vertex (see Eq.~\ref{eq:pangraph_Katz_walk},~\ref{eq:hypergraph_Katz_walk}).

\begin{figure}[ht]
    \centering\includegraphics[width=1\linewidth]{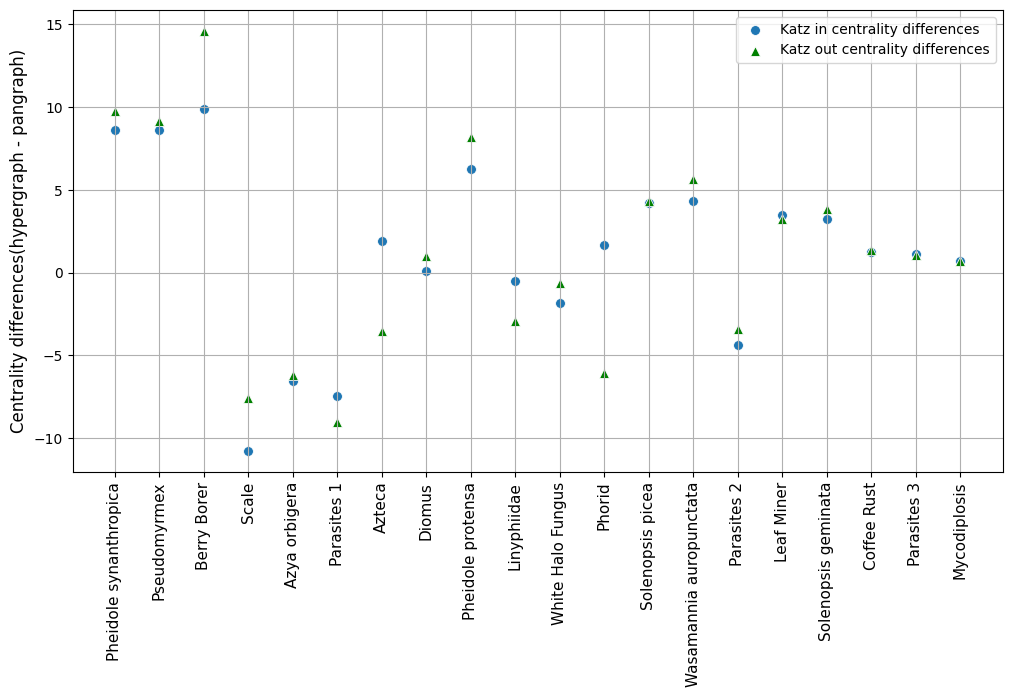}
    \caption{Differences between hypergraph and pangraph values of generalized Katz centrality for fundamental vertices}
    \label{fig:katz-diff}
\end{figure}

What are the differences between pangraph and hypergraph generalized Katz centrality scores? Fig.~\ref{fig:katz-diff} shows a plot of the differences between in- and out-centralities for hypergraph and pangraph. We can distinguish a group of vertices, such as Diomus, Coffee Rust, Parasites 3, or Mycodiplosis, where the differences between hypergraph and pangraph representations oscillate around zero. These vertices are neither directly adjacent to vertices of high centrality nor involved in numerous HOIs. As in the case of classical Katz centrality, vertices participating in or adjacent to the ones participating in numerous HOI generally have greater differences in centrality depending on the choice of graph representation.

Synthesizing the results, panedge centrality generally decreases with its depth. Furthermore, centrality values and ranks differ significantly depending on whether a pangraph or hypergraph representation is used. It is worth noting that the discrepancies between these representations are more apparent for vertices involved in higher-order interactions.

While classic Katz centrality omits the edges that are not influenced by any other vertex in the initial structure, generalized Katz centrality treats all panedges as vertices in the Levi graph. As we can see in Eq. \ref{eq:as_tw2}, the generalized Katz centrality approach induces some influences between vertices to be multiplied by $\alpha^2$ rather than $\alpha$. This property may affect the values of higher-order interaction effects. We can observe that for some vertices, centrality scores vary depending on measure choice, e.g. out-centrality for Scale or Azteca. However, an important advantage of generalized Katz centrality is the possibility of expressing centrality values for all panedges, providing a more comprehensive representation of the system's structure.

\section{Coffee agroecosystem pangraph and dihypergraph visualisations}
\label{sec:visualisations}
\begin{figure}[ht]
	\begin{center}
        \includegraphics[width=\linewidth]{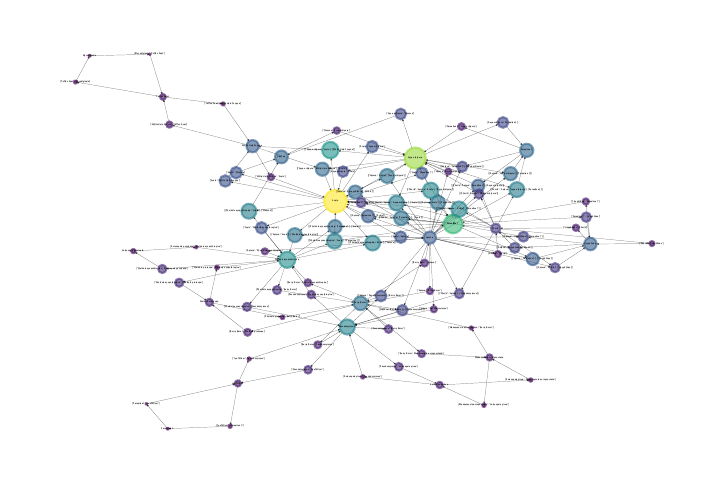}
        \caption{Levi graph of directed hypergraph representing the coffee agroecosystem. Sizes of nodes and their colors map the Katz in-centrality values $c_{\mL}^{\textrm{in}}$. See the plot's individual PDF file for higher resolution.} 
    	\label{fig:Katz_graph_hyper_in}
    \end{center}
\end{figure}

\begin{figure}[ht]
	\begin{center}
        \includegraphics[width=\linewidth]{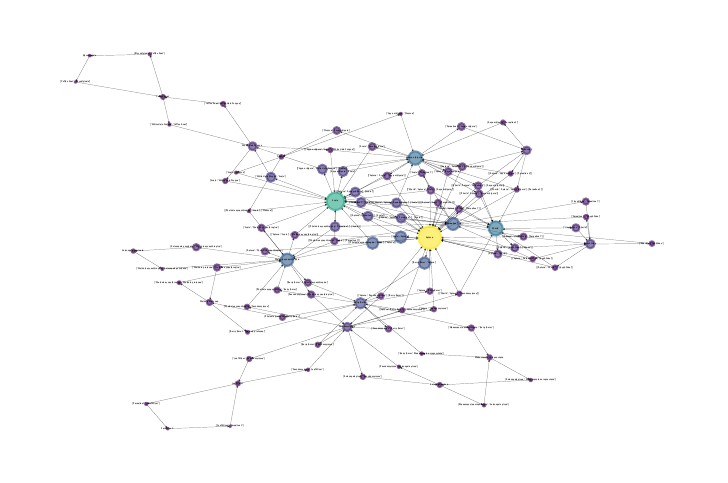}
        \caption{Levi graph of directed hypergraph representing the coffee agroecosystem. Sizes of nodes and their colors map the Katz out-centrality values $c_{\mL}^{\textrm{out}}$. See the plot's individual PDF file for higher resolution.} 
    	\label{fig:Katz_graph_hyper_out}
    \end{center}
\end{figure}

\begin{figure}[ht]
	\begin{center}
        \includegraphics[width=\linewidth]{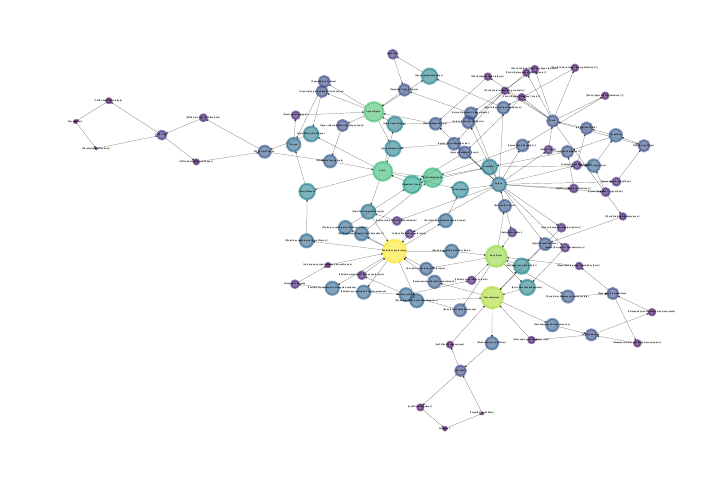}
        \caption{Levi graph of directed pangraph representing the coffee agroecosystem. Sizes of nodes and their colors map the generalized pangraph Katz in-centrality values $\tilde{c}_{\mP}^{\textrm{in}}$. See the plot's individual PDF file for higher resolution.} 
    	\label{fig:Katz_graph_pan_in}
    \end{center}
\end{figure}

\begin{figure}[ht]
	\begin{center}
        \includegraphics[width=\linewidth]{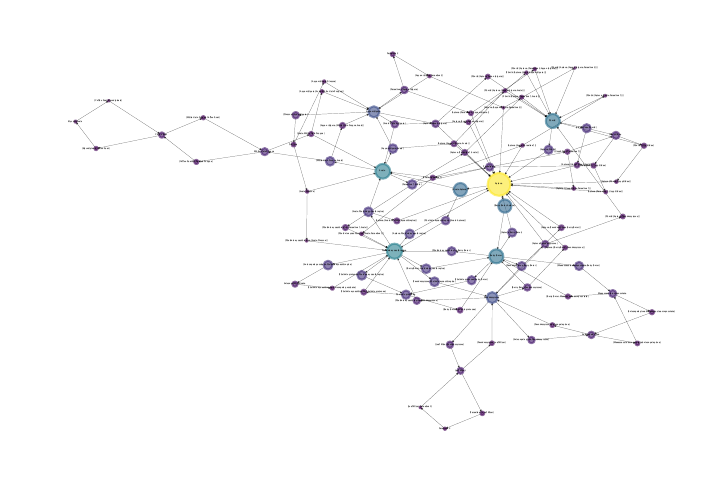}
        \caption{Levi graph of directed pangraph representing the coffee agroecosystem. Sizes of nodes and their colors map the generalized pangraph Katz out-centrality values $\tilde{c}_{\mP}^{\textrm{out}}$. See the plot's individual PDF file for higher resolution.} 
    	\label{fig:Katz_graph_pan_out}
    \end{center}
\end{figure}

Fig.~\ref{fig:Katz_graph_hyper_in} and~\ref{fig:Katz_graph_hyper_out} show the Levi graphs of the directed hypergraph and Fig.~\ref{fig:Katz_graph_pan_in} and~\ref{fig:Katz_graph_pan_out} of the pangraph representing the coffee agroecosystem. Sizes of nodes and their colors map the generalized Katz centrality values of Levi graph vertices. We present both Katz in- and out- centralities.

\end{document}


\appendix\section{Literature review}

Relevant articles:
\begin{itemize}

\item \cite{Cervantes_2021} presents an analysis of experimental data on behavior of four pollinator species in controlled conditions. The main focus is the influence of resource availability and presence of pesticide on pollinator-pollinator interactions, both between species and within one. Statistical tools show that out of the three presented possible models, one accounting for these environmental conditions is by far the best. Their influence differs between species but is clearly present. The authors stress that the environmental context in which pollinators interact with plants and with each other is the crucial part of the interactions. \\
I think this is a good example of an ecological paper describing a higher-order interaction, since the influence of other species is impacted by the environment (available amount of flowers and pesticide presence).
\begin{enumerate}
    \item adding effects from a distinct type of objects calls for a multilayer network formulation more than for a hypergraph 
    \item binary/discrete case, like presence/absence of a pesticide can be modelled through a multiplex network (with two/more copies of the underlying bipartite graph plant-pollinator)
    \item continuous case (continuous resource $R$ availability in formula (11)): a multilayer network with one layer of the bipartite graph, and another with the vertices influencing continuously the interactions in the bipartite graph
\end{enumerate}

\item \cite{Grilli2017} considers higher-order competition models. Their research stems from the lacking stability of of simple models as well as neutral models. They focus on large communities where competitors interact. They note that the impact of higher-order interactions of species is crucial on the existence of other species. They consider both deterministic and stochastic models.

Their narrative is based on a forest with $ m $ species. Given that a tree of species $ i $ dies, there is then a competition to fill up the gap created. There are two outputs,
\begin{enumerate}
    \item Deterministic model: An equivalence is shown where the competition is done in a sequence (where interactions are still done pairwise at a time, but with the winner of a game proceeding to compete with the next competitor) and a simultaneous competition (where trees from all $ m $ species compete at the same time).
    \item Stochastic model: Higher-order interactions leads to longer periods of coexistence of species.
\end{enumerate}

    \item \cite{Arruda} 
    has a good intro on dynamics on hypergraphs and linear stability analysis
    
    In the paper the hypergraph $\HH =\left\{\V,\E\right\}$ is defined in the following way. $\V=\left\{ v_i:\,\,i\in I\right\}$ is a set of verticies, $N=|\V|$ and $\E=\left\{e_j\right\}$ is a set of hyperedges, and $e_j\in 2^{\V}$ has a cardinality $|e_j|$. By $\E_i$ we also denote the set of hyperedges that contain the vertex $v_i$. The weighted adjacency matrix can be defined as

\[
A_{ij}=\sum_{e_j\in \E;\,\, i,k\in e_j;\,\,i\neq k }\frac{w_{ij}(e_j)}{|e_j|-1}
\]

where $w_{ik}(e_j)\in \RR$ is a weight.

    Let us consider a system

    \begin{equation}\label{eq:main}
\frac{dx_i}{dt}=f_i(x_i)+\sum_{e_j\in \E_i}g_j(x_{\left\{e_j\right\}}),
    \end{equation}
    where $x_i$ is the state of the vertex $v_i$, $f_i:\RR \rightarrow \RR$ is function that depends only on the state $x_i$ and $g_j(x_{\left\{e_j\right\}}):\RR^{|e_j|}\rightarrow \RR$ that takes all the states of the vertices on the hyperedge $e_j$, here denoted as $x_{\left\{e_j\right\}}$, and compute its contribution to $x_i$.

    Its linear stability in the neighbourhood of fixed point $x^*$ depends on the spectrum of matrix $M=F(x^*)+G(x^*)$ where 
    \[
    G(x^*)=\sum_{m=1}^{\max{|e_j|}}G^m(x^*), \qquad G^m_{ik}(x^*)=\sum_{e_j\in \E_i;\,\,|e_j|=m}\partial_{x_k}g_j(x_{\left\{e_j\right\}})|_{x=x^*}.
    \]



    \item \cite{Bairey2016} is the most typical example of use of a dynamical model with higher-order (non-bilinear) interactions in mathematical ecology. This is what we might want to cover - how to translate it to a weighted hypergraph properly?
    
    They used a replicator equation with non-linear fitness, see Fig.~\ref{Bairey_eq_ex} (for our reference). The model parameters are given by random two-, three- and four-dimensional tensors with entries drawn from a Gaussian distribution (mean 0, variance 1). It describes evolving fractional abundances of species whose total population is constant. The initial fractional abundances are equal ($\tfrac{1}{N}$, where $N$ is the number of species/variables).
    \begin{figure}[h!]
	\begin{center}
        \subfloat{\includegraphics[width=0.49\linewidth]{./fig/Bairey_1_equation.jpg}
		}
        \subfloat{\includegraphics[width=0.49\linewidth]{./fig/Bairey_2_example.jpg}}
        \caption{The model of \cite{Bairey2016} (left) and their examples of higher-order ecological interactions of microorganisms (right).} 
    	\label{Bairey_eq_ex}
    \end{center}
    \end{figure}
    To compare the impacts of various terms, they have run numerical simulations separately for only pairwise ($B\equiv 0$, $C\equiv 0$), then only three-way, and only four-way interactions.
    They defined a community as feasible if all the species existed at the end of the simulation with an abundance above $10^{-5}\tfrac{1}{N}$. They defined the \emph{critical strength of interactions} as corresponding to  of the communities exhibiting extinctions. 
    They found that the interaction strength at which $5\%$ of the random systems exhibit extinctions is the lower the more species there are for pairwise interactions and can be higher the more species there are for four-way interactions (Fig.~\ref{Bairey_res})

    \begin{figure}[h!]
	\begin{center}
        \subfloat{\includegraphics[width=0.49\linewidth]{./fig/Bairey_3_result.jpg}
		}
        \subfloat{\includegraphics[width=0.49\linewidth]{./fig/Bairey_4_interpret.jpg}}
        \caption{The results of \cite{Bairey2016} (left) and their description (right).} 
    	\label{Bairey_res}
    \end{center}
    \end{figure}

They remark, that local stability analysis models exclusively pairwise interactions, since the equation is linearised around a point and the high-order interactions are therefore embedded in the coefficients of the effective pairwise interactions obtained in the linearisation.

    \item 
    \cite{Young2021}
    introduces a method (a Bayesian generative model) to reconstruct possible higher-order interactions (represented using hypergraphs) from information about pariwise interactions, since actual data is often limited to those. Authors claim their method introduces hyperedges only with enough enough statistical evidence. The possible hypergraphs are evaluated using the formula \(P(H|G) = \frac{P(G|H)P(H)}{P(G)}\). \(P(H|G)\) is the probability of the hypergraph $H$ being a certain way for a given network $G$. $P(G|H)$ is called a projection component and is by analogy the probability of the network $G$ being a certain way given a hypergraph $G$. $P(G)$ is not important and serves as a normalisation constant. $P(H)$ is called a hypergraph prior. \\ 
    $P(G|H)$ follows a simple distribution, being $1$ iff two vertices appear jointly in any of the hyperedges of $H$ and $0$ otherwise. Authors remark that checking if $G$ is equal to the projection of the hypergraph $H$ is not a problem from a numerical point of view.
    $P(H)$ is based on an existing model, Poisson Random Hypergraphs Model (PRHM). The desired properties which the model fulfills are: the size of the interactions varies, not all vertices are connected by a hyperedge and some of the interactions are repeated (I'm not sure why it's good actually). 
    In this model the number of hyperedges connecting a set of vertices is a random variable following a Poisson distribution, with mean $\lambda_k$ dependent on the size of the set. The number of hyperedges connecting a set of $k$ vertices is equal to $A$ with probabily $P(A|\lambda_k) = \frac{\lambda_k^A}{A!}e^{-\lambda_k}$. All the hyperedges are modeled independently. 
    From this authors derive the formula for the probability of any hypergraph (assuming also some maximal hyperedge size), density of which is controlled by the parameters $\lambda_k$. They use what they call `a hierarchical empirical Bayes approach' and treat $\lambda_k$ as unknows drawn from prior distributions, use I believe some information theory and consider the problem of finding appropriate $\lambda_k$s solved. \\ 
    The authors prove two noteworthy properties of their model: firstly, it favors hypergraphs without repeated hyperedges, even though PRHM allows for duplicates. Secondly, the model favors sparser hypergraphs and introduces hyperedges only when it is justified. Those two properties allow the authors to conclude that such minimal hypergraphs are high-quality local maxima of $P(H|G)$. When it comes to the numerical point of view, the task is quite demanding, therefore some agorithm is proposed. In the conclusion authors say they would like to see a better one. \\
    Authors give several examples with both empirical and artificial data used, with the most attention given to an example of 613 American football games between 115 teams. Teams may play each other more often because of various reasons, mainly conferences (groups of team which all play each other) and geographical proximity. Some of the results are presented in the figures below (Fig.~\ref{Young_football_1}, ~\ref{Young_football_2}). There is also some focus on bipartite networks and the obtainted results show that more complex approach allowing higher-order interactions proved to be better in multiple cases (Fig. ~\ref{Young_bipartite}).

    \begin{figure}[h!]
	\begin{center}
        \includegraphics[width=0.9\linewidth]{./fig/Young_Football_1.png}
        \caption{Results of \cite{Young2021} about american football teams} 
    	\label{Young_football_1}
    \end{center}
    \end{figure}

    \begin{figure}[h!]
	\begin{center}
        \includegraphics[width=0.9\linewidth]{./fig/Young_Football_2.png}
        \caption{More results of \cite{Young2021} about american football teams} 
    	\label{Young_football_2}
    \end{center}
    \end{figure}

    \begin{figure}[h!]
	\begin{center}
        \includegraphics[width=0.5\linewidth]{./fig/Young_bipartite.png}
        \caption{Results of \cite{Young2021} about empirical bipartite networks} 
    	\label{Young_bipartite}
    \end{center}
    \end{figure}

\item [complex hypergraphs are equivalent to either an ubergraph or a multilayer network of two digraphs] \cite{Vazquez2022} introduces a concept of complex hyperhraphs (chygraphs). The main idea is to generalize the concept of hypergraphs beyond ubergraphs. Please note that the paper is not yet reviewed or published.
\begin{definition}[Complex hypergraph]
    A complex hypergraph (chygraph) $\chi(A,C)$ is a set of vertices $A$ and hypergraphs $C$ with vertex sets in $A \cup C$.
\end{definition}
The author gives several examples. For a graph the hypergraphs are said to be edges (which I understand as hypergraphs containing only one pairwise edge each, since they are meant to be actual hyperpgraphs?).
An ubergraph is a chygraph where the hypergraphs contain only one edge. \\
Another example is the system of scientific publications, which is represented by a chygraph $\chi(A, \{ \mathcal{H}_i(A_i \cup R_i, \{A_i, R_i\})\}) $. A publication is represented by a single hypergraph $\mathcal{H}_i$ with two edges $A_i$ (authors) and $R_i$ (references), which do not overlap. Here the difference between a chygraph and an ubergraph is visible: for an ubergraph the way to represent both references and authors would probably be similar, with two edges containing references and authors respectively. However the connection between those edges is important and the chygraph structure allows to include them in one hypergraph, which also has a clear meaning, as it represents a single publication. \\
\begin{figure}[h!]
	\begin{center}
        \includegraphics[width=0.5\linewidth]{./fig/Vazquez_example.png}
        \caption{Illustration of an example from \cite{Vazquez2022}} 
    	\label{Vazquez_example}
    \end{center}
    \end{figure}
The author also defines a matrix for chygraphs, which he decides to call the chy-adjacency matrix (although is seems more like an incidence matrix if anything):
\begin{definition}[Chy-adjacency matrix]
    Let $\chi(A, C=\{ \mathcal{H}_i(V_i, E_i)\})$ be a chygraph. Then the chy-adjacency matrix $\alpha$ is a  $|A\cup C| \times |A \cup C|$ matrix with 
    \begin{equation*}
        \alpha_{ij}=\begin{cases}
            1, \quad \text{if} \ i \in V_j ,& \\
            0, \quad \text{otherwise}
        \end{cases}
    \end{equation*}
    for $i,j \in A \cup C$.
    \end{definition}
    (to me this does not seem correct for several reasons, but maybe I'm missing something) \\
    There is also a definition of the length of the chygraph:
    \begin{definition}[Chygraph length]
        Let $\chi(A, C=\{ \mathcal{H}_i(V_i, E_i)\})$ be a chygraph and let $\Pi = \Pi_1 \cup \dots \cup \Pi_l$ be a partition of $C$ such that $\Pi_i \cap \Pi_j = \emptyset \text{ for } i \neq j$ and if $\mathcal{H}_i \in \Pi_j$, then $V_i \subset A \cup (\bigcup_{k \leq j} \Pi_k)$. The chygraph length $L(\chi)$ is the maximum $l$ between such partitions. 
        \end{definition}
    The second condition means the partition is hierarchical. The author then proceeds to analyze the mean component size. \\ 
    I think the paper seems to contains inaccuracies and does not provide much theory about chygraphs (and does not justify the need to define them).

\end{itemize}

\section{Definitions and methods}
\subsection{Hypergraphs}
Hypergraphs enable us to model systems of different nature (e.g. in social systems, neuroscience, ecology and biology) in which some connections and relationships are described by higher order interactions rather than pairwise interactions. They provide the most general and flexible way of modeling such systems involving higher order interactions. As an example of systems where higher order interactions take place we can consider a complex ecosystem in which three or more species compete for food and/or territory \cite{Benson2018} or social mechanisms, such as peer-pressure or collaborations, where the connections involve three or more individuals \cite{levine2017}. 
 
A hypergraph $H$ consists of a set $V$ of vertices representing the elements or individuals in the system and a set $E$ of hyperedges representing the higher-order interactions (interactions in groups of more than two elements) within the system and describing which vertices are involved in them. 

\begin{definition}[Hypergraph]
A hypergraph $H$ is a tuple $H=(V,E)$ where $V=\{v_i : i\in I \}$ is the set of vertices (vertices) $v_i$ and $E=\{e_j : j \in J \}$ is the set of hyperedges $e_j$ where each hyperedge $e_j$ is a nonempty subset of $V$.
\end{definition}

 We assume that a hypergraph has a finite, non-empty vertex and hyperedge sets i.e. nonempty and nontrivial hypergraph. We also assume that the hypergraph does not have a repeated hyperedge which correspond to multiple edges in a graph. \al{But there are multiple 'edges' in the ubergaph in the different sense. Example: $V=\{v_1,v_2\}$, $E=\{\{v_1,v_2\}, \{\{v_1\},\{v_2\}\}\}$.} The order of the hypergraph is $ \lvert V \rvert $ and the size of the hypergraph is $\lvert E \rvert $.  

A hyperedge $e$ with $ \lvert e \rvert =1 $ is a (self-)loop and $ \lvert e \rvert =2 $ is an edge in the classical sense. Two vertices $v_i$ and $v_j$ are adjacent if there is a hyperedge $e$ which contains both vertices, i.e. $ \{ v_i , v_j \} \subset e $. In particular, if $e=\{v \} $ is an hyperedge then $v$ is adjacent to itself. Two hyperedges in a hypergraph are incident if they share some common vertices i.e. their intersection is not empty. A hypergraph is uniform or in particular $k$-uniform if all hyperedges have the same cardinality $k$. 

Pairwise graph measures and metrics can be generalised to hypergraphs. Let $ H $ be a hypergraph. Let us consider the notion of degree of a vertex for hypergraph setting. In a graph, degree of a vertex $v$ is defined as the number of adjacent vertices to $v$ or equivalently the number of edges containing $v$. However, for a hypergraph, these two definitions are not equivalent as one hyperedge may contain more than one adjacent vertices to $v$. Therefore, the notion of degree of a vertex can be defined in two different ways. Namely, we distinguish these two definitions as incidence degree \aadd{$deg_i(v)$} which is defined as the number of hyperedges containing the given vertex \aadd{$v$} and adjacency degree \aadd{$deg_a(v)$} which is defined as the number of adjacent vertices to the given vertex, namely
\begin{equation}
\aadd{deg_i(v):=|\{e\in E:\,\,v\in e\}|,\qquad deg_a(v):=|\{w\in V:\,\,\exists_{e\in E} \,v,w\in e\}|}
\end{equation}

Consider a hypergraph $H=(V,E)$ with $V=\{ v_1 , \ldots, v_n \}$ and $E=\{ e_1 , \ldots, e_m \}$ such that there is no isolated vertex. An incidence graph or a Levi graph \cite{Joslyn2017UbergraphsAD} is a bipartite graph $ G = (V \cup E, E') $ which decomposes a hypergraph into the vertex set and edge set where $ (v_i,e_j) \in E'$ if and only if $v_i \in e_j $. The incidence matrix of the hypergraph $ H $ is defined as $n \times m$ matrix $I=(I_{ij} )$ where $I_{ij} =1$ if $v_i \in e_j$ and $I_{ij} =0$ otherwise. Unlike incidence matrices of graphs, in incidence matrices of hypergraphs, the sum of the elements in a column can sum up to a value larger than 2. In fact, this value is the size of the relevant hyperedge. The transition between incidence graph and the incidence matrix is apparent, there exists an edge $ (v_i,e_j) \in E' $ if $ I_{ij} = 1$.

\textcolor{blue}{In a graph, a walk of length $ k $ is a sequence of vertices $ v_1,v_2,\ldots, v_{k+1} $ not necessarily all distinct such that for all $ i \in \{ 1,2,\ldots, k \}$ there exists $e_j \in E $ with $v_i, v_{i+1} \in e_j $ \cite{Estrada_2006}. \cite{Aksoy2020} however points out the need to distinguish between a vertex based walk and an edge based walk. With hypergraphs, there is introduced the notion of "width" of an edge in the sense of the number of vertices belonging to that edge.}

\textcolor{blue}{\begin{definition}
    An \textit{s-walk} of length $ k $ in a hypergraph $ H = (V,E) $ is a sequence of overlapping edges $e_0, e_1, \ldots, e_k$ such that $ |e_i\cap e_j | \ge s$, for all $ i \ne j$, $ i,j \in \{0,1,\ldots, k\} $.
\end{definition}}

\textcolor{blue}{Using the s-walk, other graph metrics are also defined such that connectedness, distance,eccentricity, closeness centrality. However, these are defined in terms of edges.}

Using the adjancency matrix, the centrality measure of a hypergraph can be defined. Let $A$ be an adjacency matrix of a hypergraph. We know $ A $ is a symmetric matrix and hence, there exists an orthogonal matrix $ U $ that diagonalises $ A$,
\[A = UDU^T \]
where $ D = diag(\lambda_1, \lambda2, \ldots, \lambda_n) $ and the columns of $ U $ are the corresponding eigenvectors to the eigenvalues in $D$. The number of walks of length $ k $ from vertex $ v_i $ to vertex $ v_j $ is,
\[ \mu_k (ij) = (A^k)_{ij} = \sum_{l=1}^n(UD^k)_{il}(U^T)_{lj} = \sum_{l=1}^n(UD^k)_{il}u_{jl} = \sum_{l=1}^n(u_{il}u_{jl}\lambda_l^k). \]
The number of walks of length $ k $, $W_k $, is,
\[ W_k = \sum_{ij} \mu_{ij} = \sum_{l=1}^n \left(\sum_{i=1}^n u_{il}u_{jl}\right)^2 \lambda_l^k. \]

The subgraph centrality \cite{Estrada05} of vertex $v_i$ is,
\[C_S(v_i) = \sum_{k\ge 0 } \frac{(A^k)_{ii}}{k!}\]

\subsection{Representations of higher-order interactions}
There are various ways to represent networks that involve higher-order interactions. These representations can be grouped into two categories: graph based representations and explicit higher-order representations. Graph based representations encodes all interactions with considering only edges (i.e. pairwise interactions). These include graphs, bipartite graphs (where interactions and interaction vertices are encoded as vertices in two different layers), motifs (small subgraphs with specific connectivity structures) and cliques (special type of cliques) (see Figure \ref{highorder}). 

However, in these representations some information about the higher-order interactions are not reflected completely and therefore some data is lost. For this reason, some explicit higher-order representations that are constructed using non-pairwise building blocks (e.g. simplices and hyperedges) are needed. Explicit higher order representations consist of simplicial complexes and hypergraphs (see Figure \ref{highorder}). Simplicial complexes are collection of simplices that allow us to reflect the exact nature of higher-order interactions. However, their drawback is that given a simplex all subsimplices are required in a simplical complex. This requirement results in complications in the representation. In hypergraphs, this requirement is removed and hence they provide the most efficient and general way to represent higher-order interactions.   

\subsection{Directed hypergraph}

Let us define our notion of weighted dihypergraph that, unlike its classical counterpart, allows to capture long distance interactions. We start with a classical definition of dihypergraph that can be found also here\newline

\url{https://www.pks.mpg.de/~mapcon12/Slides/Ostermeier_Mapcon12.pdf}

\begin{definition}[Classical dihypergraph]
By the notion of classical dihypergraph $\HH$ we understand a pair $\HH =\left(\V,\E\right)$ such that $\V=\left\{ v_i:\,\,i\in I\right\}$ is a set of verticies, $N=|\V|$,  and $\E=\left\{e_j\right\}$ is a set of directed hyperedges, $e_j=(\V_j^{in},\V_j^{out})$, $\V_j^{in},\V_j^{out}\subset \V$.
\end{definition}
The elements of $\V_j^{in}$ (resp. $\V_j^{out}$) we call heads (resp. tails) of an edge $e_j$. By $\E_i^{in}$, (resp. $ \E_i^{out}$) we denote the set of hyperedges that contain the vertex $v_i$ as a head (resp. tail).

\begin{definition}[Substrate digraph]
\end{definition}

\begin{definition}[Classical sub-dihypergraph]\label{def:clas_sub-dhg}
By the sub-dihypergraph $\bHH$ of $\HH$ we understand a dihypergraph $\bHH =\left(\bV,\bE\right)$ such that $\V=\bV$, $\bV_j^{x}\subset \V_k^{x}$, $x=\text{in},\text{out}$; and  
\[
\forall_{\overline{e}_j=\left(\bV_j^{in},\bV_j^{out}\right)\in \bV}\quad\exists_{e_k=\left(\V_k^{in},\V_k^{out}\right)\in \V}\qquad \bV_j^{in}\cap \V_k^{in}\neq \emptyset \neq \bV_j^{out}\cap \V_k^{out}. 
\]
\end{definition}
\begin{remark}
\begin{enumerate}
\item Note that according to definition \ref{def:clas_sub-dhg}, the condition $\bE\subset\E$ does not have to hold. Indeed, if $(\left\{v_1\right\},\left\{v_2,v_3\right\})\in \E$ then $(\left\{v_1\right\},\left\{v_2\right\})\in \bE$. In the meantime $(\left\{v_1\right\},\left\{v_2\right\})\notin \E$.

\item The inclusion holds in the sense of substrate digraphs. Namely, if $S(\HH)=\left(S(\V),S(\E)\right)$ is a substrate digraph of dihypergraph $\HH$, then $S(\bE)\subset S(\E)$.
\end{enumerate}
\end{remark}

\subsection{Weighted dihypergraph - our first approach}

\begin{definition}[Generalised weighted dihypergraph]
We say that $\tHH$ is a generalised weighted dihypergraph of a dihypergraph $\HH$ if $\tHH=\left(\tV, \tE, \tphi\right)$ where $\V=\tV$, 
\[
\tE=\left\{\left(\bHH^{out}_j,\bHH^{in}_j\right):\,\,j\in J\right\}, \qquad  \bHH^{out}_j,\bHH^{in}_j\text{ are sub-dihypergraphs of}\,\, \HH,
\]
and $\tphi:\left\{\bHH^{in}_j:\,\, j\in J\right\}\rightarrow \RR$.
\end{definition}

\subsection{Ubergraph}
An ubergraph \cite{Joslyn2017UbergraphsAD} is a generalisation of a hypergraph in which hyperedges are allowed to contain not only vertices but other edges as vertices as well. In other words, in an ubergraph, an edge (called uberedge) can consist of some vertices and some other edges. In some sense, some edges have two roles: they are edges of the graph and they can also serve as a vertex. For the formal definition of ubergraphs we need the following notation.  

Let $X$ be a finite set. We denote 
    \begin{equation*}
        \mathcal{P}(X)^k := \mathcal{P} \left( \bigcup_{i=0}^{k} P_i \right),
    \end{equation*}
where $\mathcal{P}(X)$ denotes the family of all subsets of $X$ and
    \begin{equation*}
        P_0 = X, \quad P_i =\mathcal{P} \left( \bigcup_{j=0}^{i-1} P_j \right), \quad i\geq 1. 
    \end{equation*}
As a consequence of this iterative process, we have $\mathcal{P}(X)^0 =\mathcal{P}(X)$ and $\mathcal{P}(X)^1 =\mathcal{P}(X\cup \mathcal{P}(X) )$ and so on. 
\begin{definition}(\cite{Joslyn2017UbergraphsAD})
    A depth $k$ ubergraph is a pair $U=(V,E)$ where 
    \begin{itemize}
        \item[1] $V$ is a non-empty set of fundamental vertices,
        \item[2] $E \subseteq \mathcal{P}(V)^k$ is a finite set of uberedges,
        \item[3] if $s$ belongs to an uberedge and $s \notin V$, then $s$ is itself an edge. 
    \end{itemize}  
\end{definition}
\begin{remark}
Note that the condition 3 requires that uberedges can only contain fundamental vertices and other edges. Consequently, in an ubergraph, there are some edges that are also vertices. To distinguish such vertices from fundamental vertices, we will call the elements of $V$ as fundamental vertices and elements of $V \cup E$ as vertices.
\end{remark}
\begin{plain}
    Every hypergraph $H=(V,E)$ is a depth-0 ubergraph. In this case, set of fundamental vertices is $V$, set of uberedges is set of hyperedges $E$ and the requirement 3 is fulfilled since all hyperedges can only contain vertices.    
\end{plain}
\begin{plain}\label{exuber}(\cite{Joslyn2017UbergraphsAD})
    Consider the ubergraph $U=(V,E)$ with fundamental vertex set $V=\{ 1,2,3 \}$ and let $e_1 =\{ 1 \} , \ e_2 =\{ 1,3\} $. The set of uberedges is given by $$E=\{ e_1 ,e_2, \{1,3, e_1\} , \{2, e_2 \}, \{ 1, \{2, e_2 \} \}  \} .$$
    Here, the set of vertices is $ \{ 1,2,3, e_1 , e_2 , e_3, e_4 , e_5 \} $. It is clear that the uberedge $e_1$ is a vertex in the uberedge $e_3 := \{1,3, e_1\}$. Similarly, the uberedge $e_2$ is contained as a vertex in the uberedge $ \{2, e_2 \}$ and $e_4 := \{2, e_2 \}$ is contained as a vertex in the uberedge $e_5 := \{ 1, \{2, e_2 \} \}=\{1, e_4 \}$. Note that $U $ is a depth 2 ubergraph. 
\end{plain}
\begin{definition}(\cite{Joslyn2017UbergraphsAD})
    The incidence matrix of an ubergraph can be defined in a similar manner to hypergraphs. Namely, the incidence matrix of an ubergraph $U=(V,E)$ is a $ (\lvert V \rvert +\lvert E \rvert) \times \lvert E \rvert $ matrix with the entries 
    \begin{equation*}
        I_{ij}=\begin{cases}
            1, \quad \text{if} \ i \in j ,& \\
            0, \quad \text{otherwise}
        \end{cases}
    \end{equation*}
    \gdel{The degree of a vertex $x \in V \cup E$ can be defined as the sum of the entries in the corresponding row of incidence matrix. Namely, the degree of $x$ is the number of uberedges containing $x$.}  
\end{definition}
\begin{plain}
    Let us construct the incidence matrix of the ubergraph presented in Example \ref{exuber}. Since the set of vertices is $ \{ 1,2,3, e_1 , e_2 , e_3, e_4 , e_5 \} $ and the set of uberedges is $\{ e_1 ,e_2, e_3 ,e_4, e_5 \} $, the incidence matrix can be given
    \begin{equation*}
        I=\begin{pmatrix}
    1 & 1 & 1 & 0 &1 \\
    0 & 0 & 0 & 1 & 0 \\
    0  &  1 & 1  & 0 & 0 \\
     0  & 0  & 1  & 0 & 0 \\
     0  & 0  & 0  & 1 & 0 \\
     0  & 0  & 0  & 0 & 0 \\
     0  & 0  & 0  & 0 & 1 \\
     0  & 0  & 0  & 0 & 0
    \end{pmatrix}.
    \end{equation*}
\end{plain}
\begin{definition}(\cite{Joslyn2017UbergraphsAD})
    Let $U=(V,E)$ be an ubergraph with fundamental set of vertices $V$ and set of uberedges $E$ such that $\lvert V \rvert =m $, $\lvert E \rvert =n $. Two vertices $u,v \in V \cup E$ are adjacent if there is an uberedge $e$ such that $u,v \in e$. For example, the vertices $1$ and $e_1 $ are adjacent in Example \ref{exuber} since they both lie in the uberedge $e_3 := \{1,3, e_1\}$. 
    
    The adjacency matrix of an ubergraph can be defined analogously to hypergraphs. Explicitly, the adjacency matrix $A=(A_{ij} )$ of ubergraph $U=(V,E)$ is a $ (n +m) \times (n +m) $ matrix such that $A_{ii} =0$ and for $i \neq j$, $A_{ij} $ is the number of uberedges that contain both $i$ and $j$. 
\end{definition}
\begin{plain}
    Let us construct the adjacency matrix of the ubergraph presented in Example \ref{exuber}. Since the set of vertices is $ \{ 1,2,3, e_1 , e_2 , e_3, e_4 , e_5 \} $, the adjacency matrix can be given
    \begin{equation*}
        A=\begin{pmatrix}
    0 & 0 & 2 & 1 &0 & 0 & 1 & 0 \\
     0 & 0 & 0 & 0 &1 & 0 & 0 & 0 \\
     2 & 0 & 0 & 1 &0 & 0 & 0 & 0 \\
     1 & 0 & 1 & 0 &0 & 0 & 0 & 0 \\
     0 & 1 & 0 & 0 &0 & 0 & 0 & 0 \\
      0 & 0 & 0 & 0 &0 & 0 & 0 & 0 \\
      1 & 0 & 0 & 0 &0 & 0 & 0 & 0 \\
     0 & 0 & 0 & 0 &0 & 0 & 0 & 0
    \end{pmatrix}.
    \end{equation*}
\end{plain}

\subsection{Metabolic graph}
 Metabolic graphs~\cite{metabolic_network} are defined to allow modelling networks of reactions in which every reaction can combine multiple vertices and some vertices can influence a reaction (think of enzymes inhibiting or stimulating a reaction).
 
\begin{definition}[Metabolic graph]
A metabolic graph is an ordered quintuplet $G=(V, H, U, \Psi_H, \Psi_U)$ where $V$ is the set of vertices (vertices), $H$ is the set of directed hyperedges, $\Psi_H$ is the function assigning weights to hyperedges, $U$ is the set of signed uberedges and $\Psi_U: U \rightarrow \{+,-\}$
\end{definition}

\subsection{Petri net}
Petri nets~\cite{Petri_thesis, Petri_Peterson_book} represent chemical reactions and other processes, such as population dynamics in ecology. They are bipartite graphs representing compounds (species, places) and reactions (transitions) as separate sets of vertices. Importantly, they were invented as an exact representation of reactions in the spirit of Markov chains, with weights being only natural numbers. Their literal translation to ecology used explicit analogs - e.g. a wolf and a rabbit enter a predatory reaction, with two rabbits leaving it. The definitions of such objects come in various flavours, as well as the names 'Petri nets' and 'reaction networks'. 

\begin{definition}[Reaction network]
    A reaction network is a tuple $N = ( S , T , s, t )$, where $P$ and $T$ are disjoint finite sets of species and transitions, respectively. Maps $s,t: S \rightarrow \mathbb{R}^{\|S\|}$ define the weights with which each species participate in a transition as a substrate or a product, respectively.
\end{definition}
A reaction network/Petri does not include a representations of the influence of vertices on edges/reactions.

\begin{definition}[Open Petri net]
    An open reaction network~\cite{Baez_open_petri_2017} is a reaction network $R$ with a list of inputs $X \in  \mathbb{R}^{\|S\|}$ and outputs $Y \in  \mathbb{R}^{\|S\|}$ from outside the system.
\end{definition}

Petri nets are equivalent to directed hypergraphs, which redefine transitions as hyperedges. 

\begin{figure}
    \centering
    \includegraphics{fig/higher order repr.jpg}
    \caption{Representations of higher-order interactions. This figure is taken from \cite{BATTISTON20201}.}
    \label{highorder}
\end{figure}

\subsection{Examples of hypergraphs}
Let us present some concrete examples to hypergraphs.
\begin{plain}\label{ex21}
The simplest example of a hypergraph is a graph which consists of a vertex set $V$ and an edge set $E$ where edges are connecting two vertices, hence only pairwise interactions are taken into account.
\end{plain}

\begin{plain}\label{ex22}
Let us present an abstract example. Consider the hypergraph $H=(V,E)$ with $V=\{a,b,c,d,e\}$ and $E=\{ e_1 =\{a\},e_2 = \{a,b\},e_3 = \{a,b,c\}, e_4 = \{b,c,d,e\} \}   $. Here the hyperedges $\{a,b,c\}$ and $\{b,c,d,e\}$ represent higher-order (non-pairwise) interactions, $\{a,b\}$ is an edge in the classical sense and $\{a\}$ is a loop. The degree of the vertex for example $a$ is three since it is $H(a)=\{ \{a\}, \{a,b\}, \{a,b,c\} \} $ i.e. there are three hyperedges containing $a$.
\end{plain}

\subsection{Matrix representations of hypergraphs}
Consider a hypergraph $H=(V,E)$ with $V=\{ v_1 , \ldots, v_n \}$ and $E=\{ e_1 , \ldots, e_m \}$ such that there is no isolated vertex. Then, the incidence matrix of the hypergraph $H$ is defined as an $n \times m$ matrix $I=(I_{ij} )$ where $I_{ij} =1$ if $v_i \in e_j$ and $I_{ij} =0$ otherwise. Unlike incidence matrices of graphs, in incidence matrices of hypergraphs, the sum of the elements in a column can sum up to a value larger than 2. In fact, this value is the size of the relevant hyperedge. Moreover, similar to graphs, the degree of a vertex $v_i$ is equal to the sum of the inputs on the $i$-th row of the incidence matrix.  

For example, the incidence matrix of the hypergraph in Example \ref{ex22} is
\begin{equation*}
    I=\begin{pmatrix}
    1 & 1 & 1 & 0 \\
    0 & 1 & 1 & 1 \\
    0  &  0 & 1  & 1 \\
     0  & 0  & 0  & 1 \\
     0  & 0  & 0  & 1
    \end{pmatrix},
\end{equation*}
where $v_1 =a, \ v_2 =b, \ v_3 =c, \ v_4 =d,\ v_5 =e $.

Adjacency matrix of a graph or a hypergraph completely encodes the connectivity of the graph or hypergraph. Namely, adjacency matrix of a hypergraph is an $n \times n$ matrix $A=(A_{ij} )$ where $A_{ii} =0$ and for $i \neq j$, $A_{ij} $ is the number of hyperedges that contain both $v_i$ and $v_j$. Moreover, it can be represented in terms of the incidence matrix as
\begin{equation*}
    A=II^T -D,
\end{equation*}
where $D$ is the diagonal matrix whose diagonal entries are the number of hyperedges a vertex belongs to. 

As an example, let us construct the adjacency matrix of the hypergraph in Example \ref{ex22} that can be given by 
\begin{equation*}
    A=\begin{pmatrix}
    0 & 2 & 1 & 0 & 0 \\
    2 & 0 & 2 & 1 & 1 \\
    1  &  2 & 0  & 1 & 1 \\
     0  & 1  & 1  & 0 & 1 \\
     0  & 1  & 1  & 1 & 0
    \end{pmatrix}.
\end{equation*}
As for hypergraphs in which hyperedges are weighted, the adjacency matrix can be introduced as 
\begin{equation*}
    A=IWI^T -D, 
\end{equation*}
where $I$ is the incidence matrix, $W$ is the diagonal matrix with the weights of the hyperedges along the diagonal, and $D$ is the diagonal matrix whose diagonal entries are the number of hyperedges a vertex belongs to.


\subsection{Questions}
\begin{itemize}
    \item does weighted have a unique equivalent weighted network (as in 'Dynamical systems on hypergraphs' by Carletti et al.)? Is it useful just for Laplacians or for other purposes too?
\end{itemize}

\section{Ecological examples}
\subsection{Competition}
A study of the competition between plants~\cite{Mayfield2017} provided empirical evidence that "higher-order interactions strongly influence species’ performance in natural plant communities, with variation in seed production (as a proxy for per capita fitness) explained
dramatically better when at least some higher-order interactions are considered."

The model (see Fig.~\ref{fig:Mayfield_1}) predicts a plant fecundity through an exponential dependence on the total abundance of plants in the area as well as on products of abundances of two plants. As these two species can be different from the influenced species in question, it constitutes a higher-order interaction. 

In ubergraph picture this higher-order interaction would contain two vertices influencing a self-loop of a vertex.

The predictive power of the model has gained significantly when such higher-order interactions were present (see Fig.~\ref{fig:Mayfield_2}).

\begin{figure}[h!]
	\begin{center}
        \subfloat{\includegraphics[width=0.6\linewidth]{./fig/Mayfield_1.png}}
        \subfloat{\includegraphics[width=0.4\linewidth]{./fig/Mayfield_1_1.png}}
        \caption{Impact on fecundity (number of offspring) of plants from the abundances of other plants - in \cite{Mayfield2017}.} 
    	\label{fig:Mayfield_1}
    \end{center}
    \end{figure}  

\begin{figure}[h!]
	\begin{center}
        \includegraphics[width=\linewidth]{./fig/Mayfield_2.png}
        \caption{The goodness-of-fit results of the nonlinear model of plant competition in \cite{Mayfield2017}.} 
    	\label{fig:Mayfield_2}
    \end{center}
    \end{figure}

Other competitive examples include e.g. secretion of chemical compounds affecting many other species (antibiotics). [TO DO]

\subsection{Scavengers}
A 2014 review~\cite{Moleon_scavenging} of ecological literature describing scavenging highlights the intertwined interactions between the live prey, its predators and scavengers. In particular, scavenging indirectly affects the population dynamics of consumed organisms. The familiar model in which vertices exercise influence on edges in a digraph (see Fig.~\ref{Moleon}) shows a clear case for a hypergraph application.

\begin{figure}[h!]
	\begin{center}
        \includegraphics[width=0.7\linewidth]{./fig/Moleon_1_model.jpg}
        \caption{A model of four-way live prey-predator-carrion-scavenger interactions in \cite{Moleon_scavenging}, adapted from \cite{Getz_2011}.} 
    	\label{Moleon}
    \end{center}
    \end{figure}

  A recent model of the  scavenger-predator-prey-carrion interaction~\cite{Mellard2021} employs non-polynomial four-way interaction terms between them (see Fig.\ref{Mellard_fig}). Predators and scavengers affect each other by changing the time each spends handling its prey. In short, the abundances of prey $R$, predator $P$, carrion $C$ and scavenger $S$ change according to
\begin{align}
    \dot{P}&=P[-m_P + R k_P(R,C,S) +C q_P(R,C,S)] \\
    \dot{S}&=P[-m_S + R k_S(R,C,P) +C q_S(R,C,P)] \\
    \dot{R}&=g(R) - R[P k_P(R,C,S) - S k_S(R,C,S)] \\
    \dot{C}&=P k'_P(R,C,S) + S k'_S(R,C,P) - P q'_P(R,C,S) -S q'_S(R,C,P) - C\delta ]
\end{align}
where $k_i, q_i,k'_i, q'_i$ with $i \in {P,S}$ are rational functions of indicated variables, resulting from the functional responses of predators attacking prey to the abundance of prey and predators~\cite{holling_1959}. All other symbols: $m_P, m_S, \delta$ refer to constants.

\begin{figure}[h!]
	\begin{center}
        \subfloat{\includegraphics[width=0.49\linewidth]{./fig/Mellard_1_model.jpg}}
        \subfloat{\includegraphics[width=0.49\linewidth]{./fig/Mellard_2_results.jpg}}
        \caption{Left: an explicit dynamical model of four-way live prey-predator-carrion-scavenger interactions in \cite{Mellard2021}. Right: the effects of scavengers on predator kill rates as observed in real ecosystems.} 
    	\label{Mellard_fig}
    \end{center}
    \end{figure}

\subsection{Lichens and other composite organisms}

\section{Other notes}

\section{Previous notes from Section 3}

\subsection{Graph theory toolbox}
In the ecological modelling a network serves as a basic tool to represent a food web. In our considerations we start with weighted multilayer digraph, which in classical theory not only bears the message on trophic relationship but allows to indicate e.g. the trophic level by the position in the multistructure. In this research we argue that multilayer structure can be interpreeted in the context of long-distance interection between entities in the food web. 

Let us change classical definition of multilayer weighted digraph $\mathcal{G} = (V,E,L,\phi)$, \cite[Sec.~2.1]{KivArena2014}, into the one that seem far way more complicated at the first glance, but allows for easy genaralisation.

\begin{definition}
    \label{def: WeightedDigraph}
     \textbf{A weighted multilayer digraph} is an ordered triple $\mathcal{G} = (V,E, \phi)$ where
    \begin{enumerate}
        \item $V = \{v_i\; | \; i \in I\}$, $|I|=n$, is a set of elements, called vertices;
        \item $E = \{e_k\; | \; k \in K\}\subseteq V \times V$, $|K|=M$,   is a set of ordered pairs of elements from V, called \textbf{edges};
        \item function $\phi: E \xrightarrow{} \mathbb{R}$ assigns a weight to each edge.
    \end{enumerate} 
\end{definition}

\subsection{Weighted directed hypergraph}

Hypergraphs generalise graphs by allowing edges to contain any number of vertices. To adequately describe ecosystems and generalise weighted digraphs we introduce weighted dihypergraphs~\cite{Bretto2013} straight away. They distinguish points of entry (tails) and exit (heads) among the vertices belonging to a hyperedge.
\begin{definition}
\textbf{A weighted dihypergraph} $H$ is an ordered triple $H=(V,E,\phi)$, where
\begin{enumerate}
    \item $V=\{v_i : i\in I \}$ is the set of vertices (vertices) $v_i$,
    \item $E=\{(e_j^{\mathrm{in}},e_j^{\mathrm{out}}) : j\in J \}$ is the set of directed hyperedges (hyperarcs), where each hyperedge $e$ consists of nonempty subsets of $V$ called tails and heads;
    \item $\phi: E \rightarrow \mathbb{R}$ asigns weights to hyperarcs.
\end{enumerate} 
\end{definition}

\subsection{Diubergraph}

An ubergraph~\cite{Joslyn2017UbergraphsAD} is a generalisation of a hypergraph in which hyperedges are allowed to contain not only vertices but other edges as vertices as well. In other words, in an ubergraph, an edge (called uberedge) can consist of \mdel{some }vertices and \mdel{some }other edges. \madd{This way, some edges play a second role, analogous to a digraph vertex.} For the formal definition of ubergraphs we need the following notation.  

Let $X$ be a finite set. We denote 
    \begin{equation}
        \mathcal{P}(X)^k := \mathcal{P} \left( \bigcup_{i=0}^{k} P_i \right),
    \end{equation}
where $\mathcal{P}(X)$ denotes the family of all subsets of $X$ and
    \begin{equation}
        P_0 = X, \quad P_i =\mathcal{P} \left( \bigcup_{j=0}^{i-1} P_j \right), \quad i\geq 1. 
    \end{equation}
As a consequence of this iterative process, we have $\mathcal{P}(X)^0 =\mathcal{P}(X)$ and $\mathcal{P}(X)^1 =\mathcal{P}(X\cup \mathcal{P}(X) )$ and so on. 

\begin{definition}
    A depth $k$ ubergraph is a pair $U=(V,E)$ where 
    \begin{itemize}
        \item[1] $V$ is a non-empty set of fundamental vertices,
        \item[2] $E \subseteq \mathcal{P}(V)^k$ is a finite set of \gadd{nonempty} uberedges,
        \item[3] if $s$ belongs to an uberedge and $s \notin V$, then $s$ is itself an \gadd{uber}edge. 
    \end{itemize}  
     We say that elements of $V\cup E$ are vertices, while for every $e \subset E$ elements of $e \setminus V$ are called edges. 
\end{definition}
\madd{The third condition ensures that edges participating in another uberedge already exist.}
\mat{An alternative definition introducing less symbols and consistent with layers of uberedges:}
\begin{definition}
    For a finite set $X$ we denote the family of all ordered subsets of $X$ by $\mathcal{P}(X)$. We introduce a series $P_k(X)$,
\begin{equation}
        P_0(X) = X, \quad P_1(X) = \mathcal{P}(X), \quad P_{k+1}(X) =\mathcal{P} \left( \bigcup_{i=0}^{k} P_i(X) \right), \quad i\geq 1.
\end{equation}
Elements of $P_k(X)$ are called \textbf{depth-$k$ uberedges}\footnote{We correct the definition of \cite{Joslyn2017UbergraphsAD} where $P_0$ was inconsistent with the recursive formula.}.
\end{definition}
\madd{Digraph edges are depth-one uberedges.}

\begin{definition}
    For $k\geq 1$, a \textbf{depth-$k$ diubergraph} is a triple $U=(V,E, \phi)$ where 
    \begin{itemize}
        \item[1] $V$ is a non-empty set of fundamental vertices,
        \item[2] $E \subseteq P_k(V)$ is a finite set of nonempty uberedges,
        \item[3] if $s$ belongs to an uberedge and $s \notin V$, then $s$ is itself an uberedge,
        \item[4] $\phi:E\rightarrow \mathbb{R}$ assigns weights to uberedges. 
    \end{itemize}  
     We say that elements of $V\cup E$ are vertices, while for every $e \subset E$ elements of $e \setminus V$ are called edges.   
\end{definition} 
\mat{Do we find this definition of edges useful or should we rather call uberedges edges interchangeably?}
\madd{A digraph is a depth-one diubergraph.}

We note that each uberedge can be explicitly written as a nested set of posibly ordered sets of fundamental vertices from $V$. We call it a \emph{fundamental form of an uberedge}, and denote as $e(V)$. For example, let $V=\{v_1, v_2,v_3\}$, $e_1=(v_1,v_2)$ and $e_2=(v_3,e_1)$.  Then,
\begin{equation}\label{Fuedge}
    e_2(V)=(v_3, (v_1, v_2)).
\end{equation}

\subsection{Incidence graph representation of an ubergraph}
The ubergraph concept is a natural generalisation of a graph. It is also the structure we immediately imagine when we think of vertices impacting processes (edges) rather than vertices themselves. It has been explicitly drawn in numerous applied studies~\cite{Bairey2016, Moleon_scavenging}. It can also be represented through its digraph incidence (Levi \footnote{vertices from V plus U}) graph.
The incidence matrix of an ubergraph can be interpreted as an adjacency matrix of a directed graph consisting of the joint sets of vertices and uberedges.
They are connected by the relation of being a part of another.
\begin{definition}\label{uber_levi_graph}
    A digraph incidence representation of an ubergraph $\mathcal{U}=(V,E)$ is a digraph $\mathcal{G}=(V \cup E,E')$ where for $v, w \in (V \cup E)$ 
    $$(v,w) \in E' \iff v \in w. $$
 
\end{definition}

Fig.\ref{fig:PASB_ubergraph_incidence_graph} shows an example of an ubergraph and its digraph incidence representation. The fundamental vertices $V$ can be identified as the only ones having zero in-degree.
\begin{figure}[h!]
	\begin{center}
        \subfloat{\includegraphics[width=0.42\linewidth]{./fig/Azteca_example/4_node_Azteca_v_labels.png}
		}
        \subfloat{\includegraphics[width=0.5\linewidth]{./fig/Azteca_example/incidence_4_node_Azteca.png}}
        \caption{A 4-node subgraph of the coffee agroecosystem model of \cite{GOLUBSKI2016344} (left) and its digraph incidence representation (right). Colours map the depth of an uberedge, and match those used in the original paper. Fundamental vertices (depth-zero) are green, depth-one edges are black, depth-two blue and depth-three edge red.} 
    	\label{fig:PASB_ubergraph_incidence_graph_appendix}
    \end{center}
    \end{figure}

 \madd{Paths from fundamental vertices to an uberedge define their role in it. Operationally, passing through a vertex on the path adds a bracket around objects pointing to this vertex in the fundamental form of an uberedge.} Physically, these layers represent also the interaction order and its natural strength, estimated by dimensional analysis.

Such a representation is more complicated than the straightforward ubergraph definition. However, it might be used to translate any analysis or algorithm defined for ubergraphs to one performed on digraphs.

\subsection{Classes of hyper- and uberedges}\label{sec:classes_of_hyper_uber}
Hyperedges of differing cardinality can be expected to be of different physical and causal nature. The same applies to uberedges which in addition differ by the way they are composed of fundamental vertices. This leads us to define classes of hyper-/uberedges $\mathcal{\tE}_e$ as the sets of existing edges that could be obtained from a given edge $e$ by renaming the fundamental vertices. 

\begin{definition}\label{def:classes_of_hyper_uber}
Edges $e$ and $e'$ belong to the same class $\mathcal{E}_e$ $\iff \exists$ a \madd{map} $\Pi: V\rightarrow V$, such that $e(V)=e'(\Pi(V))$.
\end{definition}
In a way, an uberedge defines its class through the the way the brackets are placed in its fundamental form. Def.\ref{def:classes_of_hyper_uber} partitions the set of uberedges into equivalence classes.
\begin{plain}
	Let us consider an ubergraph $U=(\{u,v,w\},\{e_1=(u,(v,w)), e_2=((u,v),w),$ $e_3=(w,(u,v)),e_4=(u,(v,v))\})$. The edge set consists of two classes:
	$\mathcal{E}_1=\{e_1,e_3,e_4\}$ and $\mathcal{E}_2=\{e_2\}$.
\end{plain}

\subsection{Ecograph - flows of matter and their functional dependence}
\begin{definition}
\label{def: FoodWeb}
\textbf{A food web} is a connected weighted digraph $\mathcal{G} = (V,E,\phi, x)$ with flows $\phi: E \rightarrow \mathbb{R}_+ $ and biomasses $x: V \rightarrow \mathbb{R}_+ $. Vertices $V$ are subdivided into living, and non-living (detrital). 
\begin{equation*}
    V=\{v_i: i\in I_L\} \cup \{v_i: i\in I_{nL}\}, 
\end{equation*}
where $I_L=\left\{1,\ldots,l\right\}$, $I_{nL}=\left\{l+1,\ldots,n \right\}$ and $I=I_L\cup I_{nL}$.
\end{definition}

\madd{We propose an explicit structure to represent ecosystem interactions, which we call an ecograph. Ubergraphs can model all the examples from ecological literature without loss of information. However, a less general object suffices to represent them.  
All the reported ecological examples contain food web flows (or abundance changes) and interactions in which one vertex influences other interactions.}
\begin{definition}[Alternative Definition]
An ecograph is an ordered triple $\mathcal{G} = (V,U, \phi)$ where $V=\{v_i : i \in I \}$ is the set of fundamental vertices, $U=\{ u_j : j \in J\}$ is the set of uberedges such that an uberedge $u_j$ can consist of either any number of fundamental vertices (i.e. $ u_j = \{v_i, \ldots, v_j \}$) or some fundamental vertices and only one uberedge (i.e. $ u_j = \{v_i, \ldots, v_j, u_k \}$). The weights of the uberedges are assigned via the function $\phi$. 
\end{definition}

\badd{A vertex in a food web represents a single species participating in trophic relationships or a resource supplying energy or matter to the ecosystem. However, representing higher-order interactions might necessitate the inclusion of other vertices which do not serve either of these roles, although they do influence them. Such vertices could represent both biotic and abiotic external factors, e.g. luminosity, temperature or pesticide presence~\cite{Cervantes_2021, Polatto2014}. }

\badd{Since the interacitons involving these vertices are of different kind, they could be represented in a separate way. An approach similar to multilayer networks could be considered, as one of their common purposes is representation of multiple interaction types, including non-trophic ones~\cite{Kefi2016}. A simpler approach would be to incorporate the external factors as vertices on a par with those involved in trophic relationships. Any of these special vertices could be involved in multiple uberedges, although each of them would consist only of said vertex and a single edge representing the influenced interaction. This means the depth of any uberedge involving this vertex would be equal to at least $2$. 
}
\section{Backup}\label{backup}
\subsubsection{Weighted multilayer digraph}
Let us change the classical definition of weighted multilayer  digraph $\mG = (V,V_L,E,\phi)$, \cite[Sec.~2.1]{KivArena2014}, into one that is slightly more complicated, but allows for easy generalisation. Let us denote by $\mathcal{P}^*(\cdot)$ a power set in which an empty set is excluded.

\begin{definition}
    \label{def:WeightedDigraph}
     \textbf{A weighted multilayer digraph} is an ordered tuple $\mG= (V,V_L,E,\phi)$ where
    \begin{enumerate}
        \item $V = \{v_i\; | \; i \in I\}$, $|I|=n$, is a set of vertices;
        \item $L=\{L_j\subset V\; | \; j \in J\}$, $|J|=d$, is a set of layers;
        \item $E = \{(e_k^{\tin},e_k^{\out})\; | \; k \in K\}\subseteq \mathcal{P}^*(V) \times \mathcal{P}^*(V)$, $|K|=m$,   such that for any $e=(e^{\tin},e^{\out})\in E$
    \begin{equation}\label{eq:edge}
        |e^{\tin}|=|e^{\out}|=1,
        \end{equation}
        is a set of edges;
        \item function $\phi: E \xrightarrow{} \mathbb{R}$ assigns a weight to each edge.
    \end{enumerate} 
\end{definition}


\aadd{We can easily note that in this approach one vertex can belong to arbitrary many layers. If $V_L$ is a partition of $V$ into nonempty subsets then following the standard nomenclature we say that $\mathcal{G}$ is \textbf{layer-disjoint}. If there exists exactly one layer we call it a \textbf{weighted digraph} and denote by $\mG= (V,\EG,\phi)$.}

\madd{An edge $(e^{\tin},e^{\out})\in E$ represents a connection from its tail $e^{\tin}$ to its head $e^{\out}$.} Following standard definitions edges that stay within a single layer are called \textbf{intra-layer}, while those which cross layers are called \textbf{inter-layer} edges. \textcolor{red}{Consequently we define an intra-layer graph, an inter-layer graph and a coupling graph. }
\al{To be added.}

\madd{Let us summarise a few standard matrices that characterise digraphs. 
The interactions in a multilayer digraph can be encoded in its incoming and outgoing adjacency tensors $\adji, \adjo$ with indices $i,j\in I$ and $l,k\in J$.}
\begin{multicols}{2}
    \begin{equation}
        \mathbb{A}^{\tin}_{ij;lk}=\phi( (v_i^l, v_j^{k}) ),
    \end{equation}
\break
    \begin{equation}
        \mathbb{A}^{\out}_{ij;lk}=\phi( (v_j^l, v_i^{k}) ),
    \end{equation}
\end{multicols}
They describe the incoming and outgoing interactions of vertices $v_i^l, v_j^k\in V$ belonging to the respective layers $L_l$ and $L_k$, with $l,k \in J$.
\madd{Supra-adjacency matrices flatten the layer dimension, reducing the number of indices. They treat the multilayer digraph $(V,V_L,E,\phi)$ as a digraph $(V \times V_L, E,\phi)$. If $d_i$ denotes the number of vertices in layers below layer $i$,
\begin{multicols}{2}
    \begin{equation}
        \adji_{d_i+l\, d_j+k}=\mathbb{A}^{\tin}_{ij;lk},
    \end{equation}
\break
    \begin{equation}
        \adjo_{d_i+l\, d_j+k}=\mathbb{A}^{\out}_{ij;lk},
    \end{equation}
\end{multicols}

Let $D=\sum_{i=1}^d|V_i|$ and $D_j=\sum_{i=1}^{j-1}|V_i|$. $ D\times D$ matrix $\mathbb{A}=(\mathbb{A}^{ij})_{i,j\in J}$ is supra-adjacency matrix if it is a block matrix where for any fixed $i,j\in J$ $\mathbb{A}^{ij}$ is $|V_i|\times |V_j|$ matrix such that 
$$
\mathbb{A}=A_{ij}^{lk}
$$
\al{To be corrected.}
}

\madd{A supra-adjacency matrix describes a multilayer digraph as if it were an ordinary digraph, storing the layer information in vertex order. In the rest of the article we drop the layer indices.}

\madd{Incidence matrices $\inci, \inco: V \times E \rightarrow {0,1}$ offer an alternative definition of digraph interactions, by stating which vertices are tails or heads of which edges.}
\begin{multicols}{2}
\begin{equation}
    \inci_{ij}=\begin{cases}
        1, \quad \text{if} \quad v_j \in e_{i}^{\mathrm{in}}\\
        0, \quad \text{otherwise}.
    \end{cases}
\end{equation}
\break
\begin{equation}
    \inco_{ij}=\begin{cases}
        1, \quad \text{if} \quad v_j \in e_{i}^{\mathrm{out}} \\
        0, \quad \text{otherwise}.
    \end{cases}
\end{equation}
\end{multicols}
\madd{We store the information about hyperedge weights in a weight matrix $\Phi: E \times E \rightarrow \mathbb{R}$,}
\begin{equation}\label{eq:weight_matrix}
    \Phi(e_i, e_j)=\begin{cases}
    \phi(e_i), \quad \text{if } i = j  \quad\text{and} \quad e_i \in E \\
    1, \quad \text{if } i = j \quad \text{and} \quad e_i \in V.
    \end{cases}
\end{equation}
\madd{The adjacency matrix can then be computed as $(\inco)^T \Phi \inci$.}

\al{We have to choose a notation $I, I^{\HH}$?}
\mat{We have $I=I^{\HH}$, while for $I^U$ index i runs over $V\cup \EU$. $\leftarrow$ that was a good premonition of incidence matrix issue}


\section{Additional copy}
\subsubsection{Weighted multilayer dihypergraph}\label{sec:H_graph}
One can also allow each edge to have more than one head and one tail. A \textbf{weighted multilayer dihypergraph} $\mH = (V,V_L,\EH,\phi)$ satisfies all conditions of Definition \ref{def:WeightedDigraph} except \eqref{eq:edge}. In order to distinguish elements from a set $\EG$ from those in $\EH$ we call the latter hyperedges. This notion has been already used in ecological context in the undirected version, see \cite{Bretto2013}.

Let us summarise standard objects defined for dihypergraphs that will be generalised in the next subsection. Unlike in a digraph, there may be several different hyperedges all containing the same vertex $v_i$ as a head and the same $v_j$ as a tail. Thus, there is no one-to-one correspondence between a dihypergraph and its adjacency matrix $A=(A_{ij})_{i,j\in I}$~\cite{BATTISTON20201},
    \begin{equation}
        A_{ij}=\sum_{\left\{e:\,\, v_i \in e^{\mathrm{out}}, \, v_j \in e^{\mathrm{in}}\right\}} \phi(e).
    \end{equation}
\aadd{In the theory of hypergraphs there are some generalisations know as adjacency tensor, degree normalized k-adjacency tensor, eigenvalues normalized k-adjacency tensor etc. To avoid this ambiguity it the paper we use the notation related to incidence matrix.}     
\al{We have two topics to understand.\newline
1. What is the relation between incidence matrix ($\mathcal{I}$), adjacency matrix ($A$) and Laplacian $\mathcal{L}$ in the theory of dihypergraphs (with positive weights).
2. How to combine all notions with arbitrary sign weights.}

\al{In the unnormalised case (unnormalised Laplacian) we have, \cite{Mugnolo}:
\begin{equation}
\mathcal{L}=\mathcal{I}\mathcal{I}^T=D-A
\end{equation}
In the normalised graph case we have, \cite{Li2012Dilaplacian}:
\begin{equation}
\mathcal{L}=D^{-\frac{1}{2}}(D-A)D^{\-\frac{1}{2}}.
\end{equation}
In the unnormalised digraph case we have, \cite{Boley2016}
\begin{equation}
\mathcal{L}=D^{out}-A
\end{equation}
In the normalised digraph case we have [broken Latex equation, commented out], 
\cite{Li2012Dilaplacian}: }

On the contrary, the \emph{incidence matrix} $I^{\HH}=(I_{ij}^{\HH})_{i\in I\,j\in K}$ uniquely determines the hypergraph,
\begin{equation}\label{eq:hypergraph_incidence}
    I_{ij}^{\HH}=\begin{cases}
        \phi(e_{j}), \quad \text{if} \quad v_i \in e_{j}^{\mathrm{in}} \\
        -\phi(e_{j}), \quad \text{if} \quad v_i \in e_{j}^{\mathrm{out}} \\
        0, \quad \text{otherwise}.
    \end{cases}
\end{equation}
\mat{when checking direction convention, we go with $A_{ij}$, from $j$ to $i$, I had a problem with incidence like above. Formulas like $I^T I$ do not produce anything reasonable for directed hypergraphs - you could go from tail to tail, or head to head.
It seems we do need two incidence matrices: 'in' and 'out':
$I_{in}=I[\text{entries}>0]$,$I_{out}=I[\text{entries}<0]$, then $A=-I_{out}^T I_{in}$}

\al{The matrix that you define below is called signless incidence matrix ref: Mugnolo eq. (2.3) but for $V\times V$. In my opinion $I^{\mathcal{U}}$ should be defined on $V_{{\mathcal{U}}}$. So for hypergraph we obtain classical def.}

\begin{figure}[h!]
	\begin{center}
        \includegraphics[width=\linewidth]{./fig/Laying_out_motivation.jpg}
        \caption{Laying out our motivation} 
    	\label{motivation}
    \end{center}
    \end{figure}

    \begin{figure}[h!]
	\begin{center}
        \includegraphics[width=\linewidth]{./fig/Narrative.jpg}
        \caption{Narrative ideas} 
    	\label{motivation}
    \end{center}
    \end{figure}